%% file: main.tex
\documentclass[12pt]{article}
\usepackage[a4paper, margin=1in]{geometry} 
\usepackage{setspace}
\usepackage[version=3]{mhchem} 
\usepackage{amsmath}
\usepackage{ulem}
\usepackage{authblk}
\usepackage{graphicx}
\usepackage{cite}
\usepackage[table]{xcolor}
\usepackage[utf8]{inputenc}
\usepackage{etoolbox}
\usepackage[T1]{fontenc}
\usepackage[colorlinks, linkcolor=blue,citecolor=blue, urlcolor=blue,bookmarks=true]{hyperref} 

\hypersetup{
colorlinks = true, 
urlcolor   = black,
linkcolor  = black, 
citecolor  = black 
}

\usepackage{xr}
\usepackage{xcite}
\makeatletter
\newcommand*{\addFileDependency}[1]{
\typeout{(#1)}
\@addtofilelist{#1}
\IfFileExists{#1}{}{\typeout{No file #1.}}
}\makeatother

\definecolor{darkgreen}{rgb}{0.0, 0.42, 0.24}

\newcommand*\revision[1]{\color{black}}

\definecolor{hcyan}{rgb}{0.0, 0.8, 0.8}
\definecolor{dorange}{rgb}{0.8, 0.4, 0.0}

\title{
Reproducibility of fixed-node diffusion Monte Carlo across diverse community codes: The case of water-methane dimer
}

\author[1]{Flaviano Della Pia$^*$}
\author[1]{Benjamin X. Shi$^*$}
\author[2,3]{Yasmine S. Al-Hamdani}
\author[2,3]{Dario Alfè}
\author[4]{Tyler A. Anderson}
\author[5]{Matteo Barborini}
\author[6]{Anouar Benali}
\author[7]{Michele Casula}
\author[8]{Neil D. Drummond}
\author[9]{Mat\'{u}\v{s} Dubeck\'{y}}
\author[10]{Claudia Filippi}
\author[11]{Paul R. C. Kent}
\author[12]{Jaron T. Krogel}
\author[13]{Pablo López Ríos}
\author[14]{Arne Lüchow}
\author[6]{Ye Luo}
\author[1]{Angelos Michaelides}
\author[15]{Lubos Mitas}
\author[16]{Kousuke Nakano}
\author[17]{Richard J. Needs}
\author[18]{Manolo C. Per}
\author[19]{Anthony Scemama}
\author[14]{Jil Schultze}
\author[10]{Ravindra Shinde}
\author[10]{Emiel Slootman}
\author[20]{Sandro Sorella$^\ddagger$}
\author[21]{Alexandre Tkatchenko}
\author[22]{Mike Towler}
\author[4]{C. J. Umrigar}
\author[23]{Lucas K. Wagner}
\author[24]{William A. Wheeler}
\author[25]{Haihan Zhou}
\author[2,3]{Andrea Zen$^\dagger$}

\affil[1]{Yusuf Hamied Department of Chemistry, University of Cambridge, Cambridge CB2 1EW, United Kingdom}
\affil[2]{Department of Earth Sciences, University College London, London WC1E 6BT, United Kingdom}
\affil[3]{Dipartimento di Fisica Ettore Pancini, Università di Napoli Federico II, Monte S. Angelo, I-80126 Napoli, Italy}
\affil[4]{Laboratory of Atomic and Solid State Physics, Cornell University, Ithaca, New York 14853, United States of America}
\affil[5]{HPC Platform, University of Luxembourg, L-4365 Esch-sur-Alzette, Luxembourg}
\affil[6]{Computational Science Division, Argonne National Laboratory, Lemont, IL, United States of America}
\affil[7]{Institut de Minéralogie, de Physique des Matériaux et de Cosmochimie (IMPMC), Sorbonne Université, CNRS UMR 7590, MNHN, 4 Place Jussieu, 75252 Paris, France}
\affil[8]{Department of Physics, Lancaster University, Lancaster LA1 4YB, United Kingdom}
\affil[9]{Department of Physics, Faculty of Science, University of Ostrava, 30. Dubna 22, 701~03 Ostrava, Czech Republic}
\affil[10]{MESA+ Institute for Nanotechnology, University of Twente, Enschede 7500 AE, The Netherlands}
\affil[11]{Computational Sciences and Engineering Division, Oak Ridge National Laboratory, Oak Ridge, Tennessee 37831, United States of America}
\affil[12]{Materials Science and Technology Division, Oak Ridge National Laboratory, Oak Ridge, Tennessee 37831, United States of America}
\affil[13]{Max Planck Institute for Solid State Research, Heisenbergstr. 1, 70569 Stuttgart, Germany}
\affil[14]{Institute of Physical Chemistry, RWTH Aachen University, Landoltweg 2, 52074 Aachen, Germany}
\affil[15]{Department of Physics, North Carolina State University, Raleigh, North Carolina 27695-8202, United States of America}
\affil[16]{Center for Basic Research on Materials, National Institute for Materials Science (NIMS), 1-2-1 Sengen, Tsukuba, Ibaraki 305-0047, Japan}
\affil[17]{Theory of Condensed Matter Group, Cavendish Laboratory, J. J. Thomson Avenue, Cambridge CB3 0HE, United Kingdom}
\affil[18]{CSIRO Data61, Clayton, VIC 3168, Australia}
\affil[19]{Laboratoire de Chimie et Physique Quantiques (UMR 5626), Université de Toulouse, CNRS, UPS, 31062 Toulouse, France}
\affil[20]{International School for Advanced Studies, SISSA, 34136 Trieste, Italy}
\affil[21]{Department of Physics and Materials Science, University of Luxembourg, L-1511 Luxembourg City, Luxembourg}
\affil[22]{The Apuan Alps Centre for Physics, Vallico Sotto, Italy}
\affil[23]{Department of Physics, University of Illinois at Urbana-Champaign, Urbana, IL, 61801, United States of America}
\affil[24]{Department of Materials Science and Engineering, University of Illinois at Urbana-Champaign, Urbana, IL 61801, United States of America}
\affil[25]{Department of Physics, NC State University, Raleigh, NC, 27606, United States of America}

\begin{document}
\footnotetext[1]{These authors contributed equally. All others, except for the corresponding author, are ordered alphabetically.}
\footnotetext[2]{Corresponding author email: andrea.zen@unina.it}
\footnotetext[3]{Deceased, 10 August 2022.}
\footnotetext[4]{This manuscript has been authored by UT-Battelle, LLC under Contract No. DE-AC05-00OR22725 with the U.S. Department of Energy. The United States Government retains and the publisher, by accepting the article for publication, acknowledges that the United States Government retains a non-exclusive, paid-up, irrevocable, worldwide license to publish or reproduce the published form of this manuscript, or allow others to do so, for United States Government purposes. The Department of Energy will provide public access to these results of federally sponsored research in accordance with the DOE Public Access Plan (https://www.energy.gov/doe-public-access-plan).} 

\maketitle

\begin{abstract}
Fixed-node diffusion quantum Monte Carlo (FN-DMC) is a widely-trusted many-body method for solving the Schr\"{o}dinger equation, known for its reliable predictions of material and molecular properties.
Furthermore, its excellent scalability with system complexity and near-perfect utilization of computational power makes FN-DMC ideally positioned to leverage new advances in computing to address increasingly complex scientific problems.
%
Even though the method is widely used as a computational gold standard,
reproducibility across the numerous FN-DMC code implementations has yet to be demonstrated.
This difficulty stems from the diverse array of DMC algorithms and trial wave functions, compounded by the method's inherent stochastic nature.
This study represents a community-wide effort to assess the reproducibility of the method, affirming that: Yes, FN-DMC is reproducible (when handled with care). 
Using the water-methane dimer as the canonical test case, we compare results from eleven different FN-DMC codes and show that the approximations to treat the non-locality of pseudopotentials are the primary source of the discrepancies between them.
In particular, we demonstrate that, for the same choice of determinantal component in the trial wave function, reliable and reproducible predictions can be achieved by employing the T-move (TM), the determinant locality approximation (DLA), or the determinant T-move (DTM) schemes, while the older locality approximation (LA) leads to considerable variability in results.
%
{\revision{}
These findings demonstrate that, with appropriate choices of algorithmic details, fixed-node DMC is reproducible across diverse community codes—highlighting the maturity and robustness of the method as a tool for open and reliable computational science.
}
\end{abstract}


\section{Introduction}\label{sec:introduction}

The credibility of a scientific result hinges on its reproducibility; independent and equivalent experiments should lead to the same conclusion.
Achieving reproducibility is, however, not easy.
There are several historical examples from both social and natural sciences\cite{reproducibility_psychology, reproducibility_bug,reproducibility_economics,reproducibility_social_science_2018} that have served to illustrate its challenges, and substantial ongoing effort is dedicated to addressing this so-called ``reproducibility crisis''~\cite{reproducibility_crisis,reproducibility_social_science}.
The problem of reproducibility is particularly pertinent within computational experiments in the hard sciences, where
different computational codes should ideally lead to the same prediction.
Nonetheless, reproducibility can be compromised by small algorithmic differences, undocumented approximations, and undetected bugs in the simulation software or its dependencies (numerical libraries, compilers etc.). 
Determining the source of discrepancies can be difficult, e.g., due to restricted source code availability\cite{reproducibility_bioinformatics,reproducibility_bug,reproducibility_guide,reproducibility_open_source}.

Here, we consider reproducibility in the context of the many-electron Schr\"{o}dinger equation\cite{schrodinger26}, fundamental to the quantum mechanical description of matter, and its countless applications to physics, chemistry, biology, engineering, and materials science.
In this context, the topic of reproducibility has been recently addressed\cite{reproducibility_dft,nature_reproducibility_dft} in two seminal papers for density functional theory (DFT) -- the work-horse of materials science.
However, despite its widespread success, DFT often falls short of providing the necessary quantitative, and sometimes qualitative, description of key complex systems.
Fortunately, advances in hardware, algorithms, and fundamental theories are paving the way for the routine application of methods beyond the accuracy of DFT\@.
The scope of these methods has recently broadened significantly beyond simple benchmarks, towards an extensive description of molecules, surfaces and condensed phases\cite{hellgren_rpa_crystalline_2021,dmcice13_dellapia,X23_dellapia,shi2023going,Shi_surface_2023,Zen_PNAS_18} that can include complex dynamics facilitated by machine learning potentials\cite{oneill2024pair,tirelli_mlp_qmc_hydrogen,tahir2024_water_rpa,Janos_couples_cluster_water_2022,schran_coupled_cluster_water_clusters_2020,David_Ceperley_MLP_2024,Tenti_MLP_Hugoniot_2024,Nakano_VMC_force_2024}.

{\revision{}
Several quantum many-body approaches have been developed as alternatives to DFT for electronic structure calculations, particularly in systems where strong correlation or high accuracy is essential. Methods such as GW~\cite{GW_PhysRevB_1986, GW_RevModPhys_2002}, dynamical mean-field theory (DMFT)~\cite{DMFT_RevModPhys_1996, DMFT_RevModPhys_2006}, coupled cluster theory~\cite{CC_solids_Berkelbach, CC_matsci_Gruneis}, and auxiliary-field quantum Monte Carlo (AFQMC)~\cite{AFQMC_SUGIYAMA1986, AFQMC_Zhang1997, AFQMC_Zhang2003, AFQMC_Motta2018} have been successfully applied to both molecular and condensed-phase systems. 
More recently, full configuration interaction quantum Monte Carlo (FCIQMC)~\cite{FCIQMC, Booth2013Nature} and neural network-based quantum Monte Carlo~\cite{Carleo2017Science, Pfau2020FermiNet, Hermann2020PauliNet} methods have also gained attention. However, among the quantum Monte Carlo methods, real-space fixed-node diffusion Monte Carlo (FN-DMC) remains the most widely used approach in materials science and quantum chemistry, offering a compelling balance between accuracy, scalability, and methodological maturity. Its use of explicitly correlated many-body wave functions and its favorable scaling with system size make it particularly attractive for benchmarking and systematic studies. For these reasons, this work focuses on FN-DMC and its reproducibility across a variety of independently developed community codes.
}

FN-DMC is an accurate state-of-the-art computational approach for solving the Schr\"{o}dinger equation for a variety of systems, including molecules, solids, and surfaces. 
This method obtains the electronic ground-state by performing an imaginary-time evolution from a starting trial wave function $\Psi_\mathrm{T} (\mathbf{R})$. Within the Born-Oppenheimer approximation, $\mathbf{R}$ consists of the real space positions of all the electrons.
Typically, $\Psi_\mathrm{T} (\mathbf{R})$ is the product of an antisymmetric function (e.g., a Slater determinant or a sum of Slater determinants\cite{slater1929}) and a symmetric, positive function, called the Jastrow factor\cite{jastrow1955}. The Jastrow factor is explicitly dependent on electron-electron and electron-nucleus distances, and able to directly capture a significant fraction of the electronic correlation.

The FN-DMC projection is achieved with an ensemble of electron configurations, known as walkers, which evolve according to the imaginary-time Green function\cite{grimm1971}, yielding a drift-diffusion process over discrete imaginary time steps, $\tau$, to stochastically sample the ground-state wave function; the stochastic uncertainty is then inversely proportional to the square root of the
number of samples.
The main approximation in FN-DMC is that
the fixed-node wave function is constrained to have the same nodal surface as $\Psi_\mathrm{T} (\mathbf{R})$, in order to avoid the so-called {\it fermion sign problem}.~\cite{troyer2005} This introduces a variational error in the computed ground state energy. For single-reference systems, this error is typically small even for simple single determinant trial wave functions built from DFT.
FN-DMC exhibits almost perfect efficiency on modern supercomputers\cite{CASINO,turborvb,qmcpack_1} and a cubic scaling per Monte Carlo step with system size\cite{foulkes_qmc}, making it often the only computationally affordable method beyond DFT for treating large condensed phase systems with more than 100 atoms.
Over time, numerous algorithmic improvements have enhanced the accuracy, efficiency, and stability of FN-DMC\@. 
These advances have enabled the successful application of FN-DMC to a wide array of problems across the natural sciences,
including the calculation of the energies of condensed phases and large
molecules\cite{Zen_PNAS_18,dmcice13_dellapia,X23_dellapia,Alhamdani_large_molecules_2021,luo_phase_2016,santana_cohesive_2016}, the binding of molecules on
surfaces\cite{Shi_surface_2023,H2_adsorption_AlHamdani,luo_adsoprtion_2019,benali_graphenylene,Umrigar_NO2_CNT_2008,Alhamdani_water_hBN_2015,Alhamdani_water_carbon},
phase
diagrams\cite{chen_hydrogen_DMC_2014,Drummond_H_phase_diagram,Mauri_H_phase_diagram,LopezRios_PRL_hydrogen,tirelli_mlp_qmc_hydrogen,qmc_mgo_alfe_2005,alfe_2009_iron,bar+22prb},
reaction barrier heights\cite{barrier_heights_Zhou, Luchow_clusters_reaction_JCP_2001, DMC_barriers_Lucas_2017, bar+12jcp, Per2017_methanol},
spin-polarized uniform electron gas\cite{ueg_prb_lopezrios}, two-dimensional electron liquid\cite{drummond_24_electron_liquid},
lithium  systems\cite{Lithium_Mitas_2015}, electronic and optical properties of delafossites\cite{kent_agnio2_2024}, defect
formation energies\cite{defects_ertekin_2017,defects_luo_2023}, calculation of energy
derivatives\cite{moroni2014,vanrhijn2022,slootman2024}, radical stabilization energies\cite{per_radical_2020}, excited
states\cite{ZimTouZhaMusUmr-JCP-09,bar+12jctc,bar+15jctc,scemama_2018,scemama_2019,dash_2019,cuzzocrea_2022,neuscamman_excited_states_2019,scemama_double_excitation_2022,kent_excitations_2023},
training of quantum machine learning models\cite{QML_benali_2023}, electron-positron interactions\cite{barborini_positron_2022},
polymorphism\cite{Umrigar_mbenzyne_2008,ferlat2019van,BN_polymorphs_Nakano_2022}, electronic band gaps\cite{bandgap_dubecky},
Landau-level mixing in quantum dots\cite{JeoGucUmrJai-PRB-05}, localization in quantum dots and quantum
wires\cite{GhoGucUmrUllBar-NP-06,GhoGucUmrUllBar-PRB-07,GucGhoUmrBar-PRB-08,GucUmrJiaBar-PRB-09},
nearly exact density functional quantities\cite{FilGonUmr-Review-96,SavUmrGon-CPL-98} and more.
Recent progress in the use of neural networks as trial wave functions for FN-DMC\cite{NN_1,NN_2,NN_3} has served to boost its accuracy and potential future uptake even further.

%
There are numerous QMC codes currently used for research, many of which have been under development for over a decade. Each makes somewhat different algorithmic and implementation choices, such as the use of different Jastrow factors and methods for evaluating single-particle orbitals.
{\revision{} 
In this study, we compare eleven such codes and provide  
details of the algorithmic and implementation choices in Section S8 of the Supplementary Information (SI).
}
%
This diversity raises open questions on the reproducibility of FN-DMC\@.
%
If FN-DMC is to be widely accepted as a highly accurate reference method, it is important
that consistent results can be obtained from these different FN-DMC codes.
With this goal in mind, the present work represents a collaborative effort among the users and developers of eleven distinct FN-DMC codes, to rigorously assess the reproducibility of FN-DMC\@.

A key obstacle to the reproducibility of FN-DMC comes from the use of non-local pseudopotentials (NLPPs),
which increase the efficiency of the method for systems with heavy atoms.~\cite{ceperley1986,hammond1987,hammond1988}
{\revision{}
While all-electron FN-DMC calculations are possible for light atoms, the computational cost increases steeply with atomic number, scaling approximately as $O(Z^\alpha)$ with $\alpha$ between 5.5 and 6.5, depending on the method details~\cite{ceperley1986, hammond1987, Drummond2005}.  
As a result, pseudopotentials are essential for practical FN-DMC applications involving heavier elements.
}
NLPPs allow one to solve the Schr\"{o}dinger equation solely for the valence electrons, by substituting the full local nuclear potential with a smooth non-local potential near the nuclei.
In general, NLPPs hinder reproducibility in electronic structure methods, as NLPPs constructed in different ways can lead to somewhat different predictions.
NLPPs are a potential source of non-reproducibility in FN-DMC even when the same NLPPs are used,
because non-local pseudopotential operators create an additional sign problem in the projector beyond the one that is always present for fermionic calculations.  To avoid this sign problem, these operators must be ``localized'',\cite{la_mitas} or at least partially localized,\cite{casula_Tmove_2006} on a wave-function. 
A natural choice is to localize them on the trial wave-function $\Psi_\mathrm{T} (\mathbf{R})$, introducing a dependence on both the determinantal and the Jastrow components of the wave function. 
Since the Jastrow factor is different in the different codes and its parameters are stochastically optimized, yielding possible noise and reproducibility issues, some authors choose to localize only on the determinantal component.~\cite{HurChr-JCP-87,hammond1987,FlaSavPre-JCP-92,Caffarel2016,dla_zen} This removes the dependence on the Jastrow factor at the cost of losing the desirable property that the treatment of the pseudopotential is exact in the limit of exact $\Psi_\mathrm{T}$. 
%
To summarize, there are currently four localization schemes: the locality approximation (LA)\cite{Chr-JCP-91,la_mitas}, the T-move (TM) approximation\cite{casula_Tmove_2006,casula_Tmove_LRDMC_2010,AndUmr-JCP-21}, the determinant locality approximation (DLA)\cite{HurChr-JCP-87,hammond1987,FlaSavPre-JCP-92,Caffarel2016,dla_zen}, and the determinant T-move (DTM) approximation\cite{dla_zen}. 
%
These four schemes (LA, TM, DLA and DTM) result in somewhat different projected wave-functions and therefore different total energies of physical systems.

{\revision{}
As computational science matures, reproducibility and transparency are increasingly recognized as critical features of robust methodology. 
FN-DMC, while a powerful and widely used method, has historically lacked comprehensive cross-code validation. 
This work takes a step toward establishing that foundation by systematically comparing the four localization algorithms across eleven FN-DMC codes (named alphabetically): Amolqc, CASINO\cite{CASINO}, CHAMP-EU\cite{CHAMP-EU}, CHAMP-US\cite{CHAMP-US}, CMQMC, PyQMC\cite{pyqmc}, QMC=Chem\cite{qmcchem,qmcchem2}, QMCPACK\cite{qmcpack_1,qmcpack_2}, QMeCha\cite{QMeCha}, QWalk\cite{qwalk}, and TurboRVB\cite{turborvb,TurboGenius}. Different forms of Jastrow factor are necessarily tested as part of this evaluation.
}

{\revision{}
We specifically consider the cases of the total energies of a methane molecule, a water molecule, and a methane-water dimer, and the corresponding interaction energy.
We selected the water–methane dimer as a test case not only for its modest size—which allows tight statistical convergence—but also because it spans two different interaction regimes. 
It involves both weak intermolecular interactions (with a binding of only $ 27~\text{meV}$) and intramolecular energetics, enabling a sensitive probe of algorithmic consistency across codes. 
}
In addition, it is 
a prototype of more complex 
systems such as methane clathrates, important for gas storage and transportation. 
We show that consensus across all eleven codes can be made when utilizing the TM, DLA and DTM approximations, particularly following careful control of the discretized time step.

\begin{figure*}[!htp] 
\centering
    \includegraphics[width=6.in]{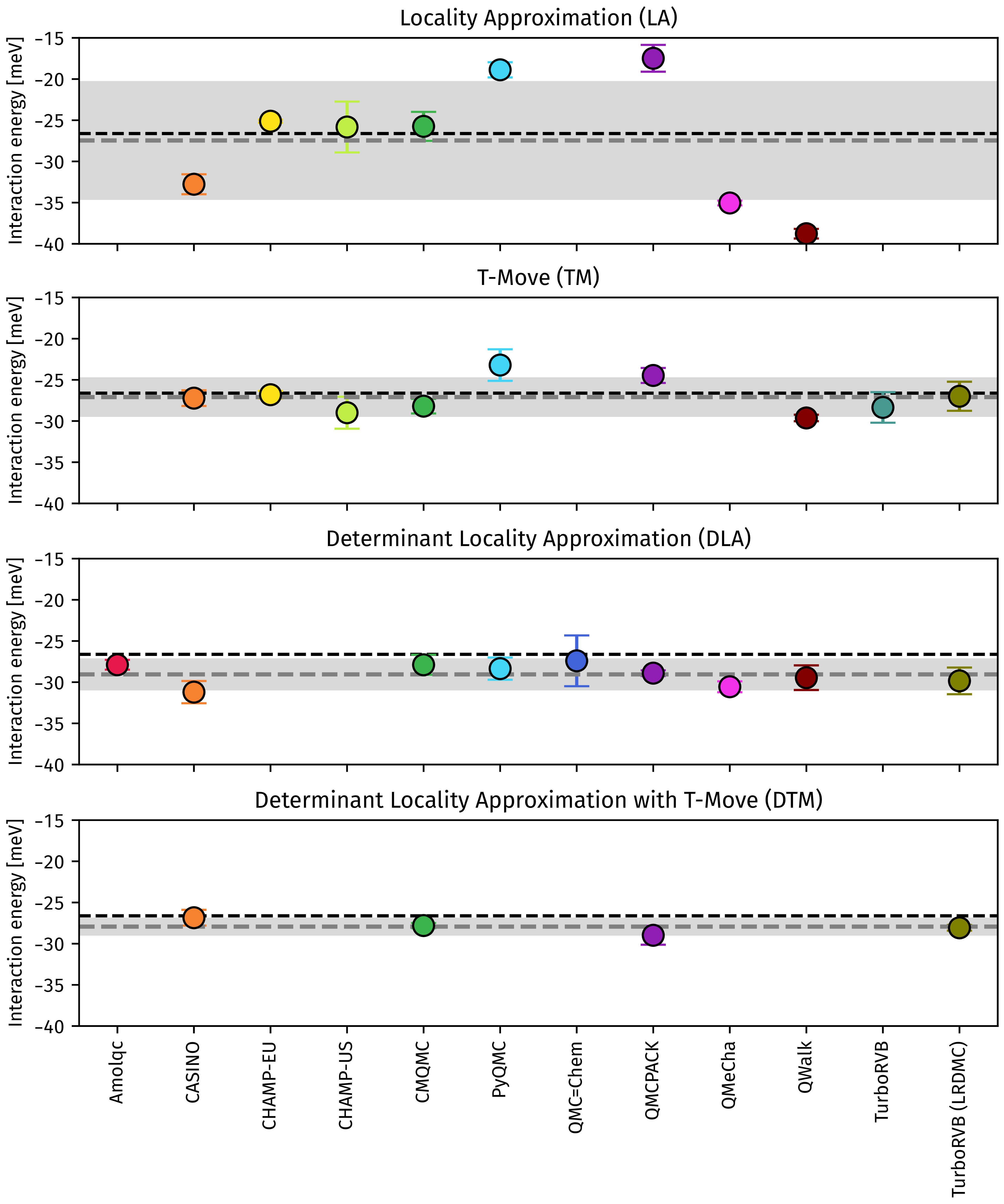} 
    \caption{FN-DMC interaction energy of the methane-water dimer with four different methods. The black dashed horizontal line indicates the reference value of $-27 \text{ meV}$ computed with CCSDT(Q)\@. The gray dashed line is the average among the interaction energies computed with different codes, and the shaded area is its statistical error bar. The energy
    differences between the various codes are much larger when the LA scheme is employed, compared to the narrower energy range obtained with TM, DLA, and DTM\@. 
    The computed averages always match the CCSDT(Q) value within the statistical error bar.}
    \label{fig:method_comparison}
\end{figure*} 

\section{Results and Discussion}\label{sec:results_and_discussion}

First, we compute the interaction energy of the methane-water dimer using the eleven codes for the four different localization schemes (where available).
The interaction energy of the methane-water dimer, 
\begin{equation}\label{eq:int_energy}
        E_\textrm{int} = E[\textrm{methane--water}] - E[\textrm{methane}] -E[\textrm{water}],
\end{equation}
is defined as the difference between the energy of the complex, $E[\textrm{methane--water}]$, minus the sum of the energies of the isolated water $E[\textrm{water}]$ and methane $E[\textrm{methane}]$  monomers 
(see the \textbf{Methods} section for details on the geometries and the DMC simulation set-up).
%
All the interaction energies are extrapolated to the zero time-step limit according to the procedure described in the SI and in Ref.~\cite{note_time_step}.

We note that two results are reported for the TurboRVB code, namely TurboRVB (DMC) and TurboRVB (LRDMC)\@.
TurboRVB (DMC) refers to the standard FN-DMC algorithm with time step discretization and available
with the T-move scheme. 
However, production simulations of FN-projection in TurboRVB are typically performed with the lattice regularized DMC (LRDMC)\cite{LRDMC_1,casula_Tmove_LRDMC_2010}, which is an alternative approach to DMC. In particular, LRDMC is based on a lattice regularization of the many-electron Hamiltonian over a spatial mesh, and the ground state is projected out via the Green function Monte Carlo method\cite{GFMC_1,GFMC_2,GFMC_3}. The zero mesh-size limit of the LRDMC prediction is equivalent to the zero time-step limit  of DMC, and is therefore also included in this work. We also note that the T-move approximation itself comes in four different versions as briefly discussed in the SI but, when presenting the TM results, we will not distinguish between them
because they differ only at finite time step, while we report here the extrapolated values at zero time step, where they are equivalent.

The computed methane-water interaction energies are shown in Fig.\ \ref{fig:method_comparison}.
We plot the FN-DMC interaction energy computed with each code with a colored circle.
In addition, the average among the interaction energies computed with different codes is reported with a gray dashed line, and its statistical error with a shaded gray region.
The average value and its statistical error are computed as the mean value and the standard deviation of the probability distribution reported in Eq.\ \ref{eq:prob_distribution}, discussed later on in the manuscript.
We compare the prediction of FN-DMC to the value computed by coupled cluster theory with single, double, triple, and perturbative quadruple excitations [CCSDT(Q)], which is expected be highly accurate for weak intermolecular interactions\cite{rezac2015} (details of the calculation are reported in Sec.~S3 
of the SI)\@.
Despite using only a single determinant in the trial wave functions and a DFT nodal surface for simplicity, broadly speaking, the FN-DMC is in excellent agreement with CCSDT(Q) (black dashed line).
However, a strikingly large spread of predictions across different codes is obtained when using the LA\@. 
In contrast, the TM, DLA, and DTM methods show a much narrower spread of the interaction energies. 

The data reported in Fig.\ \ref{fig:method_comparison} allows us to estimate a probability distribution of the interaction energy for each analyzed method.
In particular, we write the DMC energy estimated with the code $i$ and the method $\alpha$ ($\alpha=$LA, TM, DLA, DTM) as $E_{\alpha, i}$, and its statistical error bar as $\sigma_{\alpha,i}$.
Following the central limit theorem, we expect each DMC estimate to be distributed according to a normal distribution, with mean $\bar{E}_{\alpha, i}$ and standard deviation $\bar{\sigma}_{\alpha,i}$. 
Since we do not know $\bar{E}_{\alpha, i}$ and $\bar{\sigma}_{\alpha,i}$, we use here the current estimates $E_{\alpha, i}$ and $\sigma_{\alpha,i}$ and define the probability distribution of the energy $E$ for the method $\alpha$ as:
    \begin{equation}\label{eq:prob_distribution}
        P_\alpha(E) = \frac{1}{N_\alpha} \sum_{i \in \text{ codes}} 
         \frac{1}{{\sqrt{2\pi\sigma_{\alpha, i}^2}}} e^{-\frac{(E-E_{\alpha, i})^2}{2 \sigma^2_{\alpha,i}}},
    \end{equation}
where $N_\alpha$ is the number of codes for which the localization method $\alpha$ is evaluated. The mean, $\mu_\alpha$, and the variance, $\sigma_\alpha^2$, of the energy for the distribution $P_\alpha(E)$ are respectively:
\begin{equation}
    \mu_\alpha = \int E P_\alpha(E) \, dE = \frac{1}{N_\alpha} \sum_{i \in \text{codes} } E_{\alpha, i}, 
\end{equation}
and 
\begin{equation}\label{eq:sigmaalpha}
    \sigma^2_\alpha = \int (E-\mu_\alpha)^2 P_\alpha(E) \, dE = \frac{1}{N_\alpha} \sum_{i \in \text{codes} } \sigma^2_{\alpha, i} + \frac{1}{N_\alpha} \sum_{i \in \text{codes}} (E_{\alpha,i} - \mu_\alpha)^2.
\end{equation}
In particular, the variance takes into account both the statistical error bar of each FN-DMC evaluation ($\sigma_{\alpha, i}$) and its
deviation from the mean value ($E_{\alpha, i} - \mu_\alpha$).

\begin{figure}[!ht]
\centering
    \includegraphics[width=3.35in]{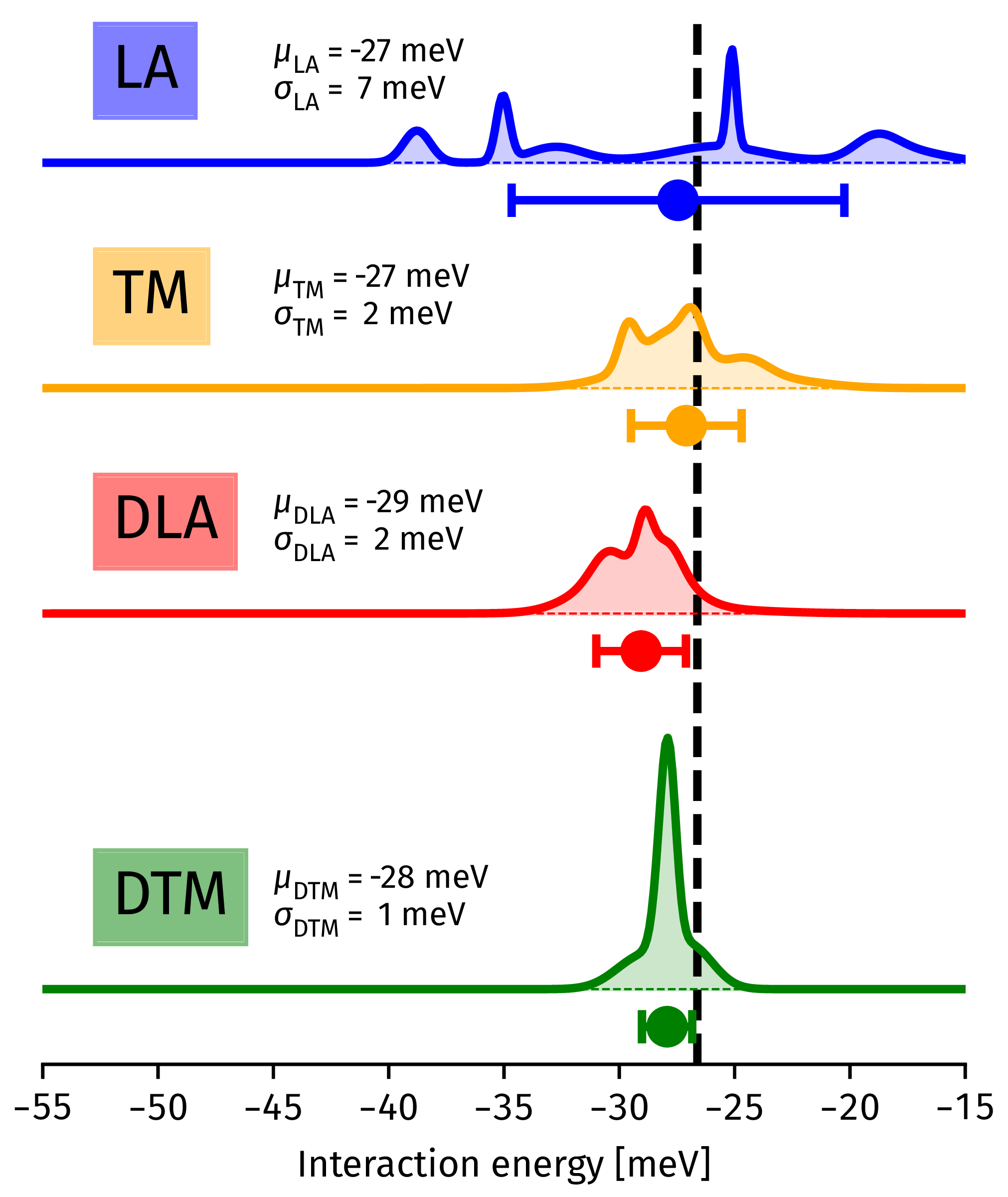}
        \caption{Probability distribution $P_\alpha(E)$ (Eq.\ \ref{eq:prob_distribution}) of the FN-DMC interaction energy of the methane-water dimer for four different schemes for treating NLPPs. The probability distribution for the LA method is spread across a large energy range of ca.\ $25 \text{ meV}$, showing the disagreement among different codes. The probability distribution is instead much narrower when the TM, DLA, and DTM algorithms are employed, implying the agreement on the final estimate of the interaction energy among different codes. The black vertical dashed line indicates the reference value computed with CCSDT(Q)\@.}
    \label{fig:prob_distribution}
\end{figure}

The probability distributions are plotted in Fig.~\ref{fig:prob_distribution}. 
%
%
When the LA is employed, the probability distribution is spread across a large energy range of $25 \text{ meV}$, with a standard deviation of $7\,$meV.
The agreement across different codes significantly improves with the TM, DLA and DTM schemes, with the probability distributions showing a quite localized peak (standard deviation of ca.\ $2\,$meV or less) centered on $-27$~meV, $-29$~meV and $-28\,$meV respectively. 
The DTM scheme gives the narrowest distribution, centered on $-28$~meV, with a standard deviation of ca.\ 1~meV, but since only four out of the eleven codes implemented DTM this is of limited significance.
%
Overall, the analysis of the probability distributions showcases that algorithms more sophisticated than LA need to be employed to guarantee reproducibility among different FN-DMC codes.

%

\begin{figure}[!ht]
\centering
    \includegraphics[width=3.35in]{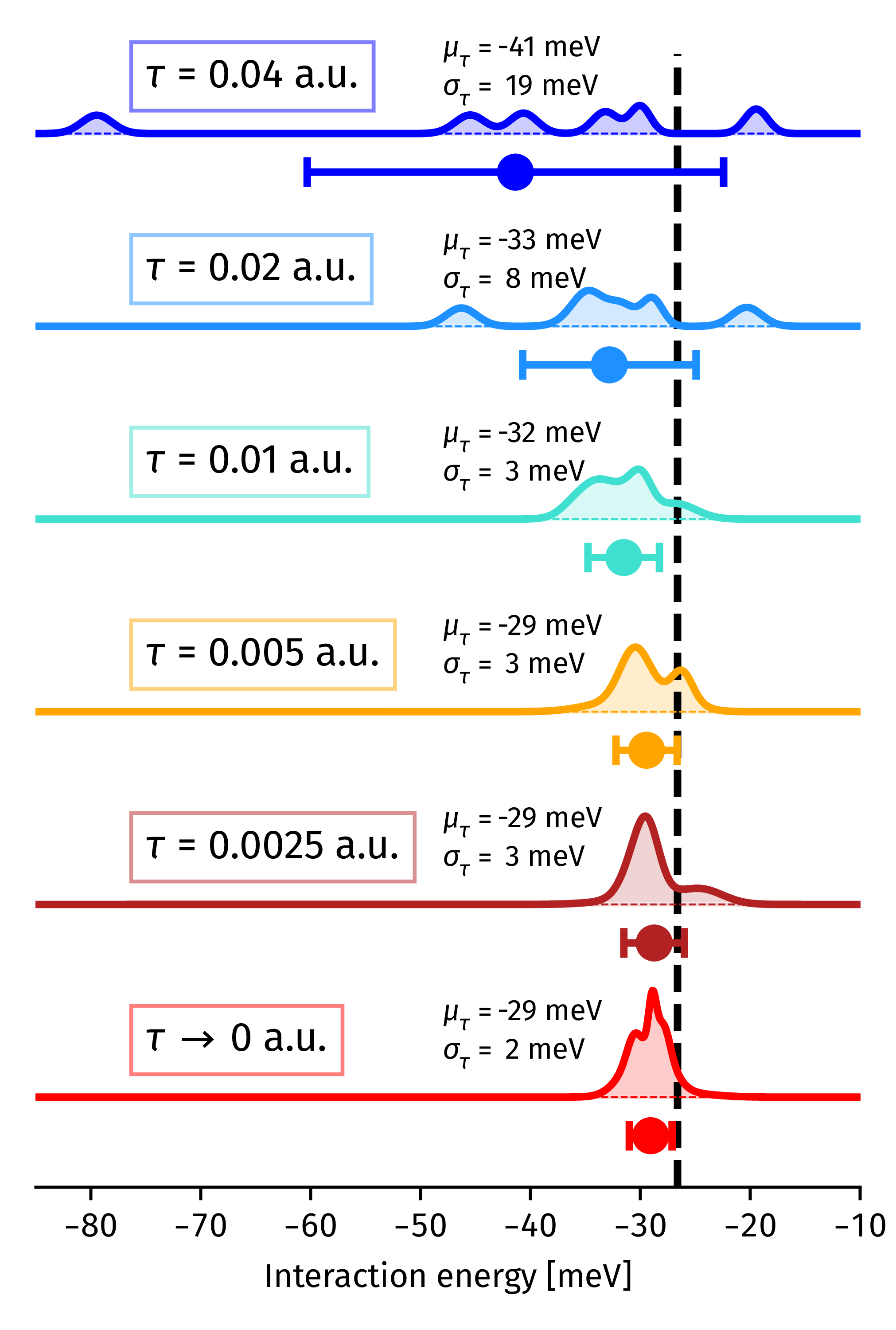}
        \caption{Convergence with respect to the simulation time step of the probability distribution, as defined in Eq.\
        \ref{eq:prob_distribution}, for the DLA\@. The probability distribution is spread over a large energy range of over $20 \text{ meV}$ at
        large time steps ($\tau > 0.01 \text{ a.u.}$), while a narrow distribution is achieved only for the smallest
        time step ($\tau = 0.0025 \text{ a.u.}$). The black vertical dashed line indicates the reference value computed with
        CCSDT(Q)\@.}
    \label{fig:timestep_convergence}
\end{figure}

A key factor in DMC is the convergence with respect to the simulation time step.  The projection is only accurate for sufficiently
small time step, requiring calculations at various time steps $\tau$ to be performed and extrapolated to  the limit $\tau \to 0$.
The required time step depends on both the system being studied and the accuracy of the trial wave function.  
%
%
%
For this reason, we also analyze the dependence of the probability distribution $P_\alpha(E)$ on the simulation time step and report it in Fig.\ \ref{fig:timestep_convergence}. 
In particular, we consider the case of the DLA, for which we have computed the interaction energy with several codes at multiple time steps ($\tau = 0.04, 0.02, 0.01, 0.005, 0.0025 \text{ a.u.}$). 
%
%
We notice that, for a large time step $\tau = 0.04 \text{ a.u.}$, the DLA energy predictions are spread across a large energy range of over $60 \text{ meV}$. 
Decreasing the time step leads to a significant reduction in the distribution's variance. 
At the time step of $\tau = 0.0025 \text{ a.u.}$, the probability distribution becomes very narrow, indicating agreement among different codes. 
We highlight here that the converged time step is system-dependent, and the time step behavior is highly sensitive to different codes and approximations, as shown in the SI\@. 
Therefore, an analysis of the convergence with respect to the simulation's time step is important to achieve a converged and reproducible FN-DMC energy, and a fair comparison across different packages.

\begin{figure}[h!]
\centering
\includegraphics[width=6in]{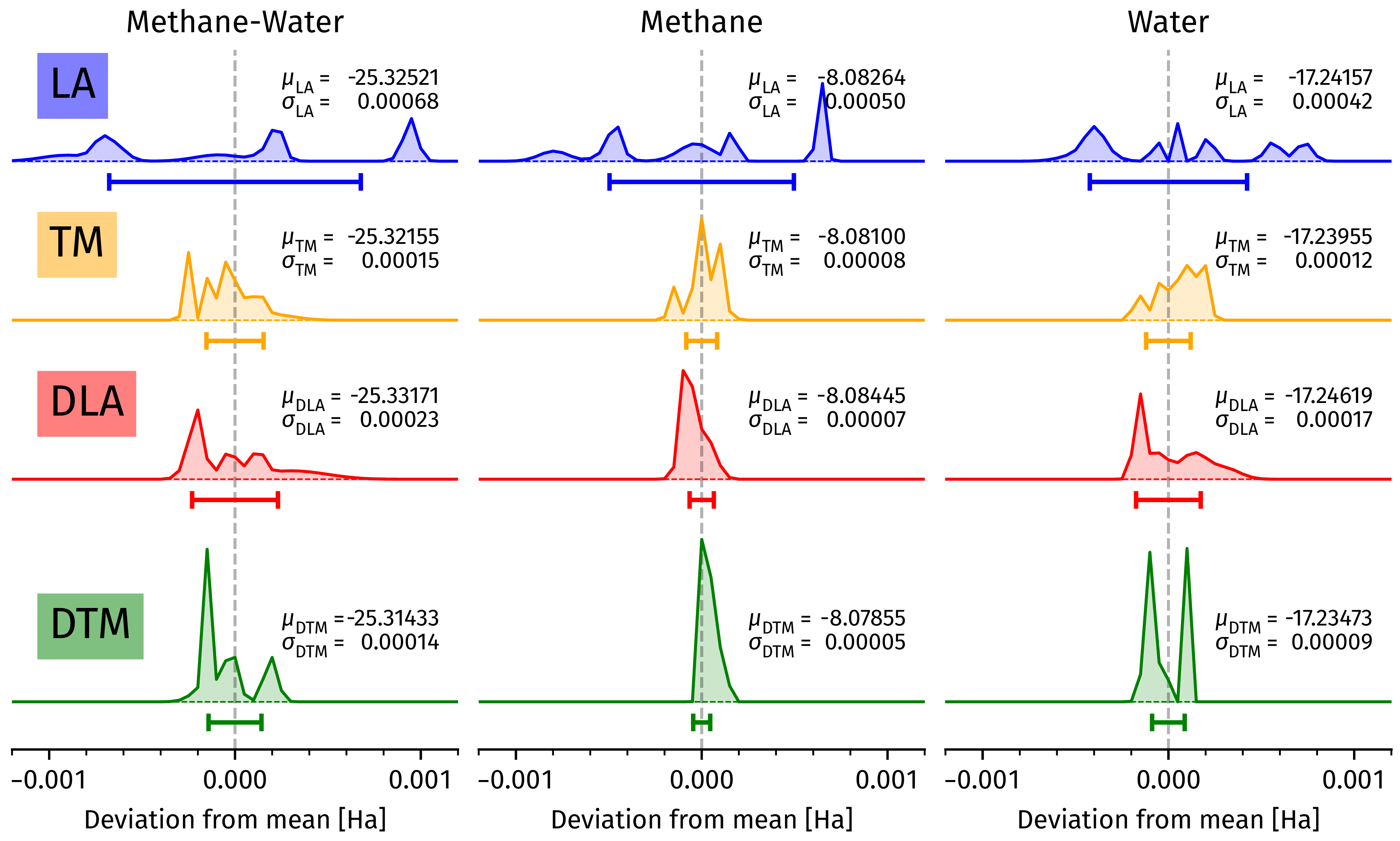}
\caption{\label{fig:totalenergy_distribution}
Probability distribution $P_\alpha(E)$ (Eq.\ \ref{eq:prob_distribution}) of the FN-DMC total energy (Hartree) of the methane-water dimer (left), methane (middle), and water (right), for four different schemes to treat NLPPs. The bars under the distributions indicate the standard deviation.
}
\end{figure}

Finally, we focus on the FN-DMC total energies  of the methane-water dimer and its constituent monomers, which are the fundamental quantities entering the computation of the interaction energy. In Fig.~\ref{fig:totalenergy_distribution}, we report the probability distribution $P_\alpha(E)$ of the total energies extrapolated to zero-time step. As in the case of the interaction energy, we find that the total energies computed in the TM, DLA, and DTM approximations differ much less among the codes than when the LA is employed. Their distributions are  significantly narrower, displaying standard deviations in a range from 2.5 to 10 times smaller than the LA case (e.g., in the water molecule $\sigma_\text{LA}\sim 2.5 \sigma_\text{DLA}$, and in the methane monomer $\sigma_\text{LA}\sim 10 \sigma_\text{DTM}$). 
Moreover, the standard deviations $\sigma_\alpha$s of the TM, DLA and DTM total energy distributions are close to the theoretical minimum allowed by the precision of the performed FN-DMC simulations, as $\sigma_\alpha$s are mostly determined by the stochastic error associated to the FN-DMC energy evaluations (between $10^{-4}$ and $10^{-5}$ Hartree, see SI), so the first term on the right hand side of Eq.~\ref{eq:sigmaalpha}.
This behavior is expected for the DLA and DTM schemes that depend only on the determinant part of the wave functions (identical in all calculations). Remarkably, despite using different Jastrow factors, all codes yield very similar extrapolated total energies even with the TM scheme, which has the desirable property of treating the pseudopotential exactly in the limit of an exact $\Psi_\mathrm{T}$.

%
%

%
%

\section{Methods}

The interaction energy of the methane-water dimer is computed by subtracting the isolated molecule energies 
from the methane-water complex, 
as defined in Eq.~\ref{eq:int_energy}.
The geometry of the dimer (shown in the SI) was obtained from Ref.~[\citenum{ZSGMA}]. The geometries of the monomers are the same as in the dimer.
%
%
In this study, in order to try to achieve consistent results, all eleven codes were required to use the same correlation consistent effective core potential (ccECPs)\cite{ccECP1, ccECP2} and the corresponding triple-zeta basis set (ccECP-ccpVTZ), as well as a Slater-Jastrow wave function with a single Slater determinant whose orbitals are obtained from DFT calculations using the Perdew-Zunger parametrization\cite{PZ-LDA_PRB81} of the local-density approximation.
For the methane-water dimer, this was sufficient to obtain accurate results.
Some of the codes exchanged wave function data via the TREXIO library.~\cite{TREXIO}
{\revision{}
This choice ensures that any observed variation is due to implementation-level or algorithmic factors rather than differences in the choice of geometry, pseudopotential, basis set, or single-particle orbitals.
}
Every code implements a slightly different parametrization of the Jastrow factor, but all codes include in the Jastrow factor an electron-electron (e-e), an electron-nucleus (e-n), and an electron-electron-nucleus (e-e-n) term.
The variational parameters of the Jastrow factor have been optimized by minimizing either the variational energy or the variance, according to the recommended scheme within each code.
%
The time steps employed in each simulation are in the range 0.001 to 0.1 a.u. 
The final estimates reported in Fig.\ \ref{fig:method_comparison} were extrapolated to the $\tau \to 0$ limit using the procedure described in the SI\@.
Further details specific to each code, the schemes used to deal with the localization error, the time step extrapolation, and the tests on the size consistency error are reported in the SI\@.

\section{Summary and Conclusions}

In this work, we investigated the reproducibility of FN-DMC calculations across 11 popular QMC codes which differ in the details of the algorithms used.
This study represents a significant collaborative effort, involving more than 300 FN-DMC calculations, spanning 11 codes, multiple DMC time steps, and different pseudopotential localization schemes. Our results establish FN-DMC as a robust reference method by demonstrating its reproducibility.

{\revision{}
In particular, we conducted a thorough analysis of two key obstacles to FN-DMC reproducibility, namely the use of NLPPs and finite time-step bias.
We systematically compared four localization schemes, LA, TM, DLA, and DTM, for the interaction energy of the methane-water dimer and the total energies of the methane and water molecules and of the dimer. 
We found that agreement in the interaction energy across all eleven codes is achieved in the limit of zero time step when employing the TM, DLA, and DTM approximations.
Notably, we achieve agreement within a standard deviation of $3$ meV on the interaction energy of the methane-water complex, approximately two hundred thousand times smaller than the total energy of the dimer.
Larger discrepancies are observed with the LA scheme.
Agreement in total energies across codes is also achieved, at sub-millihartree precision.
In particular, the total energies with the  TM, DLA, and DTM schemes have a standard deviation among the codes which is smaller than $6$ meV.
This agreement further reinforces the reproducibility of FN-DMC.
}

{\revision{}
}

{\revision{}
Looking ahead, extending this cross-code effort to periodic solids would be a natural next step. 
However, such systems introduce additional layers of complexity—including basis set periodization, Brillouin zone sampling, and finite-size corrections—that go beyond the scope of this initial benchmark. 
Moreover, as only a subset of the participating codes currently support periodic boundary conditions, we deliberately focused here on molecular systems in open boundary conditions to establish a controlled but challenging comparison for FN-DMC reproducibility.
}

\section*{Supplementary Material}
See the supplementary material for comprehensive details on the computational setup used in this study, including the geometry of the systems, descriptions of the trial wave functions, and specific parameters for each of the 11 FN-DMC codes. 
Additional data are provided on time-step convergence studies, localization error analysis, interaction and total energy comparisons. 
The file also includes technical implementation notes from each code, information on Jastrow factor optimization, and complete tables of all raw FN-DMC energies and statistical uncertainties used to generate the figures in the main text.

\section*{Acknowledgements}
M.B.\ acknowledges the computational resources from the HPC facilities of the University of Luxembourg~\cite{VBCG_HPCS14} {\small (see \href{http://hpc.uni.lu}{hpc.uni.lu})}.
M.D. acknowledges financial support from the European Union under the LERCO project number CZ.10.03.01/00/22\_003/0000003, via the Operational Programme Just Transition, and computational resources from the IT4Innovations National Supercomputing Center (e-INFRA CZ, ID: 90140).
M.C.\ acknowledges access to French computational resources at the CEA-TGCC center through the GENCI allocation number A0150906493.
J.T.K., P.R.C.K., Y.L. and L.M. were supported by the U.S. Department of Energy, Office of Science, Basic Energy Sciences, Materials Sciences and Engineering Division, as part of the Computational Materials Sciences Program and Center for Predictive Simulation of Functional Materials.
Part of the work of L.M. has been supported also by
U.S. National Science Foundation grant DMR-2316007 and employed resources at NERSC at early stages for this project.
M.C., E.S., R.S., and C.F. acknowledge partial support by the European Centre of Excellence in Exascale Computing TREX --- Targeting Real Chemical Accuracy at the Exascale. 
This project has received funding in part from the European Union's Horizon 2020 --- Research and Innovation program --- under grant agreement no.~952165.
E.S., R.S., and C.F performed the calculations on the Dutch national supercomputer Snellius with the support of SURF Cooperative.
K.N.\ acknowledges financial support from the JSPS Overseas Research Fellowships and from MEXT Leading Initiative for Excellent Young Researchers (Grant No.~JPMXS0320220025) and computational resources from the Numerical Materials Simulator at National Institute for Materials Science (NIMS).
L.K.W and W.W. were supported by U.S. National Science Foundation via Award No. 1931258.
Y.S.A., A.Z. and D.A. acknowledge support from the European Union under the Next generation EU (projects 20222FXZ33 and P2022MC742) and from Leverhulme grant no. RPG-2020-038.
A.M. and B.X.S. acknowledge support from the European Union under the ``n-AQUA'' European Research Council project (Grant No.\ 101071937).
The portion of the work done by T.A.A. and C.J.U. received initial support under AFOSR (Grant No. FA9550-18-1-0095) and was completed under the Exascale Computing Project (17-SC-20-SC),
a collaborative effort of the U.S. Department of Energy Office of Science and the National Nuclear Security Administration."

This research used resources of the Oak Ridge Leadership Computing Facility at the Oak Ridge National Laboratory, which is supported by the Office of Science of the U.S. Department of Energy under Contract No. DE-AC05-00OR22725. Calculations were also performed using the Cambridge Service for Data Driven Discovery (CSD3) operated by the University of Cambridge Research Computing Service (www.csd3.cam.ac.uk), provided by Dell EMC and Intel using Tier-2 funding from the Engineering and Physical Sciences Research Council (capital grant EP/T022159/1 and EP/P020259/1), and DiRAC funding from the Science and Technology Facilities Council (www.dirac.ac.uk). This work also used the ARCHER UK National Supercomputing Service (https://www.archer2.ac.uk), the United Kingdom Car Parrinello (UKCP) consortium (EP/ F036884/1).
FDP, BXS, and AM acknowledge EuroHPC Joint Undertaking for awarding the project ID EHPC-REG-2024R02-130 access to Leonardo at CINECA, Italy.

\bibliographystyle{ieeetr}
\bibliography{ref} 

\end{document}


\footnotetext[1]{These authors contributed equally. All others, except for the corresponding author, are ordered alphabetically.}
\footnotetext[2]{Corresponding author email: andrea.zen@unina.it}
\footnotetext[3]{Deceased, 10 August 2022.}
\footnotetext[4]{This manuscript has been authored by UT-Battelle, LLC under Contract No. DE-AC05-00OR22725 with the U.S. Department of Energy. The United States Government retains and the publisher, by accepting the article for publication, acknowledges that the United States Government retains a non-exclusive, paid-up, irrevocable, worldwide license to publish or reproduce the published form of this manuscript, or allow others to do so, for United States Government purposes. The Department of Energy will provide public access to these results of federally sponsored research in accordance with the DOE Public Access Plan (https://www.energy.gov/doe-public-access-plan).} 

\maketitle
\newpage
\tableofcontents
\newpage

We provide here additional supporting data as well as contextual information to the main text.
All output files are provided on \DATAgithub, which contains a Jupyter Notebook file that analyzes the data.
This data can also be viewed and analyzed via a web browser using \DATAcolab.

\section{Summary and computational details}

We compare the fixed-node (FN) diffusion Monte Carlo (DMC) predictions for the interaction energy of the methane-water dimer and the total energies of the fragments and the dimer, obtained with the following 11 codes (enumerated alphabetically):
\begin{enumerate}
    \item Amolqc: \href{https://github.com/luechow-group/Amolqc}{https://github.com/luechow-group/Amolqc}
    \item CASINO: \href{https://vallico.net/casinoqmc/}{https://vallico.net/casinoqmc/}
    \item CHAMP-EU : \href{https://github.com/filippi-claudia/champ}{https://github.com/filippi-claudia/champ}
    \item CHAMP-US: \href{https://github.com/QMC-Cornell/CHAMP}{https://github.com/QMC-Cornell/CHAMP}
    \item CMQMC: \href{https://research.csiro.au/mst/tools/cmqmc/}{https://research.csiro.au/mst/tools/cmqmc/}
    \item PyQMC: \href{https://github.com/WagnerGroup/pyqmc}{https://github.com/WagnerGroup/pyqmc}
    \item QMCPACK: \href{https://qmcpack.org/}{https://qmcpack.org/}
    \item QMC=Chem: \href{https://github.com/trex-coe/qmcchem2}{https://github.com/trex-coe/qmcchem2}
    \item QMeCha: \href{https://github.com/QMeCha}{https://github.com/QMeCha}
    \item QWalk: \href{https://github.com/QWalk/mainline}{https://github.com/QWalk/mainline}
    \item TurboRVB: \href{https://turborvb.sissa.it}{https://turborvb.sissa.it}
\end{enumerate}
We discuss the individual codes and their predictions in Sec.~\ref{sec:codes_description}.

Using the fixed-node approximation (arising from the same determinantal component of the trial wave function) and the same pseudopotential (PP), we compare the results of these 11 codes obtained with four different approximations to treat the non-local terms in the PP:
\begin{enumerate}
    \item Locality approximation (LA)
    \item T-move approximation (TM)
    \item Determinant locality approximation (DLA)
    \item Determinant T-move approximation (DTM)
\end{enumerate}
%
%
These are discussed in detail in Sec.~\ref{sec:local_error}.

We use the correlation-consistent effective core potentials (ccECPs)\cite{ccECP1,ccECP2} for the H, C, and O atoms.
The determinant part of the trial wave function (see main text) is generated using PySCF \cite{sunpyscf2018, sunpyscf2020},  with density functional theory (DFT) and the Perdew-Zunger\cite{PZ-LDA_PRB81} version of the local density approximation (LDA)\@.
%
%
The Jastrow factors in the trial wave function and the details of the Green's function differ between codes and are discussed in Sec.~\ref{sec:codes_description}.
We calculate the total FN-DMC energies of the isolated water, isolated methane, and methane-water complex
for a selection of time steps in the range between $0.001$ and $0.1$ ~a.u.\ for each of the 11 codes (most of the codes use 0.0025, 0.005, 0.01, 0.02, 0.04, and 0.08 a.u.) and for all four approximations (where available).

\section{Calculating the interaction energy of the methane-water dimer}

\begin{figure*}[!h]
    \includegraphics[width=6.66in]{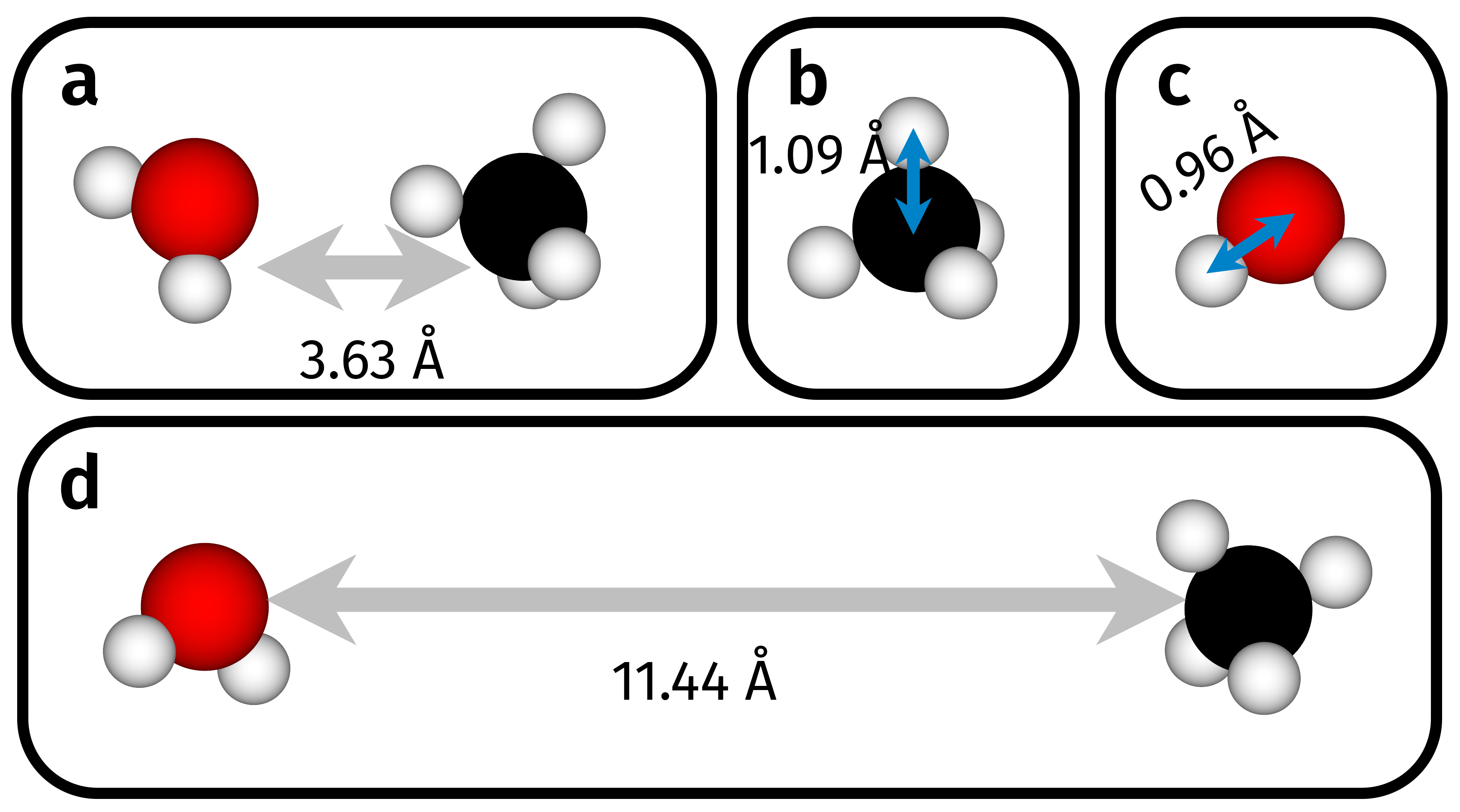}
    \caption{\label{fig:methane-water_diagram} Visualization of the (a) methane-water dimer, (b) isolated methane molecule, (c) isolated water molecule, and (d) methane$---$water systems investigated in this study.}
\end{figure*}

%
The geometry of the dimer is taken from reference \citenum{ZSGMA} and is visualized in Fig.~\ref{fig:methane-water_diagram}a.
In the dimer, the H-atom of methane points towards the lone-pair of water at a distance of $2.551\,$\AA{}.
In this work, we also consider the isolated water and methane molecules as well as the methane$---$water dimer where methane and water are  ${\sim}11\,$\AA{} apart (shown in Figs.~\ref{fig:methane-water_diagram}b, c, and d, respectively). The geometries of the monomers are the same as in the dimer.

We compute the interaction energy, $E_\textrm{int}$, by subtracting the isolated molecule energies from the methane-water complex:
\begin{equation}
    E_\textrm{int} = E[\textrm{methane--water}] - E[\textrm{methane}] -E[\textrm{water}]\,.
\end{equation}
%
To investigate the size-consistency error (SCE), we also compute the energy difference between the sum of the isolated molecules and the methane$---$water complex:
\begin{equation} \label{eq:sce}
    E_\textrm{SCE} = E[\textrm{methane$---$water}] - E[\textrm{methane}]  - E[\textrm{water}] \,.
\end{equation}

\section{\label{sec:ccsdtq_ref} CCSDT(Q) reference for methane-water dimer interaction energy}

We use the coupled cluster method to obtain the reference interaction energy for the methane-water dimer and employ the augmented Dunning (aug-cc-pV$X$Z) and its core-valence correlated variant (aug-cc-pwCV$X$Z) for the CCSDT(Q) and CCSD(T) calculations, respectively.

The reference value of $-26.6$ meV for the interaction energy  is obtained from an all-electron CCSDT(Q) calculation extrapolated to the complete basis set limit.  To obtain this estimate, we start from the basis-set extrapolated estimate of $-26.4$ meV obtained using the all-electron CCSD(T) values computed with the quadruple-zeta and quintuple-zeta basis sets.  The extrapolation is accurate as evidenced by the fact that extrapolation using triple-zeta and quadruple-zeta basis sets changes the estimate by only $0.1\,$meV.  The counterpoise-corrected and non-counterpoise-corrected estimates differ by $0.3\,$meV and the value of  $-26.4$ meV is obtained by averaging them.  Finally, a higher-order correction (of about $-0.2\,$meV), obtained from CCSD(T) and CCSDT(Q) calculations using the double-zeta basis, is added to get the final estimate of $-26.6$ meV. 

For the extrapolation to the complete basis set limit, we employ the following two-point-extrapolation formulae:
\begin{equation}
    E^\textrm{CBS}_\textrm{corr} = \dfrac{X^\beta E^X_\textrm{corr} - Y^\beta E^Y_\textrm{corr}}{X^\beta - Y^\beta},
\end{equation}
\begin{equation}
E^\textrm{CBS}_\textrm{HF} = E^X_\textrm{HF} - \dfrac{E^Y_\textrm{HF} - E^X_\textrm{HF}}{\exp  (-\alpha \sqrt{Y}  ) -  \exp(-\alpha \sqrt{X})} \exp (-\alpha \sqrt{X} ),
\end{equation}
for the correlation and Hartree-Fock components of the energy, with $X$ and $Y=X+1$ denoting the (zeta) size of the basis set.
We use $\alpha = 5.79$ and $\beta = 3.05$ as given in Ref.~\citenum{neeseRevisitingAtomicNatural2011}.

\section{\label{sec:local_error} Localization error with non-local pseudopotentials}

The pseudopotentials used have angular non-locality, which results in an additional non-local component in the Green's function and an additional sign problem.
Separating the local and non-local components of the Hamiltonian,
$\hat{H} = \hat{H}_{\rm L} + \hat{V}_{\rm NL}$, the importance sampled Green's function is
\beq
G(\Rvecp,\Rvec,\tau)=\frac{\Psi(\Rvecp)}{\Psi(\Rvec)} \langle \Rvecp | {\rm e}^{\tau(\ET - \hat{H}_{L} - \hat{V}_{\rm NL})} | \Rvec \rangle .
\eeq
Using the Suzuki-Trotter expansion for small $\tau$, the Green's function can also be split into parts that use the local and the non-local components of the pseudopotential,
\beq
G(\Rvecp,\Rvec,\tau) &\approx& \int d\Rvecpp \; G_{L}(\Rvecp,\Rvecpp,\tau) \; T(\Rvecpp,\Rvec,\tau),
\eeq
where
$G(\Rvecp,\Rvec,\tau)$ is the usual importance-sampled Green's function that contains the local component of the
pseudopotential and
\beq
T(\Rvecp,\Rvec,\tau) = \frac{\Psi(\Rvecp)}{\Psi(\Rvec)} \langle \Rvecp | {\rm e}^{-\tau \hat{V}_{\rm NL}} | \Rvec \rangle,
\label{eq:tmov}
\eeq
generates additional non-local moves coming from the non-local component of the pseudopotential.
As can be readily seen from the short-time expansion of the exponential,
\beq
T(\Rvecp,\Rvec,\tau) \approx \delta_{\Rvecp,\Rvec}-\tau\frac{\Psi(\Rvecp)}{\Psi(\Rvec)} \langle \Rvecp | \hat{V}_{\rm NL} | \Rvec \rangle
\label{Tsign}
\eeq
can change sign not only because the wave function changes sign but also because $\langle \Rvecp | \hat{V}_{\rm NL} | \Rvec \rangle$ can be positive and lead to a negative off-diagonal term.

The sign problem that occurs when $G_\mathrm{L}(\Rvecp,\Rvec,\tau)<0$ is solved by making the fixed-node approximation.
The sign problem that occurs when $T(\Rvec',\Rvec,\tau)<0$ is solved by making either the locality approximation or some variant of the T-moves approximation.

In the size-consistent variants of the T-move algorithm,~\cite{casula_Tmove_LRDMC_2010,AndUmr-JCP-21} one exploits that
$\hat{V}_{\rm NL} = \sum^{N_{\rm elec}}_{i=1} \hat{v}^{i}_{\rm NL}$ is a one-body
operator, so that the $N$-electron non-local Green's function can be factored into $N$ one-electron non-local Green's
functions:
\beq
\label{eq:tmov_1elec}
T(\Rvecp,\Rvec,\tau) =
\prod^{N_{\rm elec}}_{i=1}
\frac{\Psi(\Rveci')}{\Psi(\Rveci)}
\langle {\bf r}_{i}' | {\rm e}^{-\tau \hat{v}^{i}_{\rm NL}} | {\bf r}_{i} \rangle
\;\equiv\; \prod^{N_{\rm elec}}_{i=1} t(\Rveci',\Rveci,\tau),
\eeq
where $\Rveci' = \left\{\rvec_{1}',...,{\rvec_{i}}',{\rvec_{i+1}},...,\rvec_N\right\}$,
$\Rveci = \left\{\rvec_{1}',...,{\rvec_{i-1}}',{\rvec_{i}},...,\rvec_N\right\}$,
and $\rvec_i$ and $\rvec_i'$ are the positions of the $i$-th electron before and after the T-move, respectively.

\subsection{Locality approximation}

The locality approximation (LA)~\cite{la_mitas} replaces the Hamiltonian,
$\hat{H} = \hat{H}_{\rm L} + \hat{V}_{\rm NL}$, by the effective LA Hamiltonian, $\hat{H}^\textrm{LA} = \hat{H}_{\rm L} + \hat{V}_{\rm NL}^\textrm{LA}$, where the non-local pseudopotential is localized on the trial wave function as:
\begin{equation}
    \hat{V}_\textrm{NL}({\bf R}) \to {V}_\textrm{NL}^\textrm{LA}({\bf R}) =
\frac{\langle {\bf R} | \hat{V}_\textrm{NL} | \Psi_T \rangle}{\langle {\bf R} | \Psi_T \rangle}
\,.
\end{equation}
%
A drawback of the locality approximation is that its fixed-node energy, $E_\textrm{FN}^\textrm{LA}$, is no longer an upper bond to the ground-state energy, $E_\textrm{GS}$, of the true Hamiltonian $\hat{H}$.
Furthermore, the LA leads to negative divergences of the effective potential at the nodes of $\Psi_T$, which create severe numerical instabilities during practical DMC simulations.\cite{casula_Tmove_2006}

\subsection{T-move approximation}

Casula's T-move (TM) approximation~\cite{casula_Tmove_2006} cures the main problems of the LA approximation by localizing only the sign-violating terms (Eq.~\ref{Tsign}).
The TM approximation can be shown to yield a variational FN energy, namely, an upper bound to the ground-state energy of the true Hamiltonian~\cite{GFMC_1,casula_Tmove_2006}. Furthermore, the additional electron moves coming from the non sign-violating terms reduce the probability of encountering population explosions that come from walkers that have very negative local energies staying at the same location for multiple Monte Carlo generations.

There are four variants of the T-move algorithm in the literature.  In the original T-move (TM) approximation~\cite{casula_Tmove_2006}, the non-local terms in the pseudopotential result in at most one electron making a T-move.  Consequently, it suffers from the drawback that, at nonzero $\tau$, the algorithm behaves increasingly like the locality approximation as the system size increases.
Casula et al.~\cite{casula_Tmove_LRDMC_2010} proposed two variants of the algorithm that cure this problem by allowing all the electrons to make a T-move.
Anderson and Umrigar~\cite{AndUmr-JCP-21} introduced an additional accept-reject step to ensure the desirable property that the algorithm yields exact expectation values at nonzero $\tau$ when $\Psi_T$ is exact. The computational cost per Monte Carlo step is essentially unchanged.

In addition to the favourable properties of the T-moves approximation noted above, it is in fact slightly more efficient than the locality approximation because the additional moves reduce the autocorrelation time of the energy~\cite{AndUmr-JCP-21}.  Finally, we note that all four variants of T-moves yield the same energy extrapolated to $\tau=0$, but, they differ at finite $\tau$.

\subsection{Determinant locality approximation}

The determinant locality approximation (DLA)~\cite{HurChr-JCP-87,hammond1987,FlaSavPre-JCP-92,Caffarel2016,dla_zen} is in fact the oldest approximation for treating non-local pseudopotentials.  Instead of using the full trial wave function $\Psi_T$ to localize the pseudopotential, it employs just the determinantal part of the wave function, the advantage being that the integral over the sphere can be performed analytically.  As such, it is an approximation to the LA.  However, recently some authors have pointed out~\cite{dla_zen} that it can enhance the reproducibility of results from different codes by removing the dependence on the choice of Jastrow factor.  However this comes at the cost of reduced accuracy of all observables for typical choices of $\Psi_T$, and, a failure to recover exact expectation values of observables, other than the energy, in the exact $\Psi_T$ limit.

%

\subsection{Determinant T-move approximation}

The determinant T-move (DTM) approximation~\cite{dla_zen} uses the determinantal part of the wave function for localizing the sign-violating part of the Green's function, while using the full $\Psi_T$ for the non sign-violating part.  It has the same advantage and disadvantage relative to the TM approximation that the DLA has to the LA approximation.


\section{\label{sec:time_step_lim}Reaching the zero time-step limit}

In a DMC calculation, one employs a short-time approximation of the imaginary time Green's function used to project the FN wave function, and the results must be extrapolated to the zero time-step limit.
%
%
In this limit, the only biases in the DMC energy are due to the FN approximation and the localization error (if using a PP)\@.
Since we are using the same determinantal component of the wave function $\Psi_\mathrm{T}$, only the different choices of Jastrow factor creates some discrepancy between the extrapolated results of different codes when using the LA and TM pseudopotential localization algorithms.

%
The equation we use to extrapolate towards the zero time-step limit is of the form:
\begin{equation}
    E(\tau) = A + B\tau + C\tau^2 + D\tau^3,
\end{equation}
where A, B, C, and D are fit parameters, with A being the value in the limit of $\tau \to 0$.
Sometimes the quadratic form ($D=0$) is sufficient for a good fit, while other times the cubic term is needed.

For the interaction energy, we have chosen to extrapolate the interaction energy directly instead of extrapolating the separate total energy terms. The time-step dependence of the interaction energy is typically smoother than the dependence on the total energy, thanks to error cancellation, so a polynomial function with a lower degree can typically be used and the extrapolation is statistically more robust.

\begin{figure*}[!h]
    \includegraphics[width=6.66in]{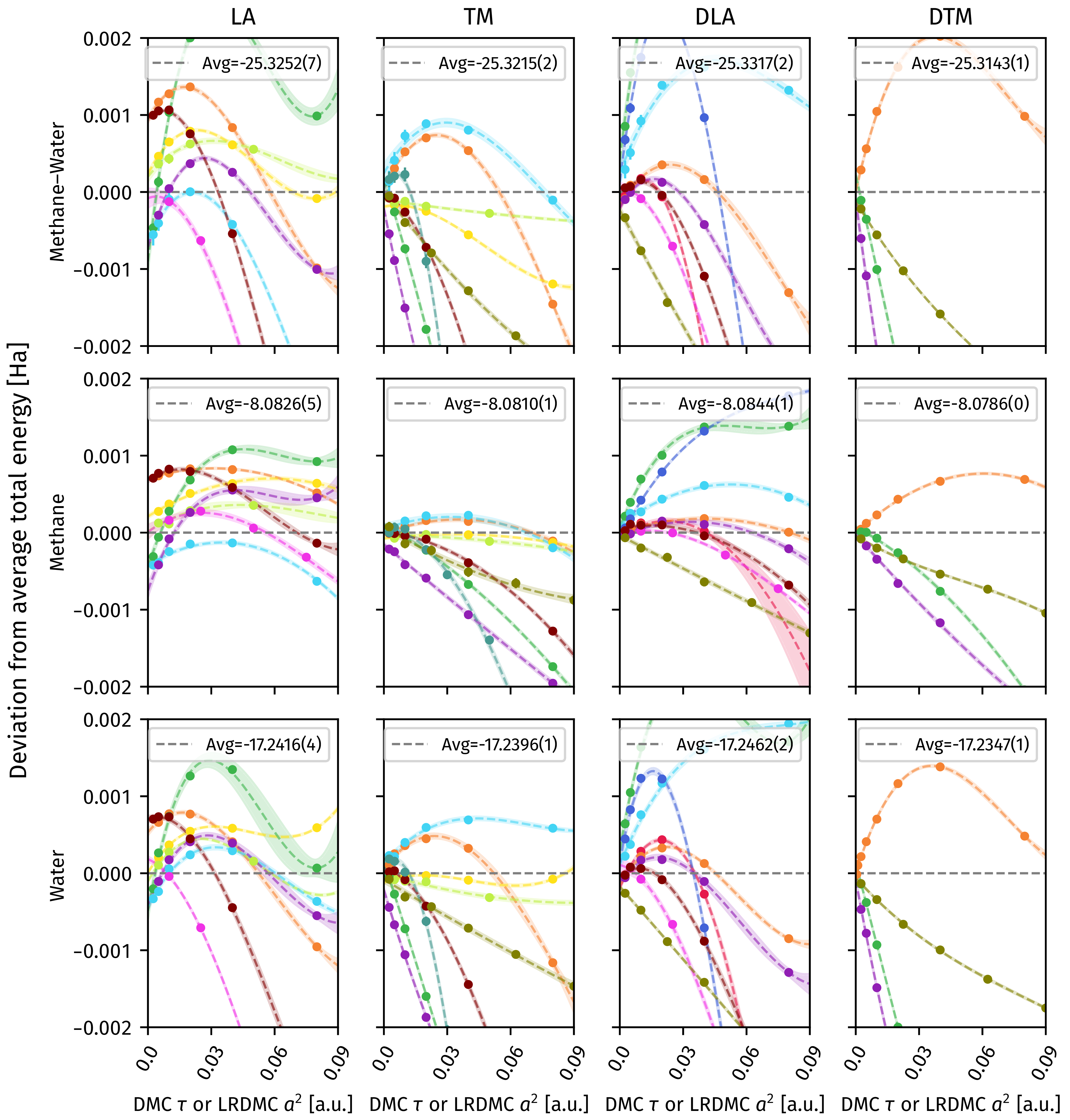}
    \caption{\label{fig:Tot_ene_timestep_convergence} Comparison of the time-step dependence of the total energies of the methane-water dimer, isolated methane molecule, and isolated water molecule for  the 11 codes and the LA, TM, DLA, and DTM pseudopotential localization schemes.}
\end{figure*}

For TurboRVB calculations in the lattice regularized diffusion Monte Carlo (LRDMC) flavor, the energies need to be extrapolated to the $a \rightarrow 0$ limit, where $a$ is the lattice spacing used for the Laplacian discretization of the Hamiltonian. We employ a randomized mesh as described in Ref.~\citenum{casula_Tmove_LRDMC_2010} and \citenum{NakanosmartlatticeregularizationinLRDMC} to reduce the discretization bias. The fitting function used for the extrapolation is:
\begin{equation}
E(a^2) = A + B \cdot a^2 + C \cdot a^4.
\end{equation}
We notice how in the systems studied here the $a$-dependence of the LRDMC energies has a very small quartic component, allowing one to easily extrapolate to the zero lattice spacing with a very few points. This can be appreciated from Fig.\ref{fig:Tot_ene_timestep_convergence}, where the lattice spacing dependence of the total ground state energies has been converted into effective time step, using the relation $a^2=\tau$, to make a direct comparison with the DMC extrapolations possible. The latter relation can be straightforwardly derived by equating the lattice spacing $a$ with the spread of the Gaussian distribution, solution of the purely diffusion equation with time step $\tau$. The same mapping has been used in Figs.~\ref{fig:int_ene_timestep_convergence} and \ref{fig:Size_consist_error}.

For all codes, we obtain excellent fits to the interaction and total energies, see Tables~\ref{table:Eint}, \ref{table:Em}, \ref{table:Ew}, and \ref{table:Emw}. 

We quantify the quality of the fits via the reduced-chi-squared ($\chi^2_\text{red}$) metric, which is defined as:
\begin{equation}
    \chi^2_\text{red} = \frac{\chi^2}{N_\textrm{dof}},
\end{equation}
where the number $N_\textrm{dof}$ of degrees of freedom (dof) is computed by subtracting from the number $N$ of time step calculations used in the fitting the number $k$ of variables in the extrapolation formulae (in our case it is one plus the degree $d_\text{poly}$ of the fitting polynomial, i.e. 3 and 4 for quadratic and cubic functions, respectively). Thus, $N_\textrm{dof} = N - k$ with $k=1+d_\text{poly}$.
We note in Table~\ref{table:Eint} that 
$\chi^2_\text{red}$ is found to be below 2 for most of the interaction energy extrapolations across the 11 codes and 4 localization schemes, indicating good fits. 

To further quantify the accuracy of these fits, 
we calculate the root mean squared residual (RMSR) between the fitted and the computed DMC values of the interaction energy for all 11 codes:
\begin{equation}\label{eq:rmsr}
    \textrm{RMSR} =  \sqrt{ {1\over N} \sum_{\tau}^{N} (E^{\tau}_\textrm{pred.} - E^{\tau}_\textrm{calc.})^2}.
\end{equation}
These RMSR values are generally below $2\,$ meV and thus errors from the extrapolation are expected to be below this values.

%

In Table~\ref{table:Eint}, we report the extrapolated zero-time-step limit $E_\textrm{int}^{\tau \to 0}$ and the smallest time-step estimate $E_\textrm{int}^{\tau_\textrm{min}}$ of the interaction energy, and find that the differences 
$\Delta E_\text{int}^{\tau_\text{min},0} = E_\textrm{int}^{\tau_\textrm{min}} - E_\textrm{int}^{\tau \to 0}$ 
are generally quite small, most of the times within $\pm 3$~meV, indicating that only small extrapolations beyond the fitted data are required.

\begin{table}[h!]
\caption{\label{table:Eint}
Across the 11 codes and 4 localization algorithms, we report 
the smallest time step $\tau_\textrm{min}$ and the corresponding estimate of the interaction energy
$E_\textrm{int}^{\tau_\textrm{min}}$, with its stochastic error,
the zero time step limit estimate of the {\bf interaction energy} 
$E_\textrm{int}^{\tau \to 0}$ and its stochastic error,
and their difference 
$\Delta E_\text{int}^{\tau_\text{min},0}$. 
We also report the degree $d_\text{poly}$ of the fitting polynomial function, 
the number $N$ of time step calculations used in the fitting, 
the reduced-chi-squared ($\chi^2_\text{red}$) metric, 
and the root mean squared residual (RMSR) of the fitted interaction energy curves. 
The unit of reported energies is meV.
}
{\small 
\input{Tables/Table_Eint.tex}
}
\end{table}

\begin{table}[h!]
\caption{\label{table:Em}
Across the 11 codes and 4 localization algorithms, we report 
the smallest time step $\tau_\textrm{min}$ and the corresponding estimate of the total energy of the {\bf methane} molecule
$E_\textrm{tot}^{\tau_\textrm{min}}$, with its stochastic error,
the zero time step limit estimate of the interaction energy 
$E_\textrm{tot}^{\tau \to 0}$ and its stochastic error,
and their difference 
$\Delta E_\text{tot}^{\tau_\text{min},0}$. 
We also report the degree $d_\text{poly}$ of the fitting polynomial function, 
the number $N$ of time step calculations used in the fitting, 
the reduced-chi-squared ($\chi^2_\text{red}$) metric, 
and the root mean squared residual (RMSR) of the fitted interaction energy curves. 
The unit of reported energies is Hartree.
}
{\footnotesize 
\input{Tables/Table_Etot_m.tex}
}
\end{table}

\begin{table}[h!]
\caption{\label{table:Ew}
Same as Table~\ref{table:Em} for the {\bf water} molecule.
}
{\footnotesize 
\input{Tables/Table_Etot_w.tex}
}
\end{table}

\begin{table}[h!]
\caption{\label{table:Emw}
Same as Table~\ref{table:Em} for the {\bf methane-water} dimer.
}
{\footnotesize 
\input{Tables/Table_Etot_mw.tex}
}
\end{table}


\clearpage

\section{Comparison of interaction energy time-step convergence}

\begin{figure*}[!ht]
    \includegraphics[width=6.66in]{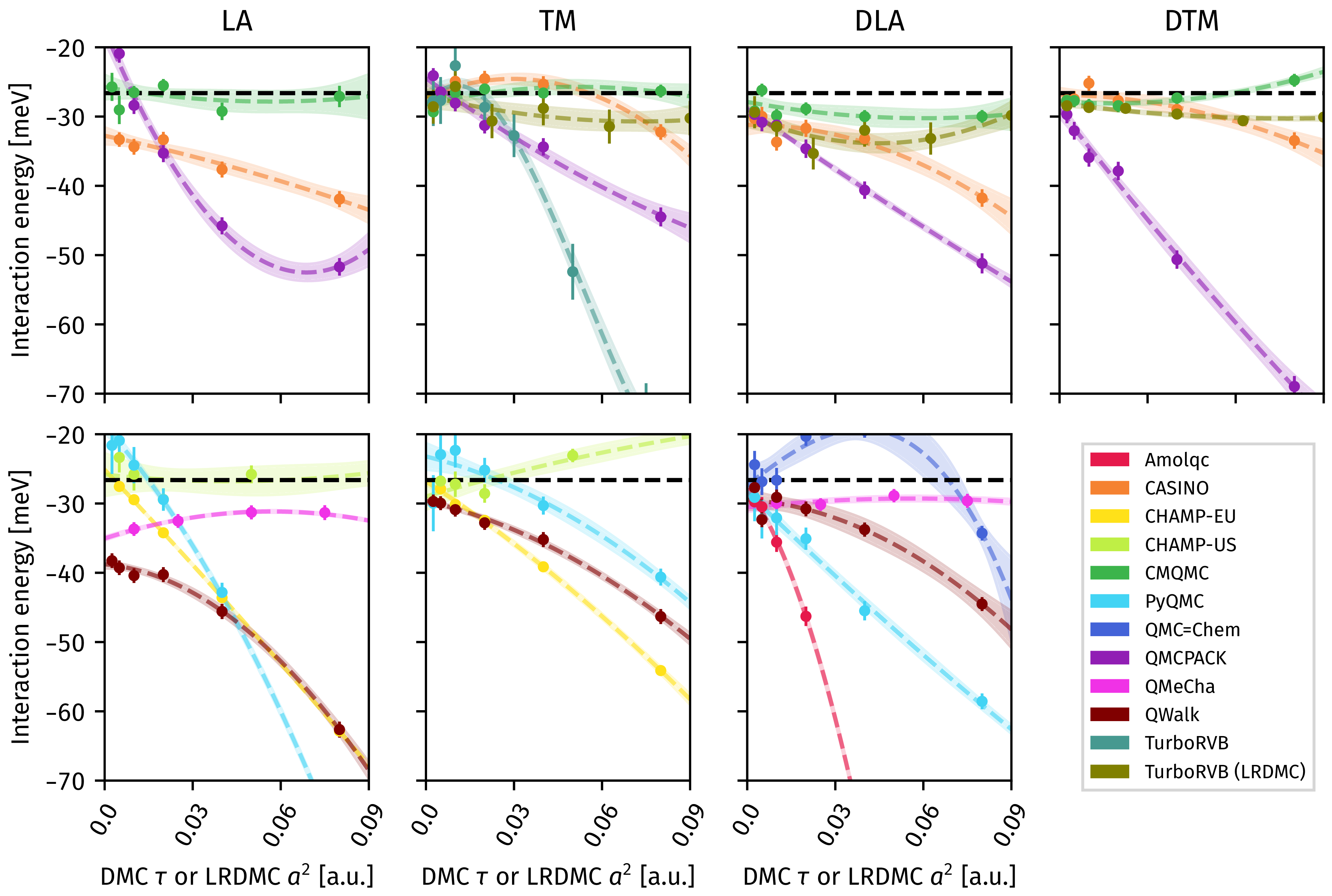}
    \caption{\label{fig:int_ene_timestep_convergence} Comparison of the zero-time-step extrapolated value of the methane-water dimer interaction energy for the 11 codes and the LA, TM, DLA, and DTM pseudopotential localization schemes.}
\end{figure*}

We compare the time-step dependence of the interaction energy for all 11 codes in Fig.~\ref{fig:int_ene_timestep_convergence}.
Some codes display a large change of almost $100\,$meV when the time step goes from 0.0025 to 0.08 a.u., while others display changes as small as $5\,$meV. 
However, except when employing the LA scheme, all codes are in good agree in the $\tau\to 0$ limit. 
Indeed, the excellent agreement we report in the main text is obtained in this limit.

The level of agreement in the  $\tau\to 0$ limit of the interaction energy across the codes, for each considered localization scheme, can be appreciated in Table~\ref{table:mean_Eint}, which reports the mean interaction energy, the number of codes providing an estimation, the maximum distance between the estimations of two different codes, the maximum distance of a code from the mean, the mean absolute deviation from the mean, and the root mean square deviation from the mean.

\begin{table}[h!]
\caption{\label{table:mean_Eint}
The table reports, for each localization scheme, the mean interaction energy (Mean), the number of codes providing an estimation (No. codes), the maximum difference between the estimations of two different codes (Max-dist), the maximum difference of the estimations from the mean (Max-error), the mean absolute deviation from the mean (MAE), and the root mean square deviation from the mean (RMSE).
The unit of the energy is meV.
The error bar associated the the Mean is estimated according to equation~(4) of the main manuscript, and it accounts for both the statistical error bar of each FN-DMC evaluation and its deviation from the mean value.
}
\input{Tables/Table_mean_Eint.tex}
\end{table}

\clearpage

\section{Comparison of total energy time-step convergence}\label{sec:total_energy_convergence}

We report here the time-step convergence of the total energies of the methane molecule, water molecule, and methane-water complex.
The time-step convergence differs again drastically among the different codes.
%
Moreover, for some codes, the time-step error of the total energy for time steps from 0.0025 to 0.08 a.u. is even larger than that of the interaction energy, namely, as large as 5 mHartrees (1 mHa${\sim}27\,$meV).
These changes in total energy appear to be consistent across the three systems for the same code, demonstrating that there is some cancellation of errors which causes a somewhat lower dependence of the interaction energy on the time step.
%

\begin{figure*}[!h]
    \includegraphics[width=6.66in]{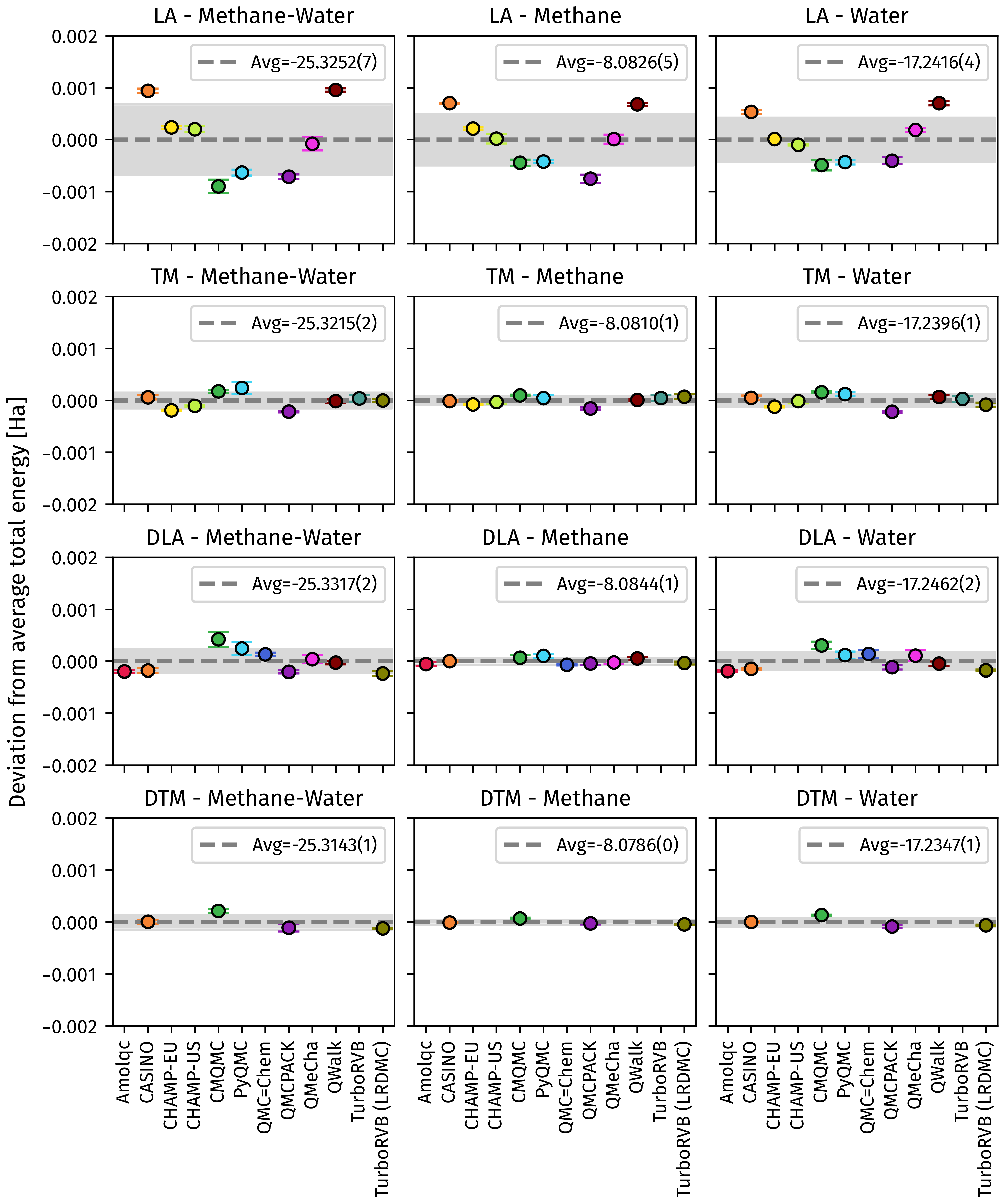}
    \caption{\label{fig:Tot_ene_method_comparison} Comparison of the zero time-step extrapolated value of the total energies for the methane-water dimer, isolated methane molecule, and isolated water molecule for the 11 codes and the LA, TM, DLA, and DTM pseudopotential localization schemes.}
\end{figure*}

In Fig.~\ref{fig:Tot_ene_method_comparison}, we compare the zero time-step extrapolations of the total energies, which are reported in Table~\ref{table:mean_Etot}, and find that, for each localization scheme, the differences among the codes are generally much less than $1\,$mHa for all three systems.
Mirroring the analysis of the interaction energy, we find that the total energies differ among the codes much less for the TM, DLA, and DTM approximations than for the LA.
%
%
%
It is quite remarkable that, despite using different Jastrow factors, all codes give very close extrapolated total energies with the TM scheme, which we recall has the desirable property of treating the pseudopotential exactly in the limit of an exact $\Psi_\mathrm{T}$.
Furthermore, the LA and TM average energies of all systems (reported in the panels of Fig.~~\ref{fig:Tot_ene_method_comparison}) are much closer to each other (by a factor of about 6) than what the DLA and DTM average values are, because of the more accurate trial wave function used for localization.

\begin{table}[h!]
\caption{\label{table:mean_Etot}
Mean total energy, in Hartree, of the water, methane, and water-methane systems using FN-DMC with either LA, TM, DLA, or DTM.
The average is performed across all codes reporting an estimation for the systems and the specific localization scheme.
The error bar is estimated according to equation~(4) of the main manuscript, and it accounts for both the statistical error bar of each FN-DMC evaluation and its deviation from the mean value.
}
\input{Tables/Table_mean_Etot.tex}
\end{table}

\clearpage

\section{\label{sec:codes_description} Computational Codes}
A description of the algorithms used in each code is described below.

\subsection{Amolqc}

\begin{figure*}[!h]
    \includegraphics[width=6.66in]{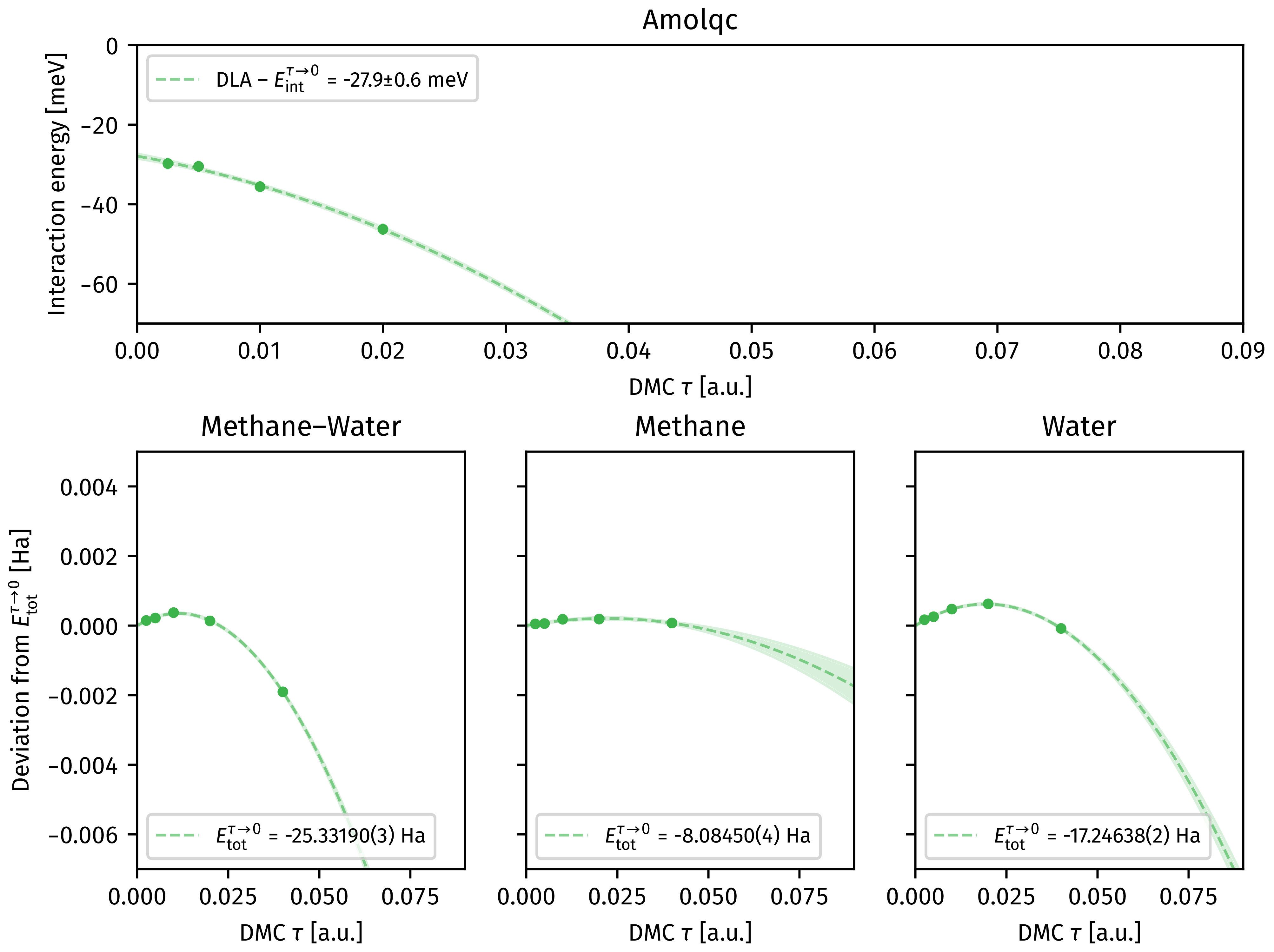}
    \caption{\label{fig:Amolqc_energy_plots} The time step dependence of the (a) methane-water interaction energy (b) methane-water dimer total energy, (c) isolated methane molecule total energy, (d) isolated water molecule total energy in the Amolqc code across the DLA algorithm(s).}
\end{figure*}

The Amolqc code allows VMC and DMC calculations with a large number of variants. Amolqc is an open-source program written in Fortran and available on github (\url{https://github.com/luechow-group/Amolqc}). It is particularly well suited for large multideterminant Slater-Jastrow trial functions. Optimization of Jastrow parameters, CI and MO coefficients is available. A variety of Jastrow factors are available and various propagators are implemented. In this work, the propagator by Umrigar, Nightingale and Runge\cite{UmrNigRun-JCP-93} has been used with minor modifications. The weighting scheme follows that by Zen et al.
\cite{ZSGMA}. The Jastrow factor in this work has e-e, e-n, and e-e-n terms up to sixth order\cite{Luchow2015, guclu_2005}. The scaled distance is of the Schmidt-Moskowitz type. The Jastrow parameters are optimized with respect to the energy. In Table \ref{tab:VMC_amolqc}, the VMC energies and the variances of the local energy are shown for the computed systems and wave functions. In Figure \ref{fig:Amolqc_energy_plots}, the individual results for the DLA calculations are plotted.

\begin{table}
\caption{\label{tab:VMC_amolqc}Total energy ($E^{tot}$) and variance of the local energy ($\sigma^2$) for the computed systems and wave functions using VMC.}
\begin{tabular}{l|c|c}
\hline
Systems     & $E_{VMC}^{tot}$ (Ha) & $\sigma^2$ (Ha$^2$) \\
\hline
Water & -17.22431(3) & 0.3194(3) \\
Methane & -8.06957(3) & 0.1193(2) \\
Methane$-$Water & -25.29019(3) & 0.4829(3) \\
Methane---Water & -25.29101(3) & 0.4976(5) \\
\hline
\end{tabular}
\end{table}

\clearpage
\subsection{CASINO}

CASINO allows the user to perform QMC simulations with a number of different setups, as documented in the User's Guide \cite{casino_guide}.
Recently, CASINO's main developers have reviewed the main features of the package and some applications in Ref.~\citenum{CASINO}.
 The code allows, for instance, the use of different versions of the modifications to the Green function aimed at making the branching and drift-diffusion processes more stable, including the version from \citenum{UmrNigRun-JCP-93} and from \citenum{ZSGMA}.
It is also possible to select whether the electrons are moved one at a time (electron-by-electron), or all at once (configuration-by-configuration).
It is expected that different choices of DMC setup affect the efficiency of the simulations and the finite time step bias, but they should not affect the results in the limit of zero time step (provided there are enough data for a reliable extrapolation).
Moreover, we have to choose a parametrization of the Jastrow factor and then optimize the parameters by minimization of the variational energy or the energy variance, or a combination of the two. The parametrization of the Jastrow factor and the choice of optimization scheme can affect the infinitesimal time step extrapolation in the LA and TM localization schemes, while it does not affect the extrapolated energy results in the DLA and DTM schemes.

We provide here the details about the setup used to obtain the results reported, which is consistent with the setup used in many recently published applications.
It should be noted that we also provide the input files and the main output files in the set of documents available on \DATAgithub.

We performed electron-by-electron moves (keyword: \verb|dmc_method:1|, which is the default).
To make the DMC simulation stable we used the modifications to the Green function defined by the keyword \verb|limdmc:5|, corresponding to the modification of the drift velocity defined in \citenum{UmrNigRun-JCP-93}, with parameter $a=0.5$ (keyword: \verb|alimit:0.5|) and a modification of the branching factors according to the cutoff defined in \citenum{ZSGMA}, but only applied when the local energy is lower than the cutoff.
We used a target population of $102,400$ walkers (keyword: \verb|DMC_TARGET_WEIGHT:102400|), which is large enough to make any possible population bias negligible in the reported results (population bias is typically negligible already with a thousand of walkers).
DMC LA simulations are identified by the keywords \verb|use_detla:F, use_tmove:F|,
TM by \verb|use_detla:F, use_tmove:T|,
DLA by \verb|use_detla:T, use_tmove:F|,
and DTM by \verb|use_detla:T, use_tmove:T|.

The Jastrow factor adopted is defined in Ref.~\citenum{CASINO}, and its parameters are specified in the file \verb|correlation.data|.
Here, we have optimized the parameters of the Jastrow factor by minimizing the variational energy variance (keywork: \verb|opt_method:varmin|) of the trial wave function.
It should be noted that we performed independent optimizations for the calculations with \verb|use_detla:F| (for LA or TM calculations) and with \verb|use_detla:T| (for DLA and DTM calculations), as the latter two schemes imply a different approximation to the localization of non-local terms even at the variational Monte Carlo level of theory.
The optimization has been performed independently for each system, using a sampling of 10 million walkers (keyword \verb|VMC_NCONFIG_WRITE:10000000|).
The VMC energy and variance for each system considered are reported in Tables \ref{tab:casino_vmc_energies} and \ref{tab:casino_vmcdla_energies}.

\begin{figure*}[!h]
    \includegraphics[width=6.66in]{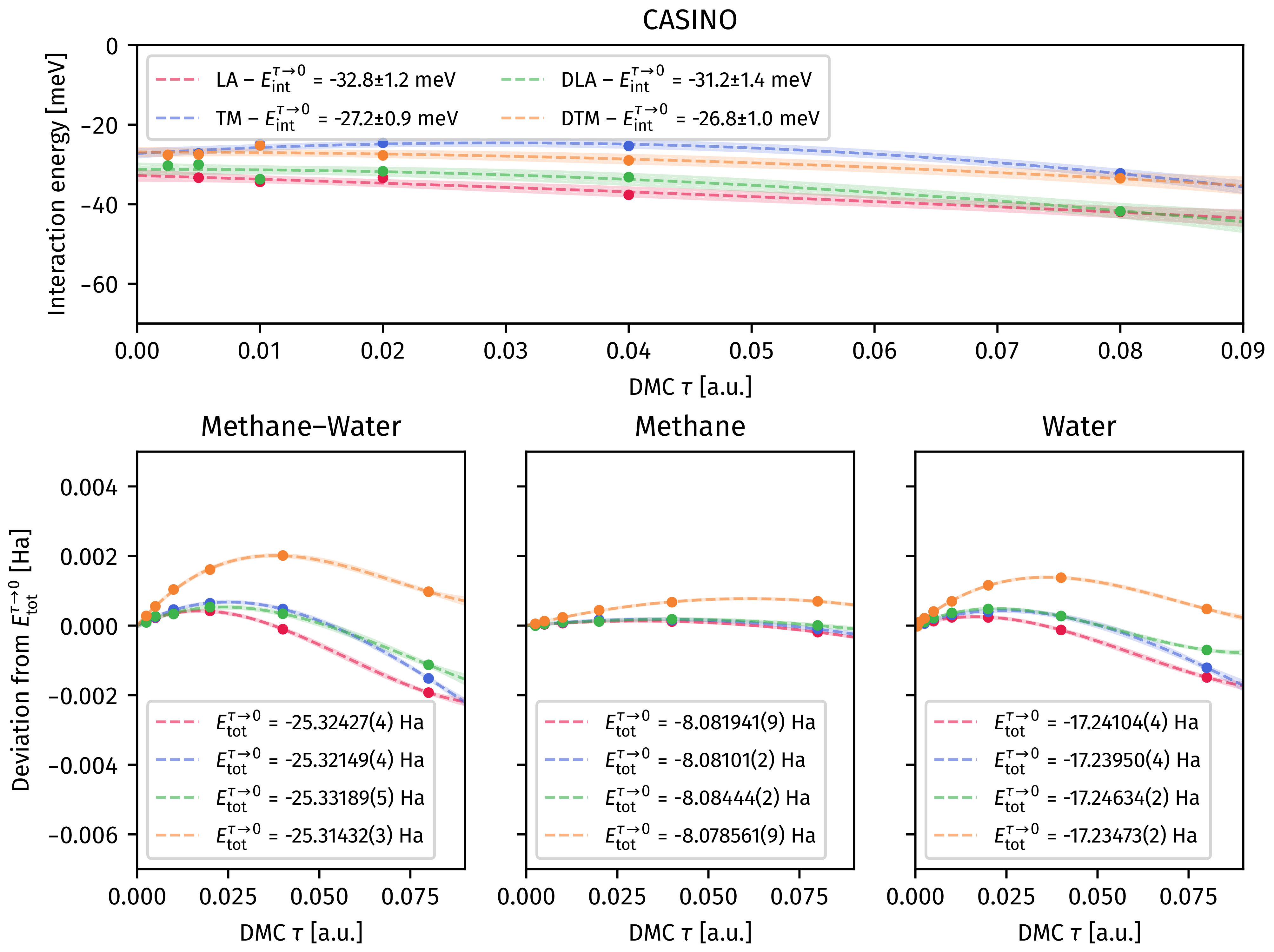}
    \caption{\label{fig:CASINO_energy_plots} The time step dependence of the (a) methane-water interaction energy (b) methane-water dimer total energy, (c) isolated methane molecule total energy, (d) isolated water molecule total energy in the CASINO code across the LA, TM, DLA, DTM algorithm(s).}
\end{figure*}

Results obtained with the above setup are reported in Fig.~\ref{fig:CASINO_energy_plots}.
%
%
%
%

\begin{table}
    \centering
    \begin{tabular}{l|c|c}
    \hline
     Systems  & $E^\textrm{tot}_\textrm{VMC}$ (Ha) & $\sigma_\textrm{VMC}^2$ (Ha$^2$) \\     \hline
     Water            & -17.2222(2) & 0.216(1) \\  
     Methane          & -8.06946(9) & 0.0796(5) \\   
     Methane$-$Water    & -25.2858(2) & 0.314(2) \\   
     Methane$---$Water  & -25.2885(2) & 0.306(1) \\   
    \hline
    \end{tabular}
\caption{Total energy ($E^\textrm{tot}$) and variance ($\sigma^2$) of the systems computed using VMC, for the wave functions used in the code CASINO for the LA and TM schemes.
}\label{tab:casino_vmc_energies}
\end{table}

\begin{table}
    \centering
    \begin{tabular}{l|c|c}
    \hline
     Systems  & $E^\textrm{tot}_\textrm{VMC-DLA}$ (Ha) & $\sigma_\textrm{VMC-DLA}^2$ (Ha$^2$) \\     \hline
     Water            & -17.2208(2) & 0.249(5) \\   
     Methane          & -8.0686(1) & 0.0854(8) \\   
     Methane$-$Water    & -25.2837(2) & 0.346(2) \\   
     Methane$---$Water  & -25.2860(2) & 0.341(3) \\   
    \hline
    \end{tabular}
\caption{Total energy ($E^\textrm{tot}$) and variance ($\sigma^2$) of the systems computed using VMC, for the wave functions used in the code CASINO for the DLA and DTM schemes (i.e., non-local pseudo potential terms are projected on the determinant, not on the entire trial wave function).
}\label{tab:casino_vmcdla_energies}
\end{table}

\clearpage

\subsection{CHAMP-EU}

\subsubsection{Code information}

The Cornell-Holland Ab-initio Materials Package - EU branch (CHAMP-EU) \cite{CHAMP-EU} is a quantum Monte Carlo suite of programs for electronic structure calculations of molecular and solid systems. The code is a sister code of the CHAMP-US and has been separately developed since 2003.

CHAMP-EU has three basic capabilities, namely, variational Monte Carlo (VMC), diffusion Monte Carlo (DMC), and optimization of many-body wave functions by VMC energy minimization for ground and excited states.
Noteworthy features are the efficient wave function optimization in a low-memory implementation for ground and multiple states of the same symmetry (both in a state-average and a state-specific fashion), the compact formulation for a fast evaluation of multi-determinant expansions and their derivatives, and the efficient computation of analytical interatomic forces in VMC and DMC.

CHAMP-EU package is an open-source package implemented in modern Fortran. The code is optimized for modern high-performance computer architectures via extensive vectorization, consideration of memory layouts and access patterns, efficient I/O using the TREXIO library \cite{TREXIO}, and implementation of modular kernels for the computation of the orbitals via the QMCkl library (\url{https://github.com/TREX-CoE/qmckl}).

\subsubsection{Computational Details}

The determinantal wave functions are stored in the TREXIO file format. As Jastrow factor, we use the exponential of the sum of three fifth-order polynomials to describe electron-electron (e-e), electron-nucleus (e-n), and electron-electron-nucleus (e-e-n) correlations~\cite{guclu_2005}. The polynomials depend on the inter-particle distances, which are rescaled as $R=[1-\exp(-\kappa r)]/\kappa$ in the e-e and e-n terms, and $R=\exp(-\kappa r)$ in the e-e-n term, where $\kappa$ is set to 0.4 a.u. We employ different electron-nucleus and electron-electron-nucleus Jastrow factors to describe the correlation of the electrons with C, O, and H. For the compound systems, we use a different e-n and e-e-n terms for the hydrogen atoms in the methane molecule and in the water molecule. The parameters of the Jastrow factors are optimized in energy minimization within VMC and are the same as used in the calculations with the CHAMP-US code. The VMC total energies and variances of the local energy are given in Table~\ref{tab:champ_EU_vmc_energies}.

For the T-move calculations, we use the size-consistent T-move algorithm~\cite{casula_Tmove_LRDMC_2010}  including an accept/reject step after each proposed T-move~\cite{AndUmr-JCP-21}. This is combined with the form for the reweighting factor from Ref.~\citenum{AndPerUmr-JCP-24} to obtain small time-step errors. In Table~\ref{tab:champ_dmc_branching}, we list the coefficients $c$ which appear in the reweighting factor and are calculated using the correlation times in VMC as described in Ref.~\citenum{AndPerUmr-JCP-24}. For the locality-approximation calculations, we employ the same reweighting factor and limit the size of the exponent to prevent  population explosions with a cutoff whose influence disappears as $\tau\to 0$ . We use a population of 100 walkers for both the TM and LA calculations. In the TM calculations, the walkers are restricted from crossing the nodes of the wave function after the drift-diffusion step, while in the LA calculations, they are allowed to cross to avoid persistent configurations.

\begin{itemize}
    \item Version of the code used for these calculations: v2.4.0 (git: 1b91436)
\end{itemize}

\begin{table}
    \centering
    \begin{tabular}{l|r|c}
    \hline
     Systems  & E$^\textrm{tot}_\textrm{VMC}$ (Ha) & $\sigma^2$ (Ha$^2$) \\     \hline
     Water            & -17.22657(1) & 0.227 \\
     Methane          &  -8.07178(1) & 0.084 \\
     Methane$-$Water    & -25.29611(1) & 0.328 \\
     Methane$---$Water  & -25.29809(1) & 0.314 \\
    \hline
    \end{tabular}
    \caption{VMC total energies and variances of the energy ($\sigma^2$) for the different systems.}
    \label{tab:champ_EU_vmc_energies}
\end{table}

\begin{table}
    \centering
    \begin{tabular}{l|c}
    \hline
     Systems  & $c$ \\     \hline
     Water            & 4.79\\
     Methane          & 3.42 \\
     Methane$-$Water    & 4.09\\
     Methane$---$Water  &  4.27\\
    \hline
    \end{tabular}
    \caption{Coefficient $c$ used in the branching factor.}
    \label{tab:champ_dmc_branching}
\end{table}

\subsubsection{Typical input file: DMC}
\begin{verbatim}[
frame=lines,
framesep=2mm,
baselinestretch=1.0,
fontsize=\footnotesize,
]{python}
%module general
    title           'H2O DMC calculation tau 0.1'
    pool            '../pool/'
    mode            'dmc_one_mpi1'
    pseudopot       ccECP
%endmodule

# Load all the input data
load trexio       ../water_c2v_LDA.hdf5
load jastrow      ../jastrow_3body.jas
load determinants ../single.det

%module blocking_dmc
    dmc_nstep     30
    dmc_nblk      100
    dmc_nblkeq    1
    dmc_nconf     100
    dmc_nconf_new 0
%endmodule

%module dmc
    tau           0.1d0
    etrial      -17.24d0
    icasula       4
    icut_e        2
    ibranching_c  4.79d0
    icross        0
%endmodule
%\end{minted}
\end{verbatim}

\begin{figure*}[!h]
    \includegraphics[width=6.66in]{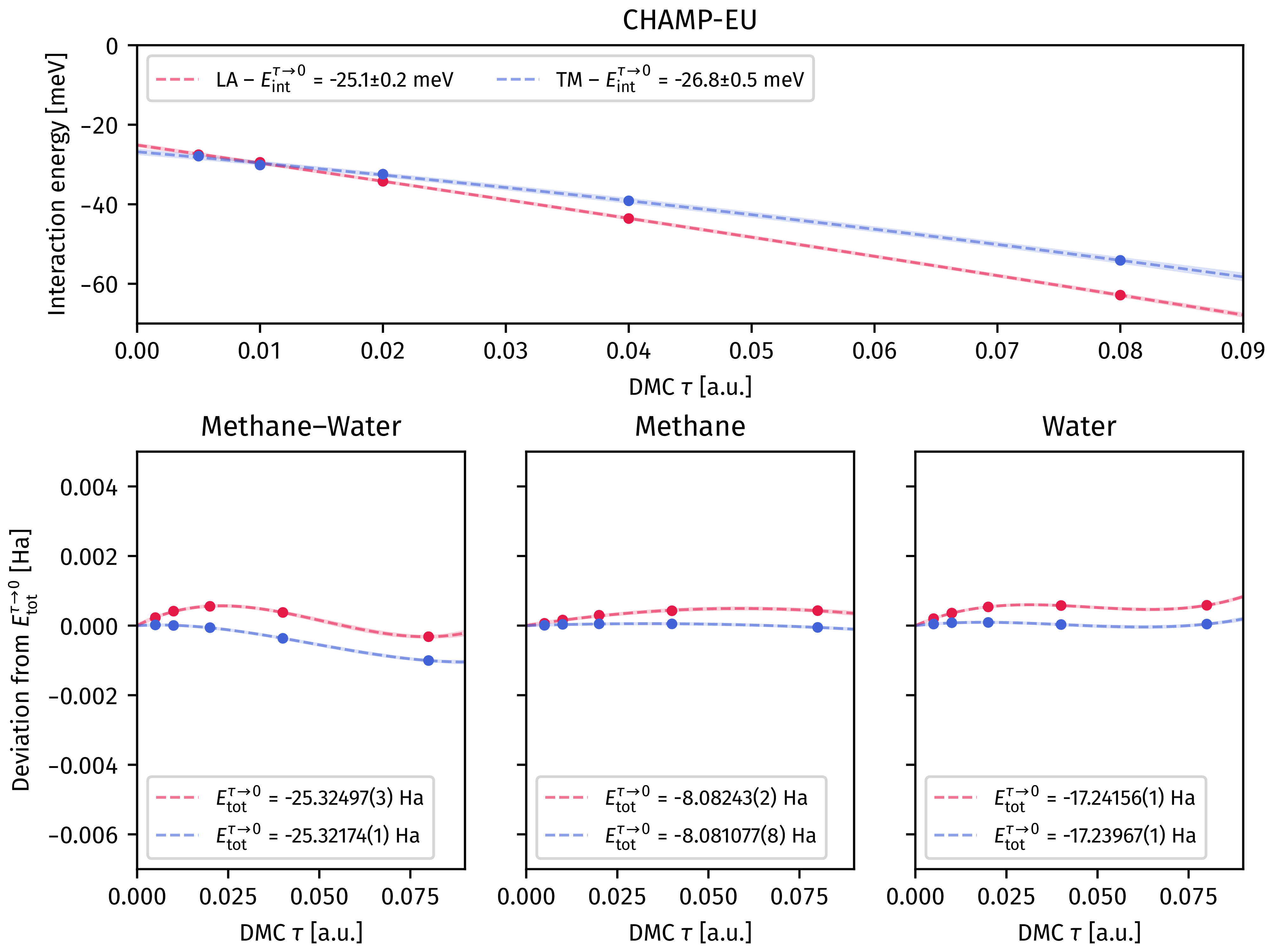}
    \caption{\label{fig:CHAMP-EU_energy_plots} The time-step dependence of the (a) methane-water interaction energy, (b) methane-water dimer total energy, (c) isolated methane molecule total energy, (d) isolated water molecule total energy for the CHAMP-EU code and the LA and TM pseudopotential localization schemes.}
\end{figure*}

\clearpage
\subsection{CHAMP-US}

The Cornell-Holland Ab-initio Materials Package (CHAMP) is a real-space quantum Monte Carlo suite of programs for electronic structure calculations, developed since 1986.
Since 2003, two branches of the program, CHAMP-US and CHAMP-EU have evolved independently, and
each has important capabilities not present in the other, though they have much in common.

CHAMP-US\cite{CHAMP-US} has been used to study the electronic structure of atoms, molecules, periodic solids, 2-dimensional
quantum dots\cite{PedUmrLip-PRB-00,ColPedLipUmr-EPJB-02,GhoUmrJiaUllBar-PRB-05,GucUmr-PRB-05,GucJeoUmrJai-PRB-05,JeoGucUmrJai-PRB-05,GhoGucUmrUllBar-NP-06,GucGhoUmrBar-PRB-08,ZenGeiRuaUmrCho-PRB-09} (possibly in a magnetic field) and quantum wires\cite{GucUmrJiaBar-PRB-09,MehUmrMeyBar-PRL-13}.  Basic physical properties that can be calculated
include ground- and excited-state energies, densities, pair densities and spin densities.  From these
physical phenomena of interest can be studied, e.g., phase transitions in solids or onset of Wigner
localization in quantum dots.  In addition, it has been used to compute nearly exact exchange and correlation
potentials and energies\cite{FilUmrGon-JCP-97,UmrSavGon-Brisbane-98,Al-ResUmr-PRA-98,SavUmrGon-CPL-98} as a benchmark for the approximate functionals used in density functional theory.

Calculations for any system of interest require choosing a functional form for the wave function, and a
program with three basic capabilities, namely wave function optimization, variational Monte Carlo (VMC)
and diffusion Monte Carlo (DMC).  We discuss these next.

{\bf Form of wave function:}  The form of the wave function is system dependent.  Here we briefly discuss
just the form used for atoms and molecules.  There, the wave function is written as a linear combination
of determinants of orbitals, multiplied by a Jastrow factor.  Linear combinations of determinants
are used to form configuration state functions (CSFs) that have the desired spin and space symmetry, thereby
reducing the number of variational parameters.  The orbitals are themselves expanded
in basis functions which are a product of a real spherical harmonic times a radial function.
The radial functions are chosen to be Slater functions for all-electron calculations and either
Gaussian functions or Gauss-Slater functions~\cite{PetTouUmr-JCP-10,PetTouUmr-JCP-11} for pseudopotential calculations.
The Gauss-Slater functions have the advantage that they have the correct exponential decay at large
distances whereas the Gaussian functions decay too quickly.
The electron-electron and the electron-nucleus cusps are always imposed exactly.
The Jastrow function has electron-electron (e-e), electron-nucleus (e-n), and electron-electron-nucleus (e-e-n)terms.
They are expressed in terms of scaled interparticle distances, that go to a constant either exponentially
or as a power law.
In this paper we used the scaling functions, $R=[1-\exp(-\kappa r)]/\kappa$ in the e-e and e-n terms, and $R=\exp(-\kappa r)$ in the e-e-n term, with $\kappa=0.4$.
One can have a Jastrow function of arbitrary order.  In this paper we use a 5th-order Jastrows, though 6th-order and 7th-order
Jastrows have only slightly greater computational expense and lead to significantly lower VMC energies and fluctuations in the local energy.
In order to have excellent size-consistency, not only in DMC but also in VMC, we treat the H atoms of water and methane as two different
atomic species, i.e., the e-e and e-e-n parts of the Jastrow factors are different.  Failure to do this
results in the VMC energy of Methane$---$Water being higher than the sum of the methane and
water energies by 2.2 mHa.

{\bf Wave function Optimization:} All the wave function paramaters (both linear and nonlinear) can be optimized either by minimizing the variance of the local energy\cite{UmrWilWil-PRL-88}
in a VMC calculation or by minimizing any linear combination the expectation value of the energy and the variance
using either the Newton method\cite{UmrFil-PRL-05} or the linear method\cite{TouUmr-JCP-07,UmrTouFilSorHen-PRL-07,TouUmr-JCP-08}.  There are four kinds of wave function parameters: Jastrow, CSF, orbital
and basis exponents.  The first three kinds optimize quickly and usually reproducibly to within statistical error, but
the basis exponent optimization can lead to multiple minima.  In this paper, since the decision was made to use a single determinant
of LDA orbitals expressed in a given basis, only the Jastrow parameters were optimized using VMC energy
minimization, but more typically we would strive for greater accuracy by optimizing also at least
the CSF and orbital parameters.

Recently developed methods for fast evaluation of large multi-determinant wave functions and their derivative~\cite{FilAssMor-JCP-16,AssMorFil-JCTC-17} are used to make the calculations efficient.

{\bf VMC:} The VMC calculations are performed using the algorithm of Ref.~\citenum{Umr-PRL-93} which enables
very large radial and angular moves with large acceptance probabilities.  This results in an autocorrelation
time, $T_{\rm corr}$ that is only a little bit larger than 1, where $\Tcorr = 2\tcorr+1$, and $\tcorr$ is the
usual definition of the integrated autocorrelation time.  Hence each sample. obtained after attempting
moves on all the electrons, is nearly independent of the previous one.
The values of the VMC energies and the variances of the local local energies are shown in
Table~\ref{tab:champ_vmc_energies1}.
They are the same within statistical error as those from CHAMP-EU because we employed identically the same optimized wave functions.
\begin{table}
    \centering
    \begin{tabular}{l|r|c}
    \hline
    Systems  & E$^\textrm{tot}_\textrm{VMC}$ (Ha) & $\sigma^2$ (Ha$^2$) \\     \hline
     Water            & -17.22656(2) & 0.227(1)\\
     Methane          &  -8.07178(1) & 0.084(1)\\
     Methane$-$Water    & -25.29608(2) & 0.328(1)\\
     Methane$---$Water & -25.29808(2) & 0.314(1)\\
    \hline
    \end{tabular}
    \caption{VMC total energies and root mean square fluctuations of the energy ($\sigma$) for the different systems.}
    \label{tab:champ_vmc_energies1}
\end{table}

{\bf DMC:} The DMC calculations are performed using the algorithm of Ref.~\citenum{UmrNigRun-JCP-93} with
some modifications\cite{AndUmr-JCP-21,AndPerUmr-JCP-24}.  First, for systems with more than a few electrons, it is more efficient to perform
the Metropolis-Hastings accept/reject step after moving each electron rather than after moving all of them.
Second, the drift step of the DMC algorithm employs the average velocity over the time step given by
Eq. 35 of Ref.~\citenum{UmrNigRun-JCP-93} with $a=0.5$.
Third, the time-step error in the total energy is reduced by using the reweighting factor of
Ref.~\citenum{AndPerUmr-JCP-24}.
The value of the parameter, $c$, in the reweighting factor is shown in Table~\ref{tab:champ_dmc_branching1}.
Further, in order to reduce the time-step error of the interaction energy, we employ a fragment based
reweighting scheme that has the desirable feature that it is exactly size-consistent, i.e., the energy of a system containing widely
separated fragments is the same as the sum of the energies of the individual fragments.
Fourth, the non-local pseudopotentials are either fully localized
using the locality approximation~\cite{la_mitas}, or partially localized using the T-moves approximation.
The particular form of the T-moves approximation used here is that of Ref.~\citenum{AndUmr-JCP-21}.
Compared to earlier T-move algorithms, this algorithm includes an additional accept-reject step after each T-move which ensures that the exact
distribution is sampled in the limit of an exact trial wave function, and results in a smaller time-step
error, particularly in expectation values of operators that do not commute with the Hamiltonian.
This additional step affects the computational cost per Monte Carlo step negligibly.
The T-move approximation is somewhat more computationally expensive per Monte Carlo step than the
locality approximation, but is actually more efficient that the locality approximation because
the reduced autocorrelation time more than compensates for the increase in computer time.

\begin{table}
    \centering
    \begin{tabular}{l|c}
    \hline
     Systems  & $c$ \\     \hline
     Water            & 4.8\\
     Methane          & 3.4 \\
     Methane$-$Water    & 4.1\\
     Methane$---$Water  &  4.2\\
    \hline
    \end{tabular}
    \caption{Coefficient $c$ used in the reweighting factor.}
    \label{tab:champ_dmc_branching1}
\end{table}

\begin{figure*}[!h]
    \includegraphics[width=6.66in]{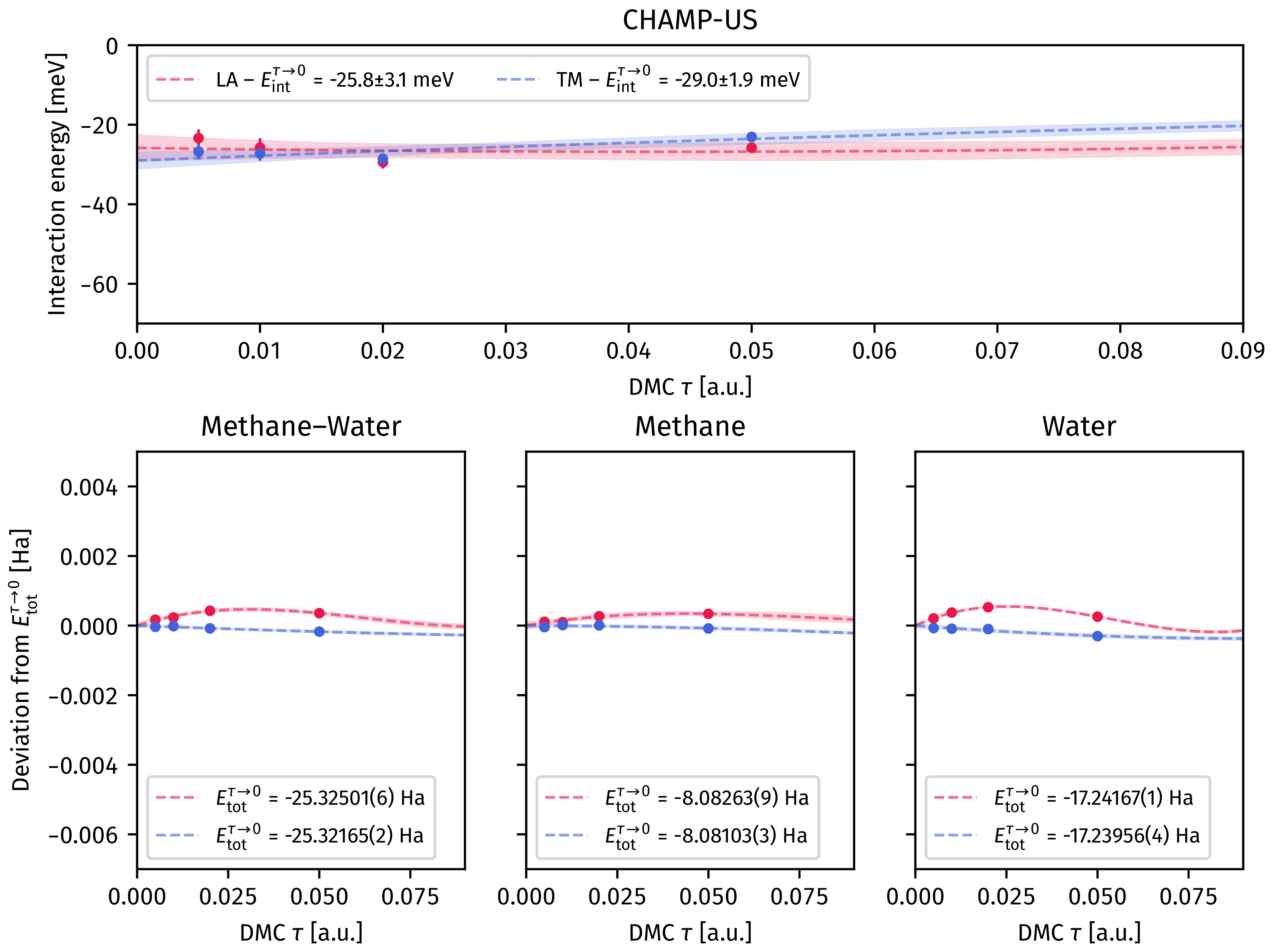}
    \caption{\label{fig:CHAMP_Cornell_Paris_energy_plots} The time step dependence of the (a) methane-water interaction energy (b) methane-water dimer total energy, (c) isolated methane molecule total energy, (d) isolated water molecule total energy in the CHAMP-US code across the LA, TM algorithm(s).}
\end{figure*}

\clearpage
\subsection{CMQMC}

The CSIRO Molecular Quantum Monte Carlo code (CMQMC) is designed to be user-friendly while also enabling straightforward implementation of new algorithms and techniques.
It is written in Fortran, and uses MPI for massive CPU-parallelism.

\begin{figure*}[!h]
    \includegraphics[width=6.66in]{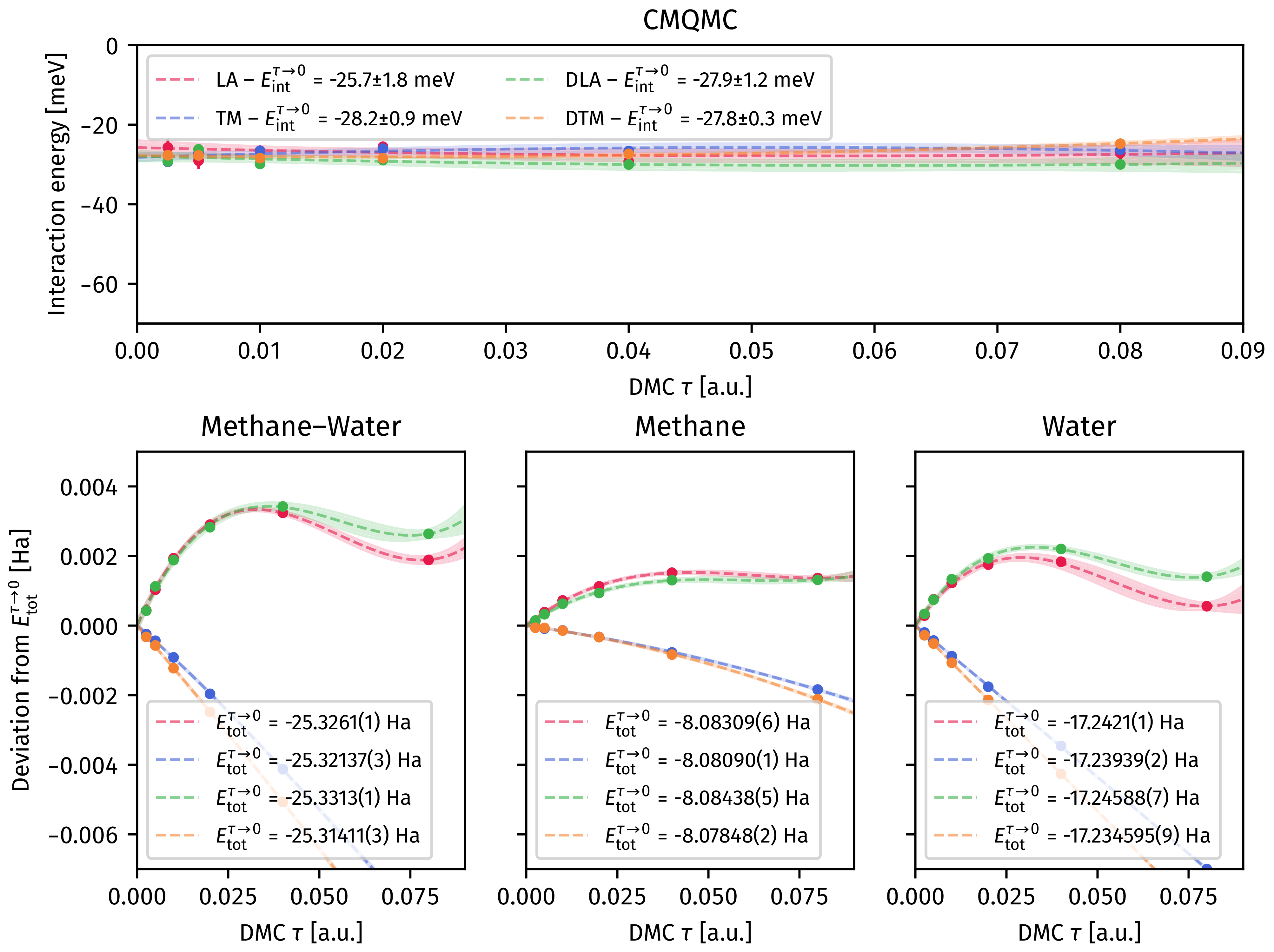}
    \caption{\label{fig:CMQMC_energy_plots} The time step dependence of the (a) methane-water interaction energy (b) methane-water dimer total energy, (c) isolated methane molecule total energy, (d) isolated water molecule total energy in the CMQMC code across the LA, TM, DLA, DTM algorithm(s).}
\end{figure*}

The DMC calculations presented here use a modification of the UNR algorithm \cite{UmrNigRun-JCP-93} that ensures exact size-consistency for all values of the time-step $\tau$. Using the notation of Ref.~\citenum{UmrNigRun-JCP-93}, we employ a modified re-weighting factor
\begin{equation}
\Delta w = \exp \left( \frac{\tilde{S}(\textbf{R}') + \tilde{S}(\textbf{R})}{2}\tau \right)	
\end{equation}
which involves the nominal time-step $\tau$ rather than an effective time-step $\tau_{\textrm{eff}}$.
Our modified branching coefficient $\tilde{S}$ is defined as
\begin{equation}
\tilde{S} = E_{\textrm{T}} - E_{\textrm{est}} + \sum_{i} \left( E_{\textrm{est},i} - E_{\textrm{L},i} \right) \frac{|\bar{\textbf{v}}_{i}|}{|\textbf{v}_{i}|}	
\end{equation}
where $E_{L,i}$ is the contribution of electron $i$ to the local energy,
\begin{equation}
E_{\textrm{L},i} = -\frac{1}{2}\frac{\nabla_{i}^{2}\Psi_{T}}{\Psi_{T}} + V_{i}^{\textrm{pp}} + \frac{1}{2}\sum_{j\neq i} \frac{1}{r_{ij}}	
\end{equation}
so that $\sum_{i}E_{\textrm{L},i} = E_{\textrm{L}}$, and similarly $E_{\textrm{est},i} = \langle E_{\textrm{L},i} \rangle$.
The quantum velocity and its modified form are the same as those used in the UNR algorithm, i.e.
\begin{equation}
\textbf{v}_{i} = \frac{\nabla_{i}\Psi_{T}}{\Psi_{T}}	
\end{equation}
\begin{equation}
\bar{\textbf{v}}_{i} = \frac{-1+\sqrt{1+2a v_{i}^{2}\tau}}{a v_{i}^{2} \tau} \textbf{v}_{i}
\end{equation}
and we use the value $a=1$ in all pseudopotential calculations.
For a composite system AB consisting of non-interacting sub-systems A and B, these modifications (together with a trial wave function satisfying $\Psi_{T}^{[AB]} = \Psi_{T}^{[A]}\Psi_{T}^{[B]}$) ensure that the re-weighting factor of the composite system is the product of the re-weighting factors of the sub-systems.
While exact size-consistency is of course only applicable to non-interacting systems, this approach significantly decreases the time-step errors associated with the interaction energies of weakly interacting systems, as shown in Fig.~\ref{fig:CMQMC_energy_plots}.
All our DMC calculations used a target population size of 16,384 walkers.
Calculations involving T-moves used the scheme labelled ``SVDMC Version 1" in Ref.~\citenum{casula_Tmove_LRDMC_2010}.

Our Jastrow factor is a sum of electron-electron, electron-nucleus, and electron-electron-nucleus terms, each constructed as compactly-supported natural polynomial expansions in the inter-particle distances with smooth Wendland function cutoffs, as described in Ref.~\citenum{swann2017}.
 Each element (H, C, O) has different electron-nucleus and electron-electron-nucleus terms, and in the composite systems the H atoms on the water molecule are treated differently to the H atoms on the methane molecule.
 In the notation of Ref.~\citenum{swann2017}, our settings are 3J.4.5, meaning that we use the three-body Jastrow factor together with a 4-point quadrature grid for treating the non-local component of the pseudopotential, and cut-offs $R$ on the pseudopotentials are defined such that $R$ is the point furthest from the nucleus which deviates by more than $10^{-5}$ from the bare Coulomb potential (local component) or zero (non-local component).
 The parameters in the Jastrow factor were optimised by minimising the variational energy using the linear method.
 The VMC energy and variance obtained for each system is shown in Tables \ref{tab:cmqmc_vmc_energies} and \ref{tab:cmqmc_vmcdla_energies}.

\begin{table}
    \centering
    \begin{tabular}{l|c|c}
    \hline
     Systems  & $E^\textrm{tot}_\textrm{VMC}$ (Ha) & $\sigma_\textrm{VMC}^2$ (Ha$^2$) \\     \hline
     Water             & -17.2218(4) & 0.285(5) \\
     Methane           &  -8.0669(4) & 0.109(2) \\
     Methane$-$Water     & -25.2882(4) & 0.387(1) \\
     Methane$---$Water & -25.2891(4) & 0.384(2)  \\
    \hline
    \end{tabular}
\caption{Total energy ($E^\textrm{tot}$) and variance ($\sigma^2$) of the systems computed using VMC, for the wave functions used in the code CMQMC for the LA and TM schemes.
}\label{tab:cmqmc_vmc_energies}
\end{table}

\begin{table}
    \centering
    \begin{tabular}{l|c|c}
    \hline
     Systems  & $E^\textrm{tot}_\textrm{VMC-DLA}$ (Ha) & $\sigma_\textrm{VMC-DLA}^2$ (Ha$^2$) \\     \hline
     Water             & -17.2248(4) &  0.36(2) \\
     Methane           &  -8.0674(3) &  0.121(1) \\
     Methane$-$Water     & -25.2906(4) &  0.444(2) \\
     Methane$---$Water & -25.2920(4) &  0.451(5) \\
    \hline
    \end{tabular}
\caption{Total energy ($E^\textrm{tot}$) and variance ($\sigma^2$) of the systems computed using VMC, for the wave functions used in the code CMQMC for the DLA and DTM schemes (i.e., non-local pseudo potential terms are projected on the determinant, not on the entire trial wave function).
}\label{tab:cmqmc_vmcdla_energies}
\end{table}

\clearpage
\subsection{PyQMC}
\begin{figure*}[!h]
    \includegraphics[width=6.66in]{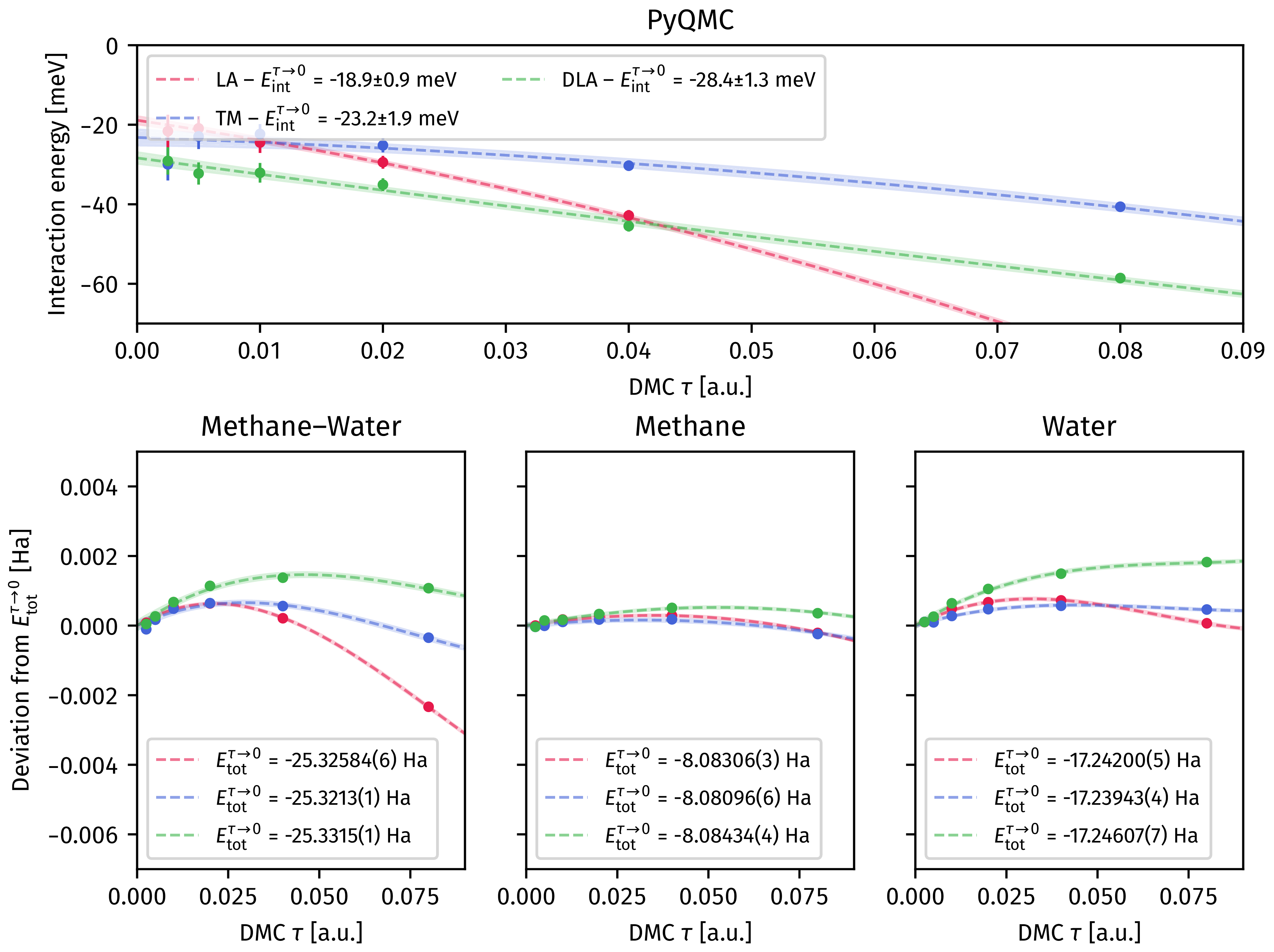}
    \caption{\label{fig:PyQMC_energy_plots} The time step dependence of the (a) methane-water interaction energy (b) methane-water dimer total energy, (c) isolated methane molecule total energy, (d) isolated water molecule total energy in the PyQMC code across the LA, TM, DLA algorithm(s).}
\end{figure*}

\begin{table}
    \centering
    \begin{tabular}{l|c|c}
    \hline
     Systems  & $E^\textrm{tot}_\textrm{VMC}$ (Ha) & $\sigma_\textrm{VMC}^2$ (Ha$^2$) \\     \hline
     Water             & -17.188608(9) &   0.35880(3) \\
     Methane           &  -8.055812(5) &  0.13465(1) \\
     Methane$-$Water     & -25.22930(1) &  0.54360(4) \\
     Methane$---$Water & -25.20694(1) & 0.58982(3)\\
    \hline
    \end{tabular}
\caption{Total energy ($E^\textrm{tot}$) and variance ($\sigma^2$) of the wave function for all systems computed using VMC, for the wave functions used in the code pyqmc..
}\label{tab:pyqmc_vmc_energies}
\end{table}

\texttt{PyQMC} is a Python-based program designed for method development. As such, it places a high premium on simplicity of implementation. By default, it implements a non-size-consistent version of the algorithm suggested by Anderson and Umrigar, and as such should not be expected to perform particularly accurately in the `size-consistent' tests. 

It is relatively simple to modify the DMC algorithm in \texttt{PyQMC}, which we did to implement the LA and DLA algorithms. The scripts are included in the Supplementary information. 

The Jastrow used in this work was of the form
\begin{equation}
    \Psi = \Psi_S e^{U_{2b}} e^{U_{gem}}.
\end{equation}
The two-body term $U_{2b}$ is expanded as 
\begin{equation}
    U_{2b} = \sum_{i\alpha m} a_m(r_{i\alpha}) + \sum_{ijn} b_n(r_{ij}),
\end{equation}
where $i,j$ refer to electron indices, and $\alpha$ refers to nuclear indices. The $a_m$ and $b_n$ functions are 
\begin{equation}
     \frac{1-p(z)}{1+\beta p(z)}, \quad z = r/r_{\rm cut}, p(z) = 6z^2 - 8z^3 + 3z^4
\end{equation}
The geminal function
\begin{equation}
    U_{gem} = \sum_{ijmn} c_{mn} \chi_m(r_i)\chi_n(r_j)
\end{equation}
is used to further improve the wave function. 

The wave functions were found using a modified stochastic reconfiguration\cite{SR1998SOR} method in which the learning rate was determined using a correlated sampling line minimization. 
We have found that minimizing a combination of energy and variance during the correlated sampling can improve the stability of the algorithm significantly.

The relatively large timestep error on the difference between the separated molecules and the complex is due to the fact that in the methane system we happen to have virtually no timestep error, in the water system, the error is positive, and in the complex it happens to be negative. Thus there is an anti-cancellation of error in this case. 
In principle one could tune the algorithm and/or wave functions to change this property but we did not do so.
We also found that for our wave functions, the LA was somewhat unstable for the separated complex at very small timesteps, often getting 'stuck,' which resulted in large statistical uncertainties.

\clearpage
\subsection{QMC=Chem}
\subsubsection{Code information}

QMC=Chem is a QMC program specifically designed for use with very large multi-determinant wave functions, particularly in a post-Full Configuration Interaction context.
It efficiently utilizes CIPSI (Configuration Interaction using a Perturbative Selection made Iteratively) wave functions generated by Quantum Package (\url{https://quantumpackage.github.io/qp2}) as trial wave functions.
When employing ECPs, QMC=Chem currently supports only the DLA.

Below is an example of a bash script for performing a DMC calculation on the methane molecule using a wave function stored in a TREXIO file.
This script assumes that Quantum Package (\url{(https://quantumpackage.github.io/qp2}) is installed on the system along with its external \texttt{qmcchem} module (available at \url{https://gitlab.com/scemama/qp_plugins_scemama}).
The DMC algorithm employed is the stochastic reconfiguration method described in \citenum{assaraf-99}.
This method represents a hybrid between pure Diffusion Monte Carlo and conventional Diffusion Monte Carlo.
It is characterized by two key parameters: the population size and the projection time.
With a single walker, the algorithm behaves as pure Diffusion Monte Carlo.
When using multiple walkers, if the projection time is set to zero, the method is equivalent to conventional DMC.
Typically, the population size required is smaller than that in conventional DMC, provided the projection times are sufficiently long.
In the calculations presented in this article, we used a population of 100 walkers and a projection time of 2.5 atomic units (equivalent to 1000 Monte Carlo steps with a time step of 0.0025).

\begin{verbatim}
frame=lines,
framesep=2mm,
baselinestretch=1.0,
fontsize=\footnotesize,
]{bash}
# Convert the TREXIO file into Quantum Package format
qp_import_trexio.py methane.h5 -o methane

# Export data for QMC=Chem
qp set_file methane
qp run save_for_qmcchem

# Set up a VMC run for 10 minutes, split into 30 second blocks with 100 walkers per core
qmcchem edit --method=VMC \
             --block-time=30 \
             --stop-time=600 \
             --walk-num=100 \
             methane

# Run the VMC
qmcchem run methane

# Set up a DMC run for one hour, split into 5 minute blocks with 100 walkers per core
qmcchem edit --method=SRMC \
             --block-time=300 \
             --stop-time=3600 \
             --sampling=Brownian \
             --projection-time=2.5 \
             --time-step=0.0025 \
             methane

# Run the VMC
qmcchem run methane

\end{verbatim}

\begin{figure*}[!h]
    \includegraphics[width=6.66in]{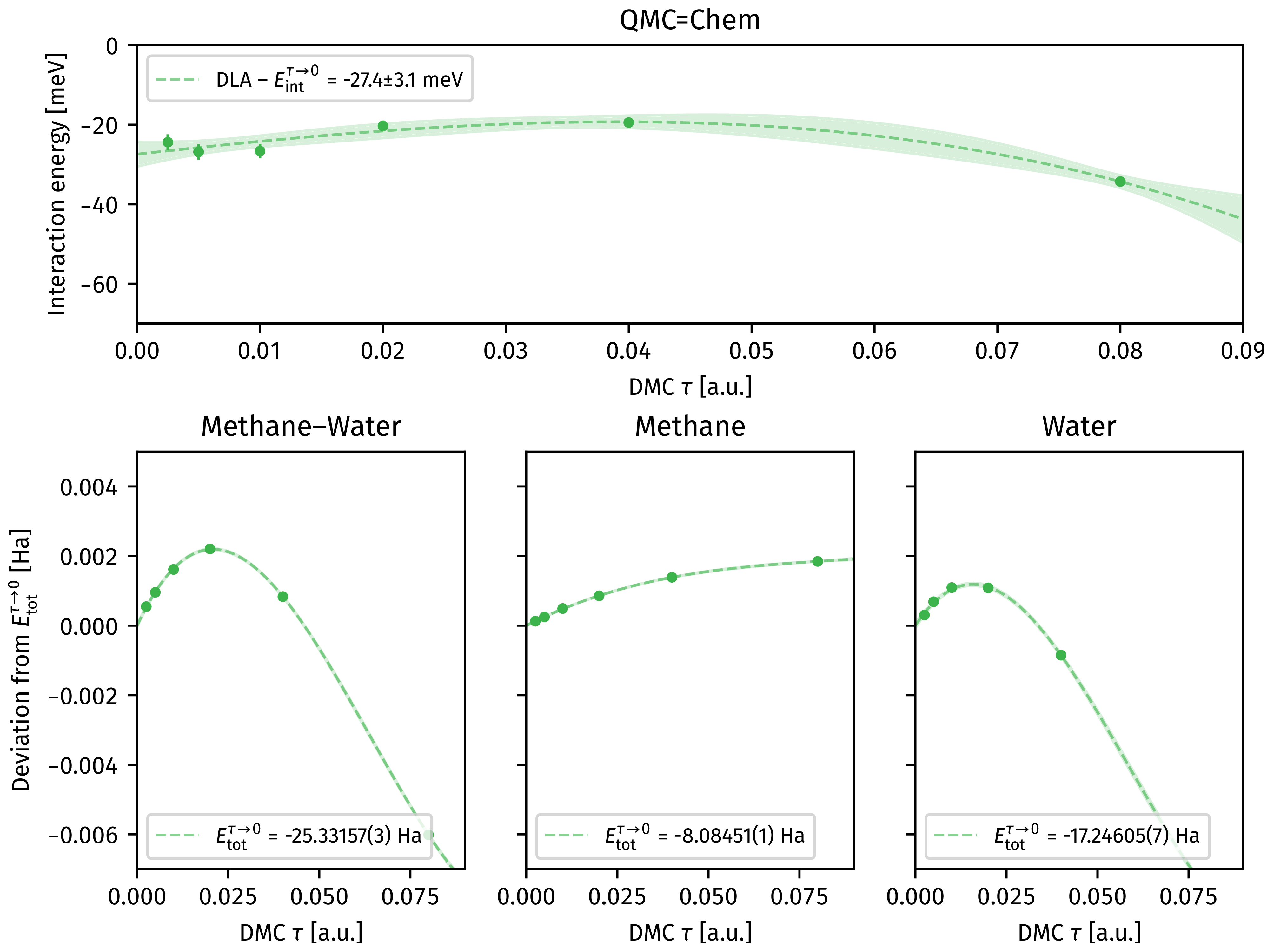}
    \caption{\label{fig:QMC=Chem_energy_plots} The time step dependence of the (a) methane-water interaction energy (b) methane-water dimer total energy, (c) isolated methane molecule total energy, (d) isolated water molecule total energy in the QMC=Chem code across the DLA algorithm.}
\end{figure*}

\begin{table}
    \centering
    \begin{tabular}{l|r|r}
    \hline
     Systems  & $E^\textrm{tot}_\textrm{VMC-DLA}$ (Ha) & $\sigma_\textrm{VMC-DLA}^2$ (Ha$^2$) \\     \hline
     Single determinant only & & \\
     Water                   & -16.936496 & 3.124(6) \\
     Methane                 &  -7.838840 & 1.471(5) \\
     Methane$-$Water           & -24.774290 & 4.611(8) \\ %
     Methane $\cdots$ Water  & -24.775334 & 4.581(8) \\ %
    \hline
     Single determinant with Jastrow & & \\
     Water                   & -17.1121(2) & 0.715(1) \\
     Methane                 &  -8.0172(1) & 0.2198(1) \\
     Methane$-$Water           & -25.1301(3) & 0.9356(7)\\
     Methane $\cdots$ Water  & -25.1271(3) & 0.940(2) \\
     \hline
    \end{tabular}
\caption{Total energy ($E^\textrm{tot}$) and variance ($\sigma^2$) of the systems computed using VMC, for the wave functions used in the code QMC=Chem for the DLA scheme (i.e., non-local pseudo potential terms are projected on the determinant, not on the entire trial wave function).
}\label{tab:qmcchem_vmcdla_energies}
\end{table}

\clearpage

\subsection{QMCPACK}

QMCPACK is a high-performance real space QMC code capable of performing molecular and solid state calculations on modern CPU and GPU machines.
The QMCPACK version used for the Jastrow optimization and for the DMC
LA, TM and DLA calculations is 3.12.0,
while for the DMC DTM calculations we used 3.17.9
(Git commit hash: 15ab1c8,
as in previous versions the DTM implementation was affected by a bug).

The orbitals of the Slater determinant have been computed using PySCF\cite{sunpyscf2018} and converted using the tools available in the QMCPACK package.
We used the default Jastrow factor implemented in QMCPACK, which includes electron-nucleus, electron-electron, and electron-electron-nucleus terms.
The electron-nucleus and electron-electron functions are both one-dimensional B-spline (tricubic spline on a linear grid) between zero and a cutoff distance.
The electron-electron-nucleus function is a polynomial expansion.
The parameters of the Jastrow factor have been optimized by minimizing the variational energy of each system, employing the linear method with line minimization via quartic polynomial fits, and performing several steps of optimization with a sampling of up to 1,000,000 configurations.
The optimization has been performed on each system both with and without the DLA approximation, and the variational energy and variance of each optimized system is given in Tables~\ref{tab:qmcpack_vmc_energies} and \ref{tab:qmcpack_vmcdla_energies}.

\begin{figure*}[!h]
    \includegraphics[width=6.66in]{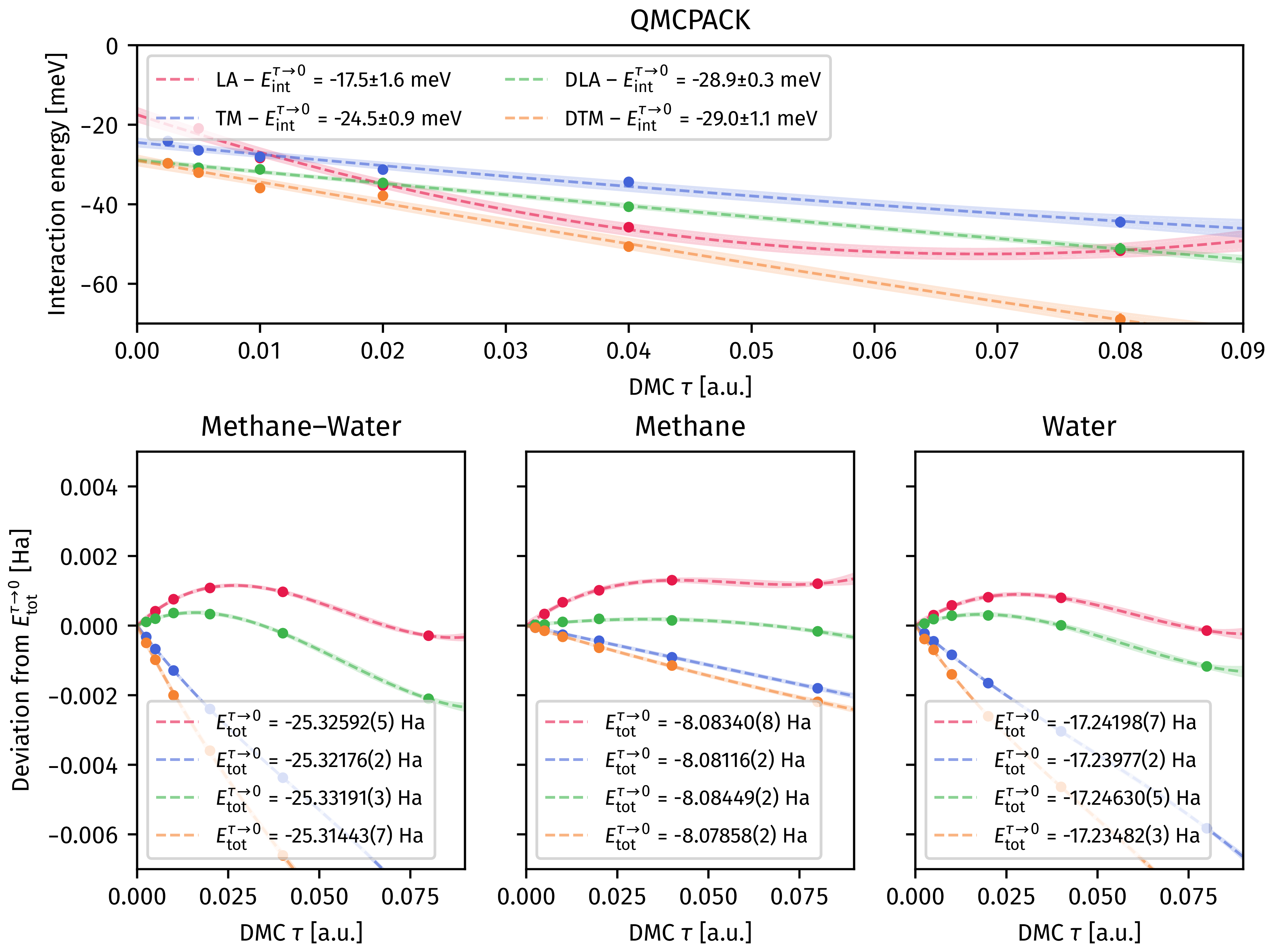}
    \caption{\label{fig:QMCPACK_energy_plots} The time step dependence of the (a) methane-water interaction energy (b) methane-water dimer total energy, (c) isolated methane molecule total energy, (d) isolated water molecule total energy in the QMCPACK code across the LA, TM, DLA algorithm(s).}
\end{figure*}

DMC simulations have been performed with a target population of 102,400 walkers, and employing the modification to the drift and branching terms suggested in Ref.~\citenum{ZSGMA}, called with the flag  \verb|ZSGMA|.
The real version of QMCPACK has been used for the DMC calculations with TM, DLA and DTM. The LA DMC simulations were performed with the complex version (fixed-phase approximation) because the real version was not stable.
The results obtained with the above setup are reported in Fig.~\ref{fig:QMCPACK_energy_plots}.

\begin{table}
    \centering
    \begin{tabular}{l|c|c}
    \hline
     Systems  & $E^\textrm{tot}_\textrm{VMC}$ (Ha) & $\sigma_\textrm{VMC}^2$ (Ha$^2$) \\     \hline
     Water             & -17.2198(3) & 0.260(2) \\   
     Methane           & -8.0658(2) & 0.1026(5) \\   
     Methane$-$Water     & -25.2808(3) & 0.382(2) \\   %
     Methane$---$Water & -25.2808(3) & 0.379(3) \\   %
    \hline
    \end{tabular}
\caption{Total energy ($E^\textrm{tot}$) and variance ($\sigma^2$) of the systems computed using VMC, for the wave functions used in the code QMCPACK for the LA and TM schemes.
}\label{tab:qmcpack_vmc_energies}
\end{table}

\begin{table}
    \centering
    \begin{tabular}{l|c|c}
    \hline
     Systems  & $E^\textrm{tot}_\textrm{VMC-DLA}$ (Ha) & $\sigma_\textrm{VMC-DLA}^2$ (Ha$^2$) \\     \hline
     Water             & -17.2195(2) & 0.321(2) \\   
     Methane           &  -8.0658(1) & 0.1155(6) \\  %
     Methane$-$Water     & -25.2795(3) & 0.461(5) \\   %
     Methane$---$Water & -25.2810(3) & 0.452(2) \\   %
    \hline
    \end{tabular}
\caption{Total energy ($E^\textrm{tot}$) and variance ($\sigma^2$) of the systems computed using VMC, for the wave functions used in the code QMCPACK for the DLA and DTM schemes (i.e., non-local pseudo potential terms are projected on the determinant, not on the entire trial wave function).
}\label{tab:qmcpack_vmcdla_energies}
\end{table}

\clearpage
\subsection{QMeCha}

The QMeCha code is an experimental code written in Fortran03 that has been developed since 2017 by Barborini and coworkers\cite{QMeCha} and is now in its release stage.

The calculations presented in this comparison, use the Slater determinant build from Kohn-Sham single particle molecular orbitals, obtained using the Perdew-Zunger exchange-correlation functional and the cc-pVTZ basis set, using the ORCA\cite{neeseORCAQuantumChemistry2020a} package.

The Jastrow factor used in the calculations is optimized using the Stochastic Reconfiguration method\cite{sor+01prb}.
The form of the Jastrow factor resembles that of the TurboRVB~{\cite{turborvb}} code introduced by Casula \textit{et al.} in ref. \citenum{AGP2003CAS}, as the linear combination of three terms
\begin{equation}
J(\textbf{R})
=
\text{exp} \left \{
\mathcal{J}^{en}(\textbf{R})
+\mathcal{J}_c^{ee}(\textbf{R})
+\mathcal{J}_{3/4}(\textbf{R})
\right \} .
\end{equation}
that are respectively, a one-body electron nuclear term that is used to remodulate the wave functions' amplitudes around the nuclei and is written as the linear combination
\begin{equation}
\mathcal{J}^{en}(\textbf{R})=\sum_{i=1}^{N_e}\sum_{a=1}^{N_n}\sum_{n=1}^{N} g^{a}_n e^{-\zeta^a_n (\textbf{r}_{i}-\textbf{R}_{a})^2},
\end{equation}
of non-normalized Gaussian functions,
the two-body cusp function, written as
\begin{equation}
\mathcal{J}_c^{ee}(\bar{\textbf{r}})=\sum_{j>i=1}^{N_e} f_{ee}(r_{ij})
\end{equation}
where
\begin{equation}
f_{ee}(r_{ij}) =
\left \lbrace
\begin{array}{ll}
-\frac{1}{4b^p(1+b^p r_{ij})}  +\sum_{n=1}^{N} g^p_n e^{-\zeta^p_n r_{ij}^2}  & indis. \\
-\frac{1}{2b^a(1+b^a r_{ij})}  +\sum_{n=1}^{N} g^a_n e^{-\zeta^a_n r_{ij}^2}  & dis. \\
\end{array}
\right . .
\end{equation}
and the dynamical three/four body Jastrow factor written as a combination of products of atomic Jastrow orbitals
\begin{equation}
\mathcal{J}_{3/4}(\textbf{R}) = \sum_{j>i=1}^{N_e} \sum_{q,p=1}^{\mathcal{Q}} \gamma_{qp} \chi_q(\textbf{r}_i) \chi_p(\textbf{r}_j).
\label{eq:jas34}
\end{equation}
Since the Jastrow factor must be symmetric with respect to the exchange of all the electrons, the $\gamma_{qp}$ parameters satisfy the condition $\gamma_{qp}=\gamma_{pq}$.

This factor slightly differs from that of TurboRVB~{\cite{turborvb}} from the fact that the one-body operator is independent for each atom non-connected by symmetry, and the three-/four- body term is built on atomic orbitals in which the angular part is normalized.

\begin{figure*}[t]
    \includegraphics[width=6.66in]{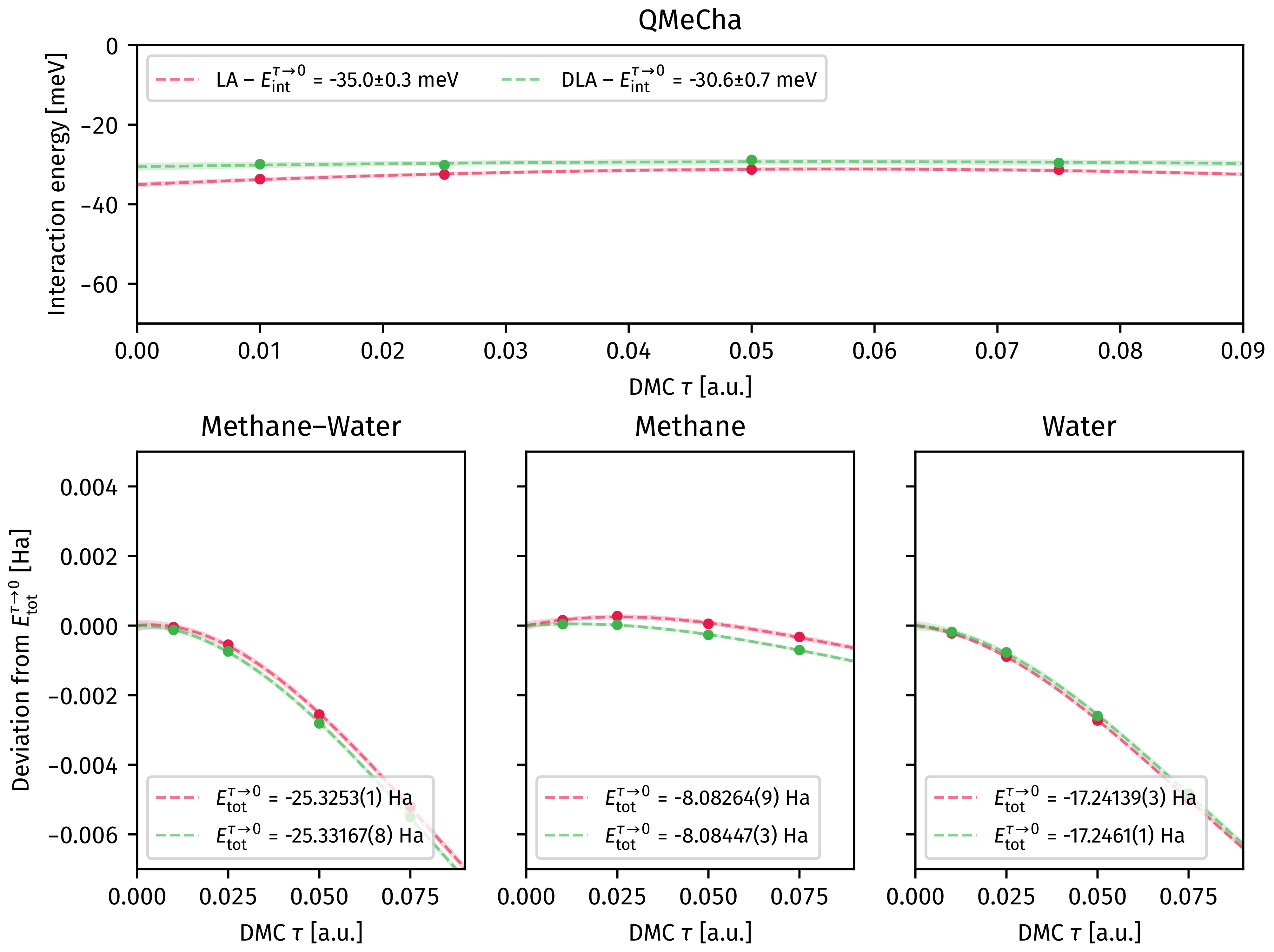}
    \caption{ The time step dependence of the (a) methane-water interaction energy (b) methane-water dimer total energy, (c) isolated methane molecule total energy, (d) isolated water molecule total energy in the QMeCha code across the LA, DLA algorithm(s).}\label{fig:QMeCha_energy_plots}
\end{figure*}

To integrate the non-local components of the pseudopotential in FN-DMC, we have used a 6-point grid with a cut-off of $10^{-5}$ Ha on the energy components.
The total number of walkers in the population is fixed at 12800 and the integration was carried out using the Stochastic Reconfiguration algorithm\cite{SR1998SOR}.
In the DLA pseudopotential integration procedure we always consider the Slater determinant and the one-body Jastrow factor, only excluding the explicit electron-electron interaction terms.

Finally, as can be seen from Figure \ref{fig:QMeCha_energy_plots}, QMeCha introduces an energy cut-off scheme, similar to that implemented by Zen et al. in ref. \cite{ZSGMA} but applied to the single particle energies as suggested in ref. \cite{Umrigar_time_step_2021}.
This approach, which uses a fixed time-step in the branching factor and in which variable single particle energies are used as references for the cut-off, greatly reduces the size-consistency error, and at least for small systems guarantees a nearly unbiased estimation of the interaction energy.

\begin{table}
    \centering
    \begin{tabular}{l|c|c|c|c}
    \hline
     Systems  & $E^\textrm{tot}_\textrm{VMC}$ (Ha) & $\sigma_\textrm{VMC}^2$ (Ha$^2$) & $E^\textrm{tot}_\textrm{VMC-DLA}$ (Ha) & $\sigma_\textrm{VMC-DLA}^2$ (Ha$^2$) \\     \hline
     Water            & -17.22583(11) & 0.24182 & -17.22737(13) & 0.34991  \\
     Methane          & -8.07010(7) &  0.09396 & -8.07093(7) &  0.10149   \\
     Methane$-$Water     & -25.29860(13)&  0.33151 & -25.30114(14) & 0.39128  \\
     Methane$---$Water  & -25.29604(14) & 0.33545 & -25.29884(15) & 0.39893  \\
    \hline
    \end{tabular}
\caption{Total energy ($E^\textrm{tot}$) and variance ($\sigma^2$) of the systems computed using VMC, for the wave functions used in the QMeCha code for the LA and DLA schemes.
}\label{tab:qmecha_vmc_energies}
\end{table}

\clearpage
\subsection{QWalk}

The QWalk code components used in the calculations presented here adhere to the original framework described in Ref.~\citenum{qwalk}. The code primarily employs the DMC process and algorithm based on the Umrigar-Runge-Nightingale scheme, as outlined in Ref.\citenum{UmrNigRun-JCP-93}, with minor modifications. The T-Moves scheme follows the early implementation introduced in 2006~\cite{casula_Tmove_2006}. All computations employed dense ECP-integration grids to facilitate improved bias cancellation across energy differences~\cite{swann2017,dubeck2019a}.

Variational Slater-Jastrow trial wave functions with fixed, tightly-converged one-particle DFT orbitals (Slater exchange and PZ81 correlation functional) from the GAMESS code \cite{GAMESS} were employed. The Schmidt-Moskowitz \cite{moskow1992q} up to a three-center Jastrow factor was used, as detailed in our review \cite{acta} (freely accessible on the Acta Physica Slovaca website). We utilized spin-restricted Jastrow electron-electron, electron-nucleus, and electron-electron-nucleus terms. Each term was expressed as a linear expansion of three polynomial Padé functions with non-linear basis-set curvature-adjusting parameters and a cutoff radius of 7.5~\AA{} \cite{acta}. In total, 31 variational parameters per atom type were used, along with 6 additional parameters for the electron-electron terms, including homogeneous and cusp contributions.

For each system, the Hessian-driven variational Monte Carlo optimization of the parametric Jastrow term was conducted over a minimum of 200 iterations, using a fixed distribution of $\sim$32,000 walkers, refreshed every 10 iterations. The optimization process utilized a cost function defined as a linear combination of energy (95\%) and variance (5\%)\cite{dbck2014pccp}.

The VMC energies and variances, for each of the systems considered, are reported in Tables \ref{tab:qwalk_vmc_energies} and \ref{tab:qwalk_vmcdla_energies}.

\begin{table}
    \centering
    \begin{tabular}{l|c|c}
    \hline
     Systems  & $E^\textrm{tot}_\textrm{VMC}$ (Ha) & $\sigma_\textrm{VMC}^2$ (Ha$^2$) \\     \hline
     Water              &  -17.22598(9) & 0.207 \\
     Methane            & -8.07042(5) & 0.083 \\
     Methane$-$Water      & -25.2933(1) &  0.311\\
     Methane$---$Water  & -25.2944(1) &  0.300 \\
    \hline
    \end{tabular}
\caption{Total energies ($E^\textrm{tot}$) and variances ($\sigma^2$) of the systems computed using VMC, for the wave functions used in the code QWalk code for the LA and TM schemes.
}\label{tab:qwalk_vmc_energies}
\end{table}

\begin{table}
    \centering
    \begin{tabular}{l|c|c}
    \hline
     Systems  & $E^\textrm{tot}_\textrm{VMC-DLA}$ (Ha) & $\sigma_\textrm{VMC-DLA}^2$ (Ha$^2$) \\     \hline
     Water              & -17.2251(1) &  0.203\\
     Methane            & -8.07039(7) &  0.082\\
     Methane$-$Water      & -25.2935(1) &  0.308\\
     Methane$---$Water  & -25.2943(3) &  0.303\\
    \hline
    \end{tabular}
\caption{Total energies ($E^\textrm{tot}$) and variances ($\sigma^2$) of the systems computed using VMC, for the wave functions used in the code QWalk code for the DLA scheme.
}\label{tab:qwalk_vmcdla_energies}
\end{table}

\begin{figure*}[!h]
    \includegraphics[width=6.66in]{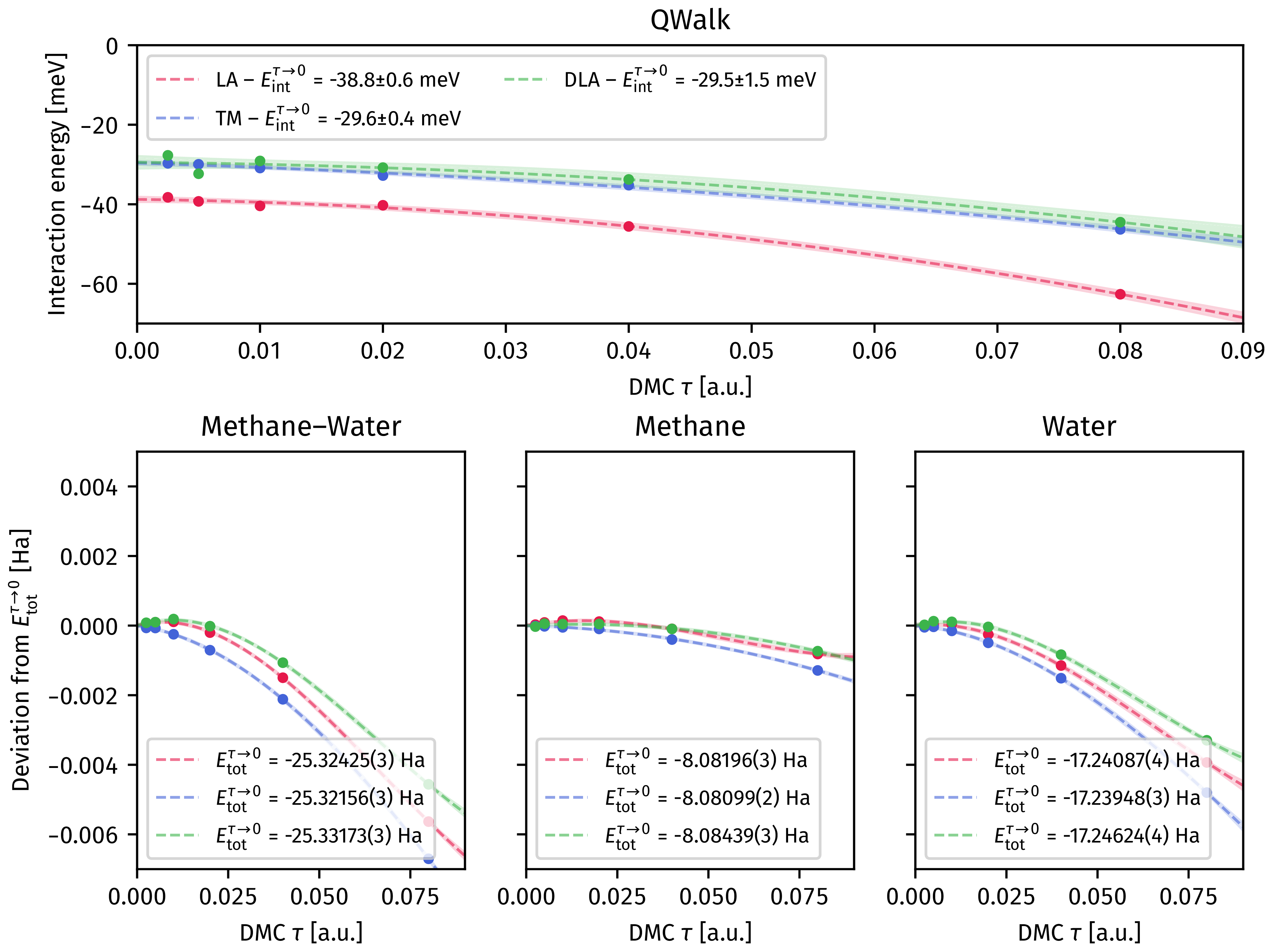}
    \caption{\label{fig:QWalk_energy_plots} The time step dependence of the (a) methane-water interaction energy (b) methane-water dimer total energy, (c) isolated methane molecule total energy, (d) isolated water molecule total energy in the QWalk code across the LA, TM, DLA algorithm(s).}
    \label{fig:qwalk_energy_plots}
\end{figure*}

Finally, for each system and timestep, the optimized trial wave function was employed in the DMC imaginary-time projection, utilizing $\sim$16,000 walkers\cite{dbck2013jctc}. The projection included 20 a.u. of thermalization followed by 4000 a.u. of production computations. Each 1 a.u. was treated as a block, consisting of (1/timestep) block-steps. For instance, in a computation with a timestep of 0.005 a.u., each block comprised 200 DMC steps. The results obtained with such a setup are reported in Fig.~\ref{fig:qwalk_energy_plots}.

\clearpage
\subsection{TurboRVB}
\subsubsection{Code information}
TurboRVB~{\cite{turborvb}} is designed to perform \textit{ab initio} QMC simulations for both molecular and bulk systems. It implements three well-established QMC algorithms: variational Monte Carlo (VMC), DMC, and LRDMC.

Trial wave functions in TurboRVB can range from a simple Slater-Jastrow form to a resonating valence bond (RVB) type, such as the Jastrow-antisymmetrized geminal power~{\cite{AGP2003CAS}} and Jastrow-Pfaffian~{\cite{Genovese2020}} wave functions. These variational \textit{ans\"atze} are optimized by robust minimization techniques available in TurboRVB, such as the stochastic reconfiguration~{\cite{SR1998SOR, SR2007SOR}}.
The adjoint algorithmic differentiation (AAD) efficiently differentiates many-body wave functions, facilitating atomic force calculations, structural optimizations, and molecular dynamics. The TurboRVB package is extended through TurboGenius and TurboWorkflows~{\cite{TurboGenius}}, allowing for automatic and high-throughput QMC calculations. These tools are implemented in Python 3 and provide a user-friendly interface for managing complicated procedures.

The TurboRVB package is an open-source project, available on GitHub under the GPLv3 license. It has been highly optimized for modern CPU- and GPU-based supercomputers, including Fugaku (RIKEN, Japan) and Leonardo (CINECA, Italy). Recently, TurboRVB has been interfaced with the TREXIO library.

The git version of the code used for DMC and LRDMC calculations is 5f3b44a (v1.0.0).

The VMC energy and variance for each system considered are reported on Table~{\ref{tab:turborvb_vmc_energies}}.

\begin{table}
    \centering
    \begin{tabular}{l|c|c}
    \hline
     Systems  & $E^\textrm{tot}_\textrm{VMC}$ (Ha) & $\sigma_\textrm{VMC}^2$ (Ha$^2$) \\     \hline
     Water             & -17.21232(2) & 0.36424(5) \\   
     Methane           &  -8.05874(2) & 0.14569(3) \\   
     Methane$-$Water     & -25.27004(2) & 0.52451(7) \\   
     Methane$---$Water & -25.27013(1) & 0.5216(2)  \\   
    \hline
    \end{tabular}
\caption{Total energy ($E^\textrm{tot}$) and variance ($\sigma^2$) of the systems computed using VMC, for the wave functions used in the code TurboRVB.
}\label{tab:turborvb_vmc_energies}
\end{table}

\subsubsection{TurboRVB-DMC}

For the TurboRVB DMC calculations reported here, we used the ``version 1'' of the size-consistent variational formulation, as detailed in Ref.~\citenum{casula_Tmove_LRDMC_2010}. In this version the T-move Green function is written as a product of single-electron contributions, and it is applied after a drift-diffusion move involving all the electrons. The weighting factors are computed according to the recipe published in Ref.~\citenum{ZSGMA}, which guarantees a size-consistent projection of the wave function. The $\alpha$ parameter of the energy cutoff $\alpha \sqrt{N/\tau}$ appearing in the weights is set to 0.4. For the branching step, we used a population made of 1408 walkers, whose number has been kept fixed throughout the simulation\cite{GFMC_2}. A branching step is performed after 4 applications of the all-electron Green function, leading to a walkers' survival rate ranging from 96.0\% to 99.6\%, according to the time step used. The remaining population bias has been cured by the ``correcting factors'' technique\cite{2017beccaqmcbook, turborvb}.

\begin{figure*}[!h]
    \includegraphics[width=6.66in]{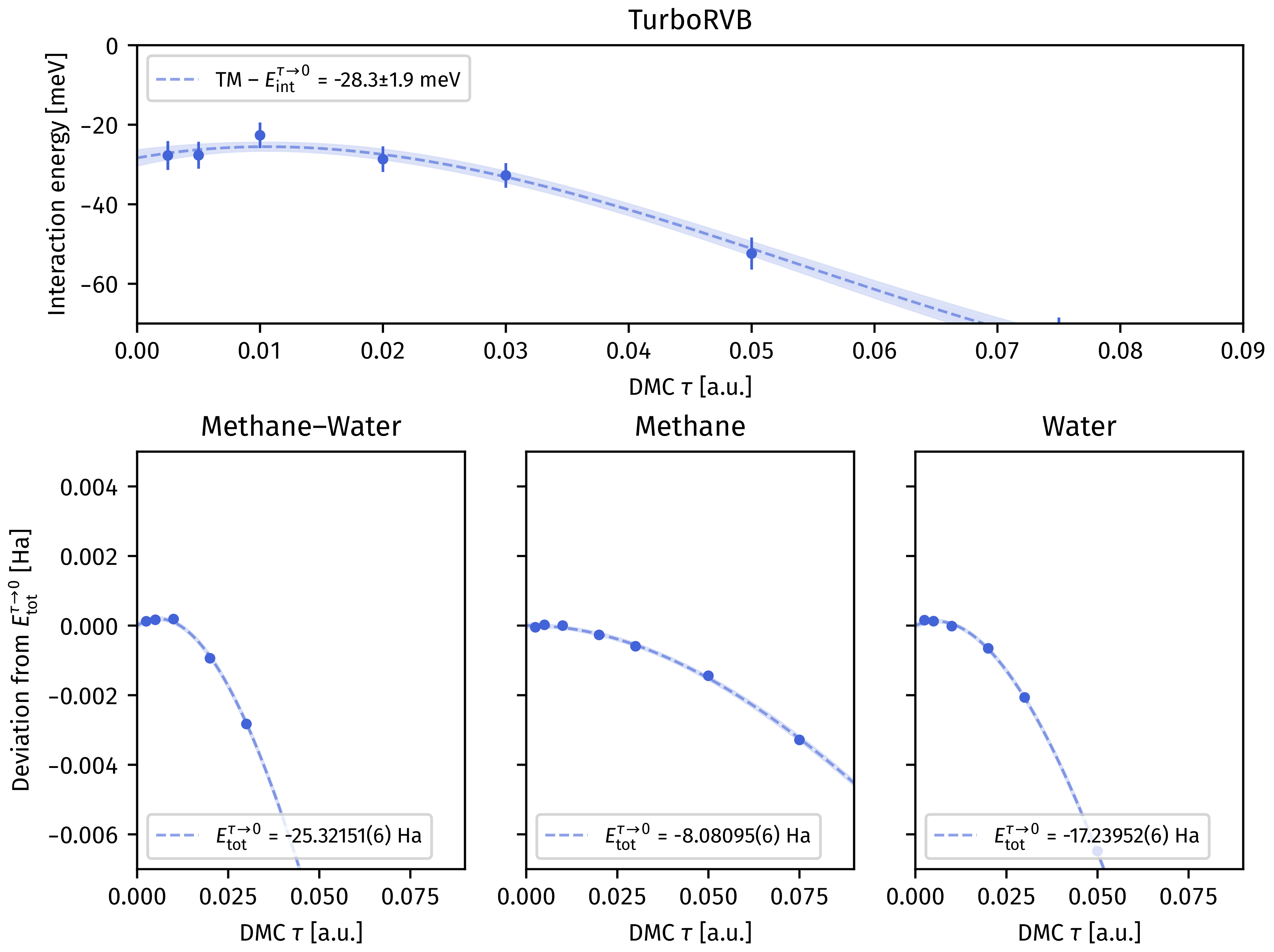}
    \caption{\label{fig:TurboRVB_DMC_energy_plots} The time step dependence of the (a) methane-water interaction energy (b) methane-water dimer total energy, (c) isolated methane molecule total energy, (d) isolated water molecule total energy in the TurboRVB-DMC code across the TM algorithm(s).}
\end{figure*}

\clearpage
\subsubsection{TurboRVB-LRDMC}

For our LRDMC calculations, we used the same branching scheme as in DMC, with a fixed number of 9216 walkers, and a projection time between two consecutive branching steps of 0.1 H$^{-1}$. The walkers' survival rate is in this case around 98 \%. The residual population bias has been corrected as in DMC.

\begin{figure*}[!h]
    \includegraphics[width=6.66in]{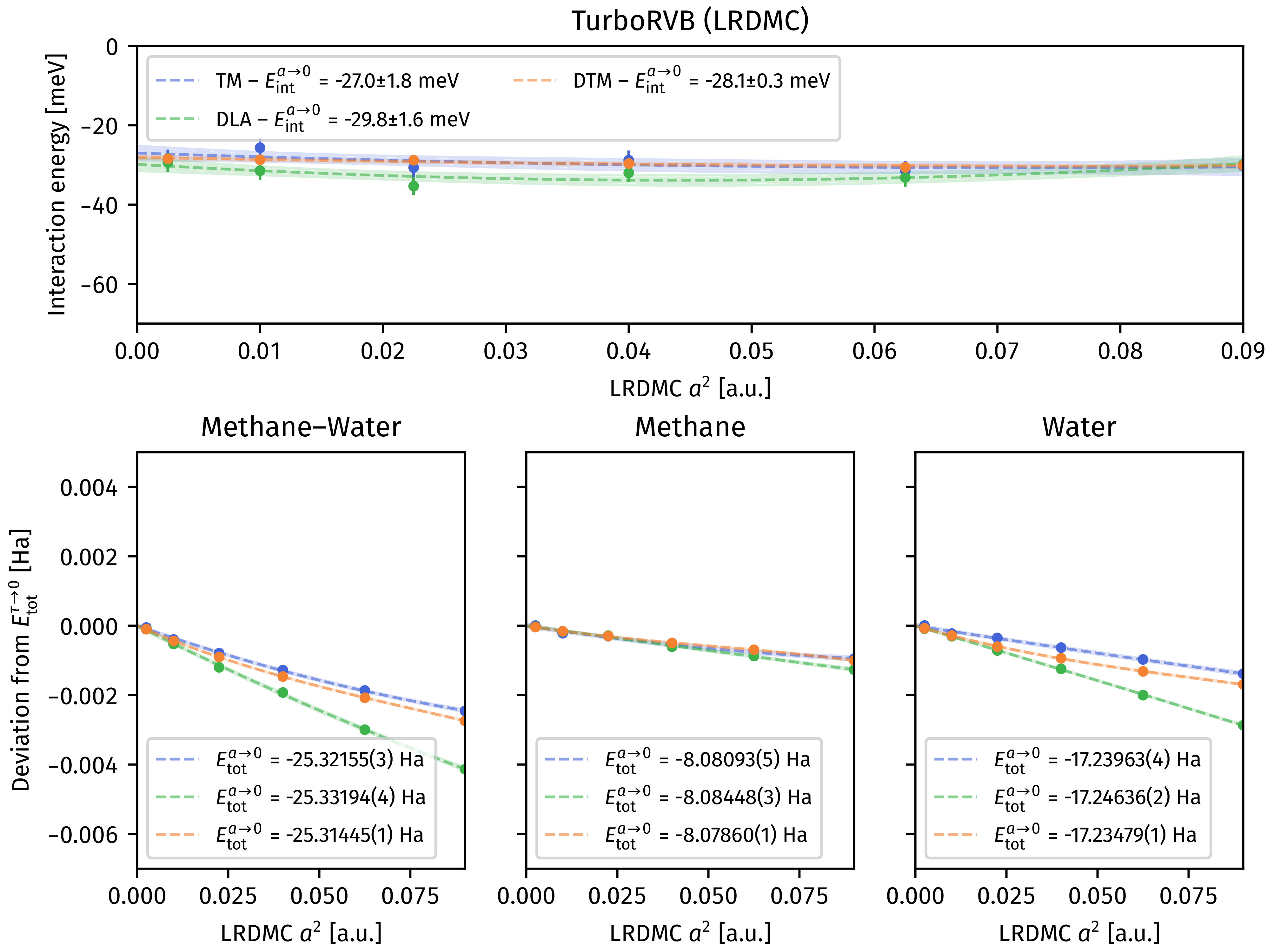}
    \caption{\label{fig:TurboRVB_LRDMC_energy_plots} The lattice-space dependence of the (a) methane-water interaction energy (b) methane-water dimer total energy, (c) isolated methane molecule total energy, (d) isolated water molecule total energy in the TurboRVB-LRDMC code across the TM, DLA, DTM algorithm(s).}
\end{figure*}

\clearpage

\section{\label{sec:dmc_validate} Validating size consistency}



The size-consistency error (Eq.~\ref{eq:sce}) is expected to converge to zero for infinitesimal time steps
if the wave function of the widely separated dimer (denoted as methane$---$water) is the product of the wave functions of the monomers. In schemes where the Hamiltonian does not depend on the Jastrow factor (DLA and DTM), one needs the determinantal component to be separable, while also the Jastrow factor should be separable in the LA and TM treatments of the pseudopotential.
%
%
%

\begin{figure*}[!ht]
    \includegraphics[width=6.66in]{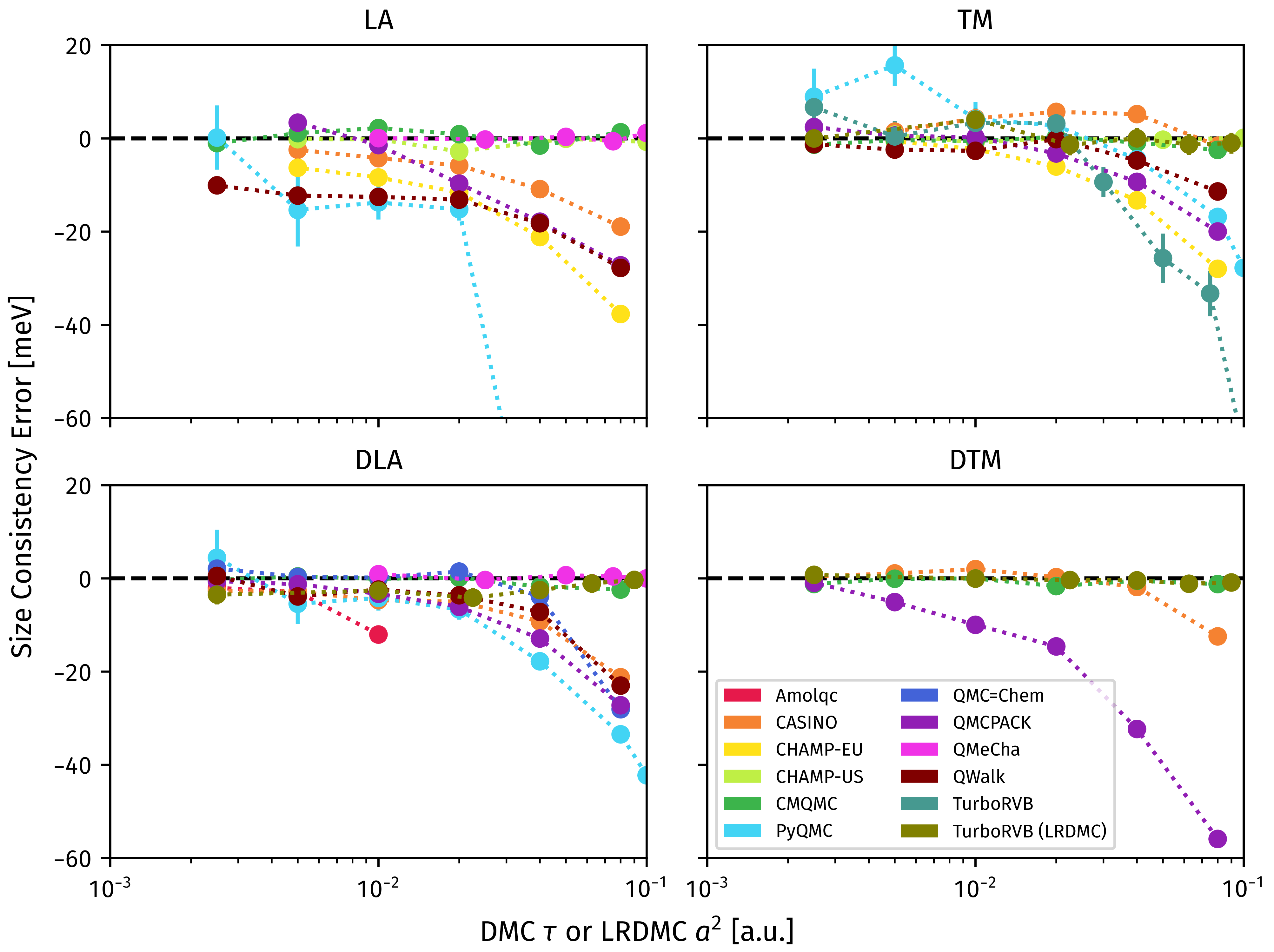}
    \caption{\label{fig:Size_consist_error} Convergence of the size-consistency error as a function of time step for the (a) LA, (b) TM, (c) DLA, and (d) DTM pseudopotential localization schemes.}
\end{figure*}

As shown in Fig.~\ref{fig:Size_consist_error},
%
the TM and DTM schemes have smaller time-step errors for the SCE than LA and DLA. Furthermore, even though not all codes employ a Jastrow factor for the dimer which is a product of the Jastrow factors of the monomers, TM yields small SCE in the $\tau\to 0$ limit for all codes.

We note that three codes (CHAMP-US, CMQMC, QMeCha) have particularly small time-step errors for SCE since they implement a scheme to ensure that the reweighting factor for the dimer is a product of the factors of the monomers. This can be done fragment by fragment (CHAMP-US~\cite{Umrigar_time_step_2021}) or electron by electron (CMQMC and QMeCha). 



\clearpage

\bibliographystyle{ieeetr}
\bibliography{ref}

%% file: Tables/Table_Eint.tex
\begin{tabular}{|lc|lr|r|r|cccc|}
\toprule
Code & Method & $\tau_\text{min}$ & $E_\text{int}^{\tau_\text{min}}$ & $E_\text{int}^{\tau\to 0}$ & $\Delta E_\text{int}^{\tau_\text{min},0}$ & $d_\text{poly}$ & $N$ & $\chi^2_\textrm{red}$ & RMSR \\
\midrule
Amolqc           &  DLA  & 0.0025 & -29.8$\pm$1.4 & -27.9$\pm$0.6 & -1.9$\pm$1.5 & 2 & 5 & 0.2 & 0.4 \\
CASINO           &  LA   & 0.0050 & -33.3$\pm$1.0 & -32.8$\pm$1.2 & -0.6$\pm$1.6 & 2 & 5 & 1.1 & 0.8 \\
CASINO           &  TM   & 0.0050 & -27.2$\pm$1.1 & -27.2$\pm$0.9 & 0.0$\pm$1.5 & 2 & 5 & 0.7 & 0.6 \\
CASINO           &  DLA  & 0.0025 & -30.3$\pm$1.4 & -31.2$\pm$1.4 & 0.9$\pm$1.9 & 2 & 6 & 1.7 & 1.2 \\
CASINO           &  DTM  & 0.0025 & -27.6$\pm$1.1 & -26.8$\pm$1.0 & -0.7$\pm$1.4 & 2 & 6 & 1.2 & 0.9 \\
CHAMP-EU         &  LA   & 0.0050 & -27.5$\pm$0.7 & -25.1$\pm$0.2 & -2.4$\pm$0.8 & 2 & 5 & 0.1 & 0.1 \\
CHAMP-EU         &  TM   & 0.0050 & -27.9$\pm$0.4 & -26.8$\pm$0.5 & -1.1$\pm$0.6 & 2 & 5 & 1.2 & 0.3 \\
CHAMP-US         &  LA   & 0.0050 & -23.4$\pm$2.2 & -25.8$\pm$3.1 & 2.5$\pm$3.8 & 2 & 5 & 2.7 & 1.8 \\
CHAMP-US         &  TM   & 0.0050 & -26.8$\pm$1.9 & -29.0$\pm$1.9 & 2.2$\pm$2.7 & 2 & 5 & 1.6 & 1.2 \\
CMQMC            &  LA   & 0.0025 & -25.7$\pm$2.0 & -25.7$\pm$1.8 & 0.0$\pm$2.7 & 2 & 6 & 2.0 & 1.5 \\
CMQMC            &  TM   & 0.0025 & -29.3$\pm$0.9 & -28.2$\pm$0.9 & -1.1$\pm$1.3 & 2 & 6 & 1.5 & 0.8 \\
CMQMC            &  DLA  & 0.0025 & -29.2$\pm$1.0 & -27.9$\pm$1.2 & -1.3$\pm$1.5 & 2 & 6 & 2.7 & 1.1 \\
CMQMC            &  DTM  & 0.0025 & -27.7$\pm$0.9 & -27.8$\pm$0.3 & 0.1$\pm$1.0 & 2 & 6 & 0.2 & 0.3 \\
PyQMC            &  LA   & 0.0025 & -21.6$\pm$4.1 & -18.9$\pm$0.9 & -2.7$\pm$4.2 & 2 & 7 & 0.2 & 0.8 \\
PyQMC            &  TM   & 0.0025 & -30.0$\pm$4.0 & -23.2$\pm$1.9 & -6.8$\pm$4.5 & 2 & 7 & 0.9 & 2.6 \\
PyQMC            &  DLA  & 0.0025 & -29.1$\pm$3.4 & -28.4$\pm$1.3 & -0.8$\pm$3.7 & 2 & 7 & 0.6 & 1.0 \\
QMC=Chem         &  DLA  & 0.0025 & -24.4$\pm$2.0 & -27.4$\pm$3.1 & 3.0$\pm$3.7 & 3 & 6 & 2.2 & 1.5 \\
QMCPACK          &  LA   & 0.0050 & -20.9$\pm$1.3 & -17.5$\pm$1.6 & -3.4$\pm$2.1 & 2 & 5 & 1.5 & 1.0 \\
QMCPACK          &  TM   & 0.0025 & -24.1$\pm$1.1 & -24.5$\pm$0.9 & 0.4$\pm$1.4 & 2 & 6 & 1.1 & 0.8 \\
QMCPACK          &  DLA  & 0.0025 & -29.7$\pm$1.2 & -28.9$\pm$0.3 & -0.8$\pm$1.2 & 2 & 6 & 0.1 & 0.3 \\
QMCPACK          &  DTM  & 0.0025 & -29.7$\pm$1.3 & -29.0$\pm$1.1 & -0.7$\pm$1.7 & 2 & 6 & 1.3 & 1.1 \\
QMeCha           &  LA   & 0.0100 & -33.7$\pm$1.0 & -35.0$\pm$0.3 & 1.4$\pm$1.0 & 2 & 5 & 0.0 & 0.1 \\
QMeCha           &  DLA  & 0.0100 & -29.9$\pm$0.9 & -30.6$\pm$0.7 & 0.6$\pm$1.1 & 2 & 5 & 0.3 & 0.3 \\
QWalk            &  LA   & 0.0025 & -38.3$\pm$1.1 & -38.8$\pm$0.6 & 0.5$\pm$1.2 & 2 & 6 & 0.5 & 0.5 \\
QWalk            &  TM   & 0.0025 & -29.7$\pm$1.0 & -29.6$\pm$0.4 & -0.1$\pm$1.1 & 2 & 6 & 0.3 & 0.4 \\
QWalk            &  DLA  & 0.0025 & -27.7$\pm$1.1 & -29.5$\pm$1.5 & 1.8$\pm$1.9 & 2 & 6 & 3.1 & 1.4 \\
TurboRVB-DMC     &  TM   & 0.0025 & -27.7$\pm$3.6 & -28.3$\pm$1.9 & 0.6$\pm$4.1 & 3 & 8 & 0.3 & 1.4 \\
TurboRVB-LRDMC   &  TM   & 0.0025 & -28.6$\pm$2.4 & -27.0$\pm$1.8 & -1.6$\pm$3.0 & 2 & 6 & 0.7 & 1.4 \\
TurboRVB-LRDMC   &  DLA  & 0.0025 & -29.4$\pm$2.3 & -29.8$\pm$1.6 & 0.5$\pm$2.8 & 2 & 6 & 0.6 & 1.3 \\
TurboRVB-LRDMC   &  DTM  & 0.0025 & -28.4$\pm$0.8 & -28.1$\pm$0.3 & -0.4$\pm$0.9 & 2 & 6 & 0.3 & 0.3 \\
\bottomrule
\end{tabular}

%% file: Tables/Table_Etot_m.tex
\begin{tabular}{|lc|lr|r|r|cccc|}
\toprule
Code & Method & $\tau_\text{min}$ & $E_\text{tot}^{\tau_\text{min}}$ & $E_\text{tot}^{\tau\to 0}$ & $\Delta E_\text{tot}^{\tau_\text{min},0}$ & $d_\text{poly}$ & $N$ & $\chi^2_\textrm{red}$ & RMSR \\
\midrule
Amolqc           &  DLA  & 0.0025 & -8.08445(3) & -8.08450(4) & 0.00005(5) & 2 & 5 & 1.3 & 0.00002 \\
CASINO           &  LA   & 0.0050 & -8.08191(2) & -8.08194(1) & 0.00003(2) & 2 & 5 & 0.2 & 0.00001 \\
CASINO           &  TM   & 0.0025 & -8.08101(2) & -8.08101(2) & 0.00000(3) & 2 & 6 & 2.3 & 0.00002 \\
CASINO           &  DLA  & 0.0025 & -8.08444(2) & -8.08444(2) & 0.00000(3) & 2 & 6 & 1.5 & 0.00002 \\
CASINO           &  DTM  & 0.0025 & -8.07851(2) & -8.07856(1) & 0.00005(2) & 2 & 6 & 0.6 & 0.00001 \\
CHAMP-EU         &  LA   & 0.0050 & -8.08237(1) & -8.08243(2) & 0.00006(3) & 2 & 5 & 3.6 & 0.00001 \\
CHAMP-EU         &  TM   & 0.0050 & -8.08107(1) & -8.08108(1) & 0.00001(1) & 2 & 5 & 1.6 & 0.00001 \\
CHAMP-US         &  LA   & 0.0050 & -8.08252(4) & -8.08263(9) & 0.00011(10) & 3 & 5 & 2.9 & 0.00003 \\
CHAMP-US         &  TM   & 0.0050 & -8.08107(3) & -8.08103(3) & -0.00004(4) & 2 & 5 & 2.0 & 0.00002 \\
CMQMC            &  LA   & 0.0025 & -8.08295(5) & -8.08309(6) & 0.00013(8) & 3 & 6 & 2.6 & 0.00003 \\
CMQMC            &  TM   & 0.0025 & -8.08095(2) & -8.08090(1) & -0.00005(2) & 2 & 6 & 0.9 & 0.00001 \\
CMQMC            &  DLA  & 0.0025 & -8.08423(2) & -8.08438(5) & 0.00015(5) & 3 & 6 & 4.4 & 0.00002 \\
CMQMC            &  DTM  & 0.0025 & -8.07854(2) & -8.07848(2) & -0.00006(3) & 2 & 6 & 1.2 & 0.00002 \\
PyQMC            &  LA   & 0.0025 & -8.08306(4) & -8.08306(3) & 0.00000(5) & 2 & 7 & 1.7 & 0.00003 \\
PyQMC            &  TM   & 0.0025 & -8.08099(4) & -8.08096(6) & -0.00004(7) & 2 & 7 & 8.4 & 0.00005 \\
PyQMC            &  DLA  & 0.0025 & -8.08438(4) & -8.08434(4) & -0.00003(5) & 2 & 7 & 3.3 & 0.00004 \\
QMC=Chem         &  DLA  & 0.0025 & -8.08439(2) & -8.08451(1) & 0.00013(3) & 3 & 6 & 0.3 & 0.00001 \\
QMCPACK          &  LA   & 0.0050 & -8.08306(2) & -8.08340(8) & 0.00033(8) & 3 & 5 & 3.5 & 0.00002 \\
QMCPACK          &  TM   & 0.0025 & -8.08121(2) & -8.08116(2) & -0.00006(3) & 2 & 6 & 2.0 & 0.00002 \\
QMCPACK          &  DLA  & 0.0025 & -8.08447(2) & -8.08449(2) & 0.00002(3) & 2 & 6 & 2.4 & 0.00002 \\
QMCPACK          &  DTM  & 0.0025 & -8.07864(2) & -8.07858(2) & -0.00006(3) & 2 & 6 & 1.7 & 0.00002 \\
QMeCha           &  LA   & 0.0100 & -8.08248(2) & -8.08264(9) & 0.00015(9) & 3 & 5 & 5.1 & 0.00002 \\
QMeCha           &  DLA  & 0.0100 & -8.08443(1) & -8.08447(3) & 0.00004(3) & 3 & 5 & 1.4 & 0.00001 \\
QWalk            &  LA   & 0.0025 & -8.08194(2) & -8.08196(3) & 0.00003(3) & 3 & 6 & 2.0 & 0.00001 \\
QWalk            &  TM   & 0.0025 & -8.08100(2) & -8.08099(2) & -0.00001(2) & 2 & 6 & 1.7 & 0.00001 \\
QWalk            &  DLA  & 0.0025 & -8.08442(2) & -8.08439(3) & -0.00003(3) & 2 & 6 & 3.8 & 0.00002 \\
TurboRVB-DMC     &  TM   & 0.0025 & -8.08100(6) & -8.08095(6) & -0.00005(8) & 3 & 8 & 1.2 & 0.00004 \\
TurboRVB-LRDMC   &  TM   & 0.0025 & -8.08093(6) & -8.08093(5) & 0.00000(8) & 2 & 6 & 0.8 & 0.00004 \\
TurboRVB-LRDMC   &  DLA  & 0.0025 & -8.08451(6) & -8.08448(3) & -0.00003(6) & 2 & 6 & 0.3 & 0.00002 \\
TurboRVB-LRDMC   &  DTM  & 0.0025 & -8.07863(2) & -8.07860(1) & -0.00004(2) & 3 & 6 & 0.5 & 0.00001 \\
\bottomrule
\end{tabular}

%% file: Tables/Table_Etot_w.tex
\begin{tabular}{|lc|lr|r|r|cccc|}
\toprule
Code & Method & $\tau_\text{min}$ & $E_\text{tot}^{\tau_\text{min}}$ & $E_\text{tot}^{\tau\to 0}$ & $\Delta E_\text{tot}^{\tau_\text{min},0}$ & $d_\text{poly}$ & $N$ & $\chi^2_\textrm{red}$ & RMSR \\
\midrule
Amolqc           &  DLA  & 0.0025 & -17.24621(3) & -17.24638(2) & 0.00017(4) & 2 & 5 & 0.5 & 0.00001 \\
CASINO           &  LA   & 0.0050 & -17.24091(2) & -17.24104(4) & 0.00013(5) & 3 & 5 & 1.1 & 0.00001 \\
CASINO           &  TM   & 0.0025 & -17.23944(2) & -17.23950(4) & 0.00006(5) & 3 & 6 & 3.4 & 0.00002 \\
CASINO           &  DLA  & 0.0025 & -17.24624(3) & -17.24634(2) & 0.00010(3) & 3 & 6 & 0.4 & 0.00001 \\
CASINO           &  DTM  & 0.0005 & -17.23475(7) & -17.23473(2) & -0.00002(7) & 3 & 8 & 0.5 & 0.00002 \\
CHAMP-EU         &  LA   & 0.0050 & -17.24136(1) & -17.24156(1) & 0.00020(2) & 3 & 5 & 0.2 & 0.00000 \\
CHAMP-EU         &  TM   & 0.0050 & -17.23963(1) & -17.23967(1) & 0.00004(2) & 3 & 5 & 1.0 & 0.00000 \\
CHAMP-US         &  LA   & 0.0050 & -17.24146(4) & -17.24167(1) & 0.00021(4) & 3 & 5 & 0.0 & 0.00000 \\
CHAMP-US         &  TM   & 0.0050 & -17.23963(4) & -17.23956(4) & -0.00007(6) & 2 & 5 & 2.0 & 0.00003 \\
CMQMC            &  LA   & 0.0025 & -17.24177(2) & -17.24206(10) & 0.00029(11) & 3 & 6 & 19.2 & 0.00005 \\
CMQMC            &  TM   & 0.0025 & -17.23959(2) & -17.23939(2) & -0.00020(3) & 2 & 6 & 1.2 & 0.00002 \\
CMQMC            &  DLA  & 0.0025 & -17.24555(2) & -17.24588(7) & 0.00034(8) & 3 & 6 & 10.4 & 0.00004 \\
CMQMC            &  DTM  & 0.0025 & -17.23487(2) & -17.23460(1) & -0.00028(2) & 2 & 6 & 0.3 & 0.00001 \\
PyQMC            &  LA   & 0.0025 & -17.24190(4) & -17.24200(5) & 0.00010(6) & 3 & 7 & 2.2 & 0.00003 \\
PyQMC            &  TM   & 0.0025 & -17.23932(4) & -17.23943(4) & 0.00011(5) & 3 & 7 & 1.6 & 0.00003 \\
PyQMC            &  DLA  & 0.0025 & -17.24597(4) & -17.24607(7) & 0.00010(8) & 3 & 7 & 5.5 & 0.00004 \\
QMC=Chem         &  DLA  & 0.0025 & -17.24574(2) & -17.24605(7) & 0.00030(8) & 3 & 6 & 9.1 & 0.00004 \\
QMCPACK          &  LA   & 0.0050 & -17.24168(2) & -17.24198(7) & 0.00030(7) & 3 & 5 & 2.5 & 0.00002 \\
QMCPACK          &  TM   & 0.0025 & -17.23999(2) & -17.23977(2) & -0.00023(3) & 3 & 6 & 0.7 & 0.00001 \\
QMCPACK          &  DLA  & 0.0025 & -17.24625(3) & -17.24630(5) & 0.00006(5) & 3 & 6 & 2.4 & 0.00002 \\
QMCPACK          &  DTM  & 0.0025 & -17.23520(2) & -17.23482(3) & -0.00039(3) & 3 & 6 & 1.1 & 0.00001 \\
QMeCha           &  LA   & 0.0100 & -17.24161(2) & -17.24139(3) & -0.00022(4) & 3 & 5 & 0.6 & 0.00001 \\
QMeCha           &  DLA  & 0.0100 & -17.24627(2) & -17.24608(10) & -0.00018(11) & 3 & 5 & 5.5 & 0.00002 \\
QWalk            &  LA   & 0.0025 & -17.24087(2) & -17.24087(4) & 0.00000(5) & 3 & 6 & 3.2 & 0.00002 \\
QWalk            &  TM   & 0.0025 & -17.23953(2) & -17.23948(3) & -0.00005(4) & 3 & 6 & 2.2 & 0.00002 \\
QWalk            &  DLA  & 0.0025 & -17.24621(2) & -17.24624(4) & 0.00002(5) & 3 & 6 & 2.6 & 0.00002 \\
TurboRVB-DMC     &  TM   & 0.0025 & -17.23936(7) & -17.23952(6) & 0.00016(9) & 3 & 8 & 0.9 & 0.00005 \\
TurboRVB-LRDMC   &  TM   & 0.0025 & -17.23963(5) & -17.23963(4) & 0.00000(7) & 2 & 6 & 0.8 & 0.00003 \\
TurboRVB-LRDMC   &  DLA  & 0.0025 & -17.24645(5) & -17.24636(2) & -0.00009(5) & 2 & 6 & 0.1 & 0.00001 \\
TurboRVB-LRDMC   &  DTM  & 0.0025 & -17.23487(1) & -17.23479(1) & -0.00007(2) & 3 & 6 & 0.7 & 0.00001 \\
\bottomrule
\end{tabular}

%% file: Tables/Table_Etot_mw.tex
\begin{tabular}{|lc|lr|r|r|cccc|}
\toprule
Code & Method & $\tau_\text{min}$ & $E_\text{tot}^{\tau_\text{min}}$ & $E_\text{tot}^{\tau\to 0}$ & $\Delta E_\text{tot}^{\tau_\text{min},0}$ & $d_\text{poly}$ & $N$ & $\chi^2_\textrm{red}$ & RMSR \\
\midrule
Amolqc           &  DLA  & 0.0025 & -25.33176(3) & -25.33190(3) & 0.00015(4) & 2 & 5 & 0.8 & 0.00002 \\
CASINO           &  LA   & 0.0050 & -25.32404(3) & -25.32427(4) & 0.00023(5) & 3 & 5 & 0.7 & 0.00001 \\
CASINO           &  TM   & 0.0050 & -25.32124(3) & -25.32149(4) & 0.00024(5) & 3 & 5 & 0.5 & 0.00001 \\
CASINO           &  DLA  & 0.0025 & -25.33179(4) & -25.33189(5) & 0.00009(7) & 3 & 6 & 1.5 & 0.00003 \\
CASINO           &  DTM  & 0.0025 & -25.31404(3) & -25.31432(3) & 0.00028(5) & 3 & 6 & 1.0 & 0.00002 \\
CHAMP-EU         &  LA   & 0.0050 & -25.32474(2) & -25.32497(3) & 0.00023(4) & 3 & 5 & 0.6 & 0.00001 \\
CHAMP-EU         &  TM   & 0.0050 & -25.32172(1) & -25.32174(1) & 0.00002(2) & 3 & 5 & 0.5 & 0.00000 \\
CHAMP-US         &  LA   & 0.0050 & -25.32484(6) & -25.32501(6) & 0.00017(8) & 3 & 5 & 0.4 & 0.00002 \\
CHAMP-US         &  TM   & 0.0050 & -25.32169(5) & -25.32165(2) & -0.00004(5) & 2 & 5 & 0.2 & 0.00001 \\
CMQMC            &  LA   & 0.0025 & -25.32567(6) & -25.32611(13) & 0.00044(14) & 3 & 6 & 6.6 & 0.00006 \\
CMQMC            &  TM   & 0.0025 & -25.32162(2) & -25.32137(3) & -0.00025(4) & 3 & 6 & 1.9 & 0.00002 \\
CMQMC            &  DLA  & 0.0025 & -25.33085(2) & -25.33128(14) & 0.00043(15) & 3 & 6 & 37.6 & 0.00007 \\
CMQMC            &  DTM  & 0.0025 & -25.31443(2) & -25.31411(3) & -0.00033(4) & 3 & 6 & 2.3 & 0.00002 \\
PyQMC            &  LA   & 0.0025 & -25.32576(14) & -25.32584(6) & 0.00008(15) & 3 & 7 & 0.4 & 0.00004 \\
PyQMC            &  TM   & 0.0025 & -25.32141(14) & -25.32131(12) & -0.00011(18) & 3 & 7 & 1.9 & 0.00010 \\
PyQMC            &  DLA  & 0.0025 & -25.33141(11) & -25.33146(13) & 0.00005(17) & 3 & 7 & 2.8 & 0.00007 \\
QMC=Chem         &  DLA  & 0.0025 & -25.33103(7) & -25.33157(3) & 0.00055(7) & 3 & 6 & 0.2 & 0.00002 \\
QMCPACK          &  LA   & 0.0050 & -25.32551(4) & -25.32592(5) & 0.00041(6) & 3 & 5 & 0.5 & 0.00001 \\
QMCPACK          &  TM   & 0.0025 & -25.32209(3) & -25.32176(2) & -0.00033(3) & 3 & 6 & 0.2 & 0.00001 \\
QMCPACK          &  DLA  & 0.0025 & -25.33180(3) & -25.33191(3) & 0.00011(4) & 3 & 6 & 0.7 & 0.00002 \\
QMCPACK          &  DTM  & 0.0025 & -25.31493(4) & -25.31443(7) & -0.00050(8) & 3 & 6 & 3.1 & 0.00004 \\
QMeCha           &  LA   & 0.0100 & -25.32533(3) & -25.32529(12) & -0.00004(13) & 3 & 5 & 4.5 & 0.00002 \\
QMeCha           &  DLA  & 0.0100 & -25.33179(2) & -25.33167(8) & -0.00013(8) & 3 & 5 & 1.9 & 0.00002 \\
QWalk            &  LA   & 0.0025 & -25.32421(3) & -25.32425(3) & 0.00004(4) & 3 & 6 & 0.9 & 0.00002 \\
QWalk            &  TM   & 0.0025 & -25.32163(3) & -25.32156(3) & -0.00007(4) & 3 & 6 & 1.1 & 0.00002 \\
QWalk            &  DLA  & 0.0025 & -25.33165(3) & -25.33173(3) & 0.00008(4) & 3 & 6 & 1.0 & 0.00002 \\
TurboRVB-DMC     &  TM   & 0.0025 & -25.32139(10) & -25.32151(6) & 0.00012(12) & 3 & 8 & 0.5 & 0.00005 \\
TurboRVB-LRDMC   &  TM   & 0.0025 & -25.32160(4) & -25.32155(3) & -0.00005(5) & 2 & 6 & 0.9 & 0.00003 \\
TurboRVB-LRDMC   &  DLA  & 0.0025 & -25.33204(4) & -25.33194(4) & -0.00010(6) & 2 & 6 & 1.4 & 0.00004 \\
TurboRVB-LRDMC   &  DTM  & 0.0025 & -25.31455(2) & -25.31445(1) & -0.00010(2) & 3 & 6 & 0.4 & 0.00001 \\
\bottomrule
\end{tabular}

%% file: Tables/Table_mean_Eint.tex
\begin{tabular}{|l c c c c c c|}
\toprule
 & Mean & No. codes & Max-dist & Max-error & MAE & RMSE \\
\midrule
LA & -27.5$\pm$7.2 & 8 & 21.3 & 11.3 & 6.1 & 7.1 \\
TM & -27.1$\pm$2.4 & 9 & 6.4 & 3.9 & 1.5 & 2.0 \\
DLA & -29.1$\pm$1.9 & 9 & 3.8 & 2.2 & 1.1 & 1.2 \\
DTM & -27.9$\pm$1.1 & 4 & 2.1 & 1.1 & 0.6 & 0.8 \\
\bottomrule
\end{tabular}

%% file: Tables/Table_mean_Etot.tex
\begin{tabular}{|l c c c|}
\toprule
 & Methane & Water & Methane-Water \\
\midrule
LA & -8.08264$\pm$0.00050 & -17.24157$\pm$0.00042 & -25.32521$\pm$0.00068 \\
TM & -8.08100$\pm$0.00008 & -17.23955$\pm$0.00012 & -25.32155$\pm$0.00015 \\
DLA & -8.08445$\pm$0.00007 & -17.24619$\pm$0.00017 & -25.33171$\pm$0.00023 \\
DTM & -8.07855$\pm$0.00005 & -17.23473$\pm$0.00009 & -25.31433$\pm$0.00014 \\
\bottomrule
\end{tabular}

%% file: main.bbl
\begin{thebibliography}{100}
\bibitem{note_time_step}
Comparable values of the interaction energy (within a statistical uncertainty
  of the order of the meV) are obtained for the smallest time step $\tau =
  0.0025 \text{a.u.}$ for the codes and localization schemes where such a time
  step is used.
\end{thebibliography}
